\documentclass[preprint, 12pt, sort&compress]{elsarticle}

\usepackage{graphicx}
\usepackage{amsmath, amssymb, bm}
\usepackage[utf8]{inputenc}
\usepackage[english]{babel}

\usepackage{subcaption}
\usepackage{indentfirst}
\usepackage{booktabs}
\usepackage{float}
\usepackage{multirow}
\usepackage{hyperref}
\usepackage{longtable}
\newcounter{savedtable}
\usepackage[nameinlink]{cleveref}
\Crefname{figure}{Fig.}{Figs.}
\numberwithin{equation}{section}
\newcommand{\pd}[2]{ \frac{\partial #1}{\partial #2} }
\newcommand{\dint}{\,\mathrm{d}}
\newcommand{\wmin}{w_\mathrm{min}}
\renewcommand{\div}{\mathrm{div}\,}

\newcommand{\bn}{\boldsymbol{n}} 
\newcommand{\bq}{\boldsymbol{q}} 

\newcommand{\bbL}{\mathbb{L}} 
\newcommand{\bbE}{\mathbb{E}} 
\newcommand{\bbI}{\mathbb{I}} 

\newcommand{\Cprime}{C'} 
\newcommand{\muprime}{\mu'} 

\journal{Geomechanics for Energy and the Environment}

\begin{document}

\begin{frontmatter}

\title{Acceleration methods for the planar 3D ILSA hydraulic fracturing model}

\author[lihsb]{V.I.~Shukalo\corref{cor1}}
\ead{vvvshukalo@gmail.com}

\author[lihsb]{A.V.~Valov}
\ead{a.valov1705@gmail.com}

\author[lihsb]{A.N.~Baykin}
\ead{alexey.baykin@gmail.com}

\cortext[cor1]{Corresponding author}

\affiliation[lihsb]{organization={Lavrentyev Institute of Hydrodynamics SB RAS},
            addressline={pr.\ Lavrentyeva 15},
            city={Novosibirsk},
            postcode={630090},
            country={Russia}}

\begin{abstract}
    Planar 3D models of hydraulic fracturing provide a practical balance between models with restrictive geometric assumptions and fully 3D simulators, capturing fractures with arbitrary planar footprints at moderate computational cost. Nevertheless, applications such as treatment design optimization and mini-frac test interpretation require large ensembles of simulations, for which the cost of planar 3D models remains a significant bottleneck. This work presents acceleration strategies for the planar 3D Implicit Level Set Algorithm (ILSA) to reduce simulation runtime while preserving numerical accuracy. A unified planar 3D ILSA scheme that consolidates the nested loops of the elastohydrodynamic solver and the front tracking algorithm into a single iterative process is introduced. A matrix splitting approach is applied to the linearized elastohydrodynamic system, moving the dense part of the elasticity operator to the right-hand side and yielding a sparse system matrix that can be solved more efficiently. Anderson acceleration is incorporated into the solution of the elastohydrodynamic system to improve convergence. A predictor--corrector scheme is additionally examined in combination with these methods. Each technique is evaluated individually and in combination on both the reference and unified schemes across five benchmark cases. Numerical experiments demonstrate that the unified scheme alone delivers an average $2.3\times$ speed-up, reaching $3.8\times$ for the Sandglass case. The combined application of all techniques achieves an average $4.2\times$ speed-up and up to $14.4\times$ for the Sandglass case. The schemes are also assessed during fracture closure, where the matrix splitting and Anderson acceleration prove most effective, together reaching up to $3.4\times$.
\end{abstract}


\end{frontmatter}

\section*{Nomenclature}
    \addcontentsline{toc}{section}{Nomenclature}

    \setcounter{savedtable}{\value{table}}
    \noindent
    \begin{longtable}{@{}p{0.18\linewidth}p{0.78\linewidth}@{}}
    $\mathcal{A}(t)$ & Fracture footprint at time $t$ \\
    $A_k$ & Contact set at iteration $k$ \\
    $\mathcal{C}(t)$ & Fracture front at time $t$ \\
    $C_L$ & Carter leak-off coefficient, m/s$^{1/2}$ \\
    $C'$ & Scaled leak-off coefficient, $C' = 2C_L$, m/s$^{1/2}$ \\
    $E$ & Young's modulus of the rock, Pa \\
    $E'$ & Plane strain modulus, $E' = E/(1-\nu^2)$, Pa \\
    $\mathbb{E}$ & Elasticity matrix \\
    $\mathbb{E}_{\text{expl}}$ & Explicit part of the elasticity matrix \\
    $\mathbb{E}_{\text{impl}}$ & Implicit part of the elasticity matrix \\
    $\mathbf{f}$ & Nonlinear fixed-point mapping \\
    $\mathbb{I}$ & Identity matrix \\
    $i, j$ & Indices of a mesh cell in the $x$- and $y$-directions \\
    $K_{I}$ & Stress intensity factor, Pa$\cdot$m$^{1/2}$ \\
    $K_{Ic}$ & Rock fracture toughness, Pa$\cdot$m$^{1/2}$ \\
    $K'$ & Scaled toughness, $K' = (32/\pi)^{1/2} K_{Ic}$, Pa$\cdot$m$^{1/2}$ \\
    $\mathbb{L}$ & Fluid flux matrix \\
    $m$ & Memory parameter of Anderson acceleration \\
    $N$ & Total number of fracture elements \\
    $\bn$ & Outward unit normal to the fracture front \\
    $p(x,y,t)$ & Fluid pressure inside the fracture, Pa \\
    $\mathbf{p}$ & Vector of discrete fluid pressures, Pa \\
    $P$ & Computational performance, $P = 1/T_{\text{run}}$, s$^{-1}$ \\
    $\bq$ & Fluid flux vector, m$^2$/s \\
    $Q(t)$ & Volumetric fluid injection rate, m$^3$/s \\
    $\mathbf{r}$ & Fixed-point residual, $\mathbf{r}(\mathbf{x}) = \mathbf{f}(\mathbf{x}) - \mathbf{x}$ \\
    $\mathbf{R}$ & Right-hand side vector of the elastohydrodynamic system \\
    $s$ & Distance from a point inside the fracture to the front, m \\
    $\mathbf{s}_k$ & Vector of front distances at iteration $k$, m \\
    $t$ & Time, s \\
    $t_0(x,y)$ & Time at which the fracture front reaches point $(x,y)$, s \\
    $\Delta t$ & Time step size, s \\
    $T_{\text{run}}$ & Total simulation time, s \\
    $T_n$ & Normal traction on the fracture surface, Pa \\
    $\mathbf{T}_n$ & Discrete normal traction vector, Pa \\
    $V$ & Normal velocity of the fracture front, m/s \\
    $w(x,y,t)$ & Fracture aperture, m \\
    $\mathbf{w}$ & Vector of discrete fracture apertures, m \\
    $w_{\mathrm{min}}$ & Minimum admissible fracture aperture, m \\
    $\mathbf{x}$ & Vector of unknowns of the fixed-point problem \\
    $x, y$ & In-plane Cartesian coordinates, m \\
    $\Delta x, \Delta y$ & Mesh cell sizes in the $x$- and $y$-directions, m \\[6pt]
    $\delta(x,y)$ & Two-dimensional Dirac delta function, m$^{-2}$ \\
    $\varepsilon_{\text{EHD}}$ & Convergence tolerance for the nonlinear elastohydrodynamic solver \\
    $\varepsilon_{\text{front}}$ & Convergence tolerance for the fracture front iterations \\
    $\mu$ & Dynamic fluid viscosity, Pa$\cdot$s \\
    $\mu'$ & Scaled fluid viscosity, $\mu' = 12\mu$, Pa$\cdot$s \\
    $\nu$ & Poisson's ratio of the rock \\
    $\sigma(y)$ & Minimum \textit{in-situ} compressive stress, Pa \\[6pt]
    $A$ & Active-constraint sub-block index (closed fracture elements) \\
    $C$ & Channel sub-block index (interior elements fully inside the fracture and not under active constraint) \\
    $T$ & Tip sub-block index (elements intersected by the fracture front) \\
    $k$ & Iteration index \\
    $\text{mod}$ & Modified algorithm \\
    $\text{ref}$ & Reference planar 3D ILSA scheme \\[6pt]
    BiCGStab & Biconjugate Gradient Stabilized method \\
    CPU & Central processing unit \\
    EHD & Elastohydrodynamic \\
    GPU & Graphics processing unit \\
    ILSA & Implicit Level Set Algorithm \\
    ILU & Incomplete LU factorization \\
    KGD & Khristianovich--Geertsma--de Klerk fracture model \\
    P3D & Pseudo-three-dimensional fracture model \\
    PKN & Perkins--Kern--Nordgren fracture model \\
    SparseLU & Sparse LU factorization solver \\
    \end{longtable}
    \setcounter{table}{\value{savedtable}}

\section{Introduction}

    Hydraulic fracturing is one of the key technologies enabling production from unconventional reservoirs. Accurate and efficient numerical simulation of hydraulic fracture propagation is essential for treatment design and field data interpretation. Optimizing a hydraulic fracturing job requires sweeping a broad parameter space, including fluid volumes, pumping rates, proppant concentrations, and pumping stage durations, which involves performing numerous simulations. The interpretation of diagnostic mini-frac tests relies on measured pressure curves to estimate reservoir properties, such as minimum \textit{in-situ} stress, closure time, leak-off coefficient, and permeability~\cite{nolte1979determination, nolte1986general, wang2017new, baykin2021mini}. Determining these parameters requires multiple pressure calculations to be matched to the field pressure history measured during the mini-frac test. Thus, for both design optimization and data interpretation, fast and reliable hydraulic fracturing simulators are essential to perform a series of simulations within a reasonable computational time.

    The first models of hydraulic fracturing were developed in the middle of the twentieth century. Among the earliest formulations are the classical KGD model~\cite{Khristianovic_KGD_1955, Geertsma_KGD_1969}, the PKN model~\cite{Nordgren_PKN_1972, Perkins_PKN_Fracture_1961}, and the radial (penny-shaped) fracture model~\cite{Geertsma_KGD_1969, Abe_Radial_1976}. These models rely on simplified geometric assumptions to enable semi-analytical solutions for fracture propagation. In the KGD model, the fracture propagates under plane-strain elastic conditions. The PKN model describes a vertical fracture of a fixed height propagating horizontally, assuming a plane-strain approximation in each vertical cross-section. The radial model simulates the propagation of a penny-shaped fracture in a homogeneous medium. The fluid flow within the fracture is assumed to be one-dimensional for all these models. Owing to their simplicity, these models are computationally very efficient. However, they are limited to relatively simple fracture geometries and homogeneous formations, and therefore cannot represent realistic geological structures or complex fracture footprints.

    To address some of these limitations, the pseudo-3D (P3D) model~\cite{Settari_P3D_1986, Palmer_P3D_1983} and its variants~\cite{Dontsov_EP3D_2015, Cohen_Stacked_height_P3D_2015, Zhang_P3D_layered_rock_2017, Zhang_P3D_layered_elasticity_2018, mclennan1985pseudo} were developed. These models extend the PKN formulation by allowing the fracture height to grow along the fracture length. P3D models significantly improve the ability to simulate fracture propagation in layered reservoirs. However, these models usually rely on local elasticity, treating each vertical cross-section independently. They also assume one-dimensional fluid flow along the fracture length, neglecting vertical pressure gradients.

    A further step toward more realistic modeling is provided by planar 3D formulations~\cite{Lecampion_Zia_Pyfrac_2020, Barree_PL3D_1983, Clifton_PL3D_1981, Peirce_Detournay_ILSA_2008}. In planar 3D models, the fracture is assumed to remain planar, but its footprint can evolve with an arbitrary shape within the fracture plane. Fluid flow within the fracture is described by the two-dimensional lubrication equation, while the mechanical response of the surrounding rock is governed by the full three-dimensional elasticity operator, which introduces a non-local relation between fracture aperture and pressure. The ability of the fracture to propagate in any direction within the plane enables the modeling of geometries that pseudo-3D approaches cannot handle. A primary example is the ``sandglass'' fracture shape, which develops when the fracture front intersects a layer of elevated compressive stress. The higher stress in this layer increases the resistance to propagation, causing the fracture front to lag behind the adjacent lower-stress formations and resulting in a characteristic non-convex footprint.

    A more comprehensive physical description leads to fully three-dimensional hydraulic fracturing models~\cite{Cherny_Lapin_non_planar_fracture_2016, Baykin_Planar3D_Biot_2018, paul20183d}. In recent studies, such approaches have increasingly incorporated additional physical mechanisms, including poroelastic effects and simultaneous propagation and interaction of multiple fractures~\cite{kumar2016three}. A phase-field variational approach to fracture modeling is considered in~\cite{yoshioka2016variational}, where it is coupled with a reservoir simulator to study the influence of natural fractures. Fully coupled three-dimensional finite element models have also been developed to simulate the growth of multiple hydraulic fractures in permeable poroelastic formations, accounting simultaneously for laminar fluid flow within the fracture, Darcy flow in the surrounding reservoir, and poroelastic deformation of the rock matrix~\cite{salimzadeh2017finite}. In addition, advanced numerical techniques such as the Generalized Finite Element Method have been used to investigate the influence of the initial fracture orientation on the propagation of multiple interacting fractures~\cite{shauer2022three}. These approaches provide a highly detailed physical description of hydraulic fracturing but involve extremely high computational cost.

    Planar 3D models offer a practical balance between models with simplified geometries and fully three-dimensional ones. By coupling 2D fluid flow inside the fracture with 3D elastic response of the surrounding rock, these models simulate fractures with arbitrary footprints. This approach avoids the strong geometric assumptions of earlier models and the high computational cost of full 3D models. Owing to the ability to capture complex fracture geometries at a moderate computational cost, planar 3D formulations have become one of the most common approaches used in modern hydraulic fracturing simulators for treatment design and analysis~\cite{chen2022review}. However, despite being significantly cheaper than fully 3D models, planar 3D simulations remain computationally demanding. Therefore, reducing the computational cost of planar 3D hydraulic fracturing models remains an important practical challenge.

    Considerable effort has been devoted to accelerating hydraulic fracturing simulations. Much of this work focuses on the coupled nonlinear elastohydrodynamic system of the planar 3D model, which dominates the total computational cost. Two principal strategies have been proposed to accelerate its solution: reducing the cost per iteration through optimized linear solves, and decreasing the total number of iterations required for convergence. Advanced preconditioning and multigrid methods, such as localized Jacobian ILU~\cite{peirce2006localized} and dual-mesh multigrid~\cite{peirce2005dual}, reduce the cost of each individual solve. Large-scale runtime reductions are also achieved via parallel implementations, including GPU-based solvers~\cite{yin2018gpu,smilovich2021parallel,yang2025efficiency}. For problems involving dense integral operators, fast multipole methods~\cite{rezaei2019applications} are used to accelerate matrix-vector products. On the iteration-count side, accelerated fixed-point algorithms, notably the Aitken and Anderson methods~\cite{aksenov2021application,derbyshev2024anderson}, reduce the number of iterations and improve stability, while time integration and front tracking have been enhanced through adaptive time-stepping strategies~\cite{Lecampion_Zia_Pyfrac_2020}, super-time-stepping Runge--Kutta--Legendre methods~\cite{chen2020explicit}, and the predictor--corrector scheme~\cite{zia2019explicit}. Separately, machine learning surrogate models have recently been proposed to replace expensive computational components, indicating significant potential despite ongoing challenges with generalization and robustness~\cite{ma2025applicationML}.

    Nevertheless, the computational cost of planar 3D simulations remains a practical bottleneck, particularly when large ensembles of simulations are required. This motivates the development of further acceleration strategies applicable within existing planar 3D frameworks.

    In this work, an acceleration strategy for the planar 3D Implicit Level Set Algorithm (ILSA) scheme~\cite{dontsov2017multiscale} is introduced. It relies on a set of conceptually simple but efficient modifications. In particular, the computational framework is reformulated into a unified planar 3D ILSA scheme, which consolidates multiple nested loops into a single algorithmic structure and consequently reduces computational overhead. Furthermore, the linearized elastohydrodynamic system is treated by a matrix splitting approach involving decomposition into a sparse implicit part and a dense explicit part, thereby improving solver efficiency. Moreover, Anderson acceleration~\cite{anderson1965iterative} is incorporated to enhance the elastohydrodynamic solver convergence. In addition, the predictor--corrector scheme~\cite{zia2019explicit} is applied to improve the initial estimate of the fracture front and reduce the number of front iterations. Each technique is simple to implement and can be integrated into existing simulators. A detailed investigation of their individual and combined effects demonstrates that the combination of the proposed strategies typically reduces simulation time by $4.2\times$, and up to $14.4\times$ for complex fracture geometries, making large-scale hydraulic fracturing simulations considerably more practical.

    The paper is organized as follows. \Cref{sec:Reference_planar_3D_ILSA_description} outlines the reference planar 3D ILSA scheme. Then \Cref{sec:Performance_enhancing_approaches} presents the performance-enhancing approaches, namely the unified scheme, the matrix splitting, and Anderson acceleration. \Cref{sec:results} demonstrates the efficiency of the proposed methods through numerical experiments. The reference and unified schemes are compared, and the individual and combined effects of the matrix splitting, Anderson acceleration, and the predictor--corrector scheme are evaluated for both formulations. Finally, the behavior of the schemes during fracture recession and closure is examined in~\Cref{sec:closure_results}.

\section{Description of the reference planar 3D ILSA scheme}\label{sec:Reference_planar_3D_ILSA_description}

    \subsection{Mathematical formulation}

        \begin{figure}[h!]
            \centering
            \includegraphics[width=0.99 \linewidth]{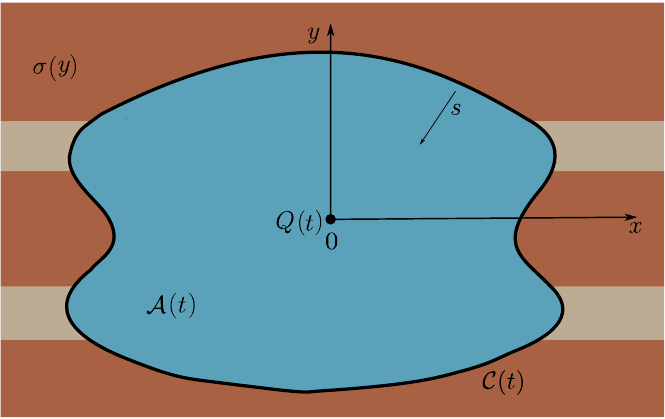}
            \caption{Schematic of a planar hydraulic fracture propagating in the $(x,y)$ plane. The surrounding rock is represented by a layered elastic medium with depth-varying \textit{in-situ} stress $\sigma(y)$. The fracture occupies footprint $\mathcal{A}(t)$, bounded by front $\mathcal{C}(t)$; $s$ is the distance to the front from a given interior point. The fluid is injected at a point source of volumetric rate $Q(t)$ located at the origin.}
            \label{fig:planar3d_fracture_scheme}
        \end{figure}

        We consider a hydraulic fracture propagating in the $(x,y)$ plane of a layered formation. \Cref{fig:planar3d_fracture_scheme} provides a schematic representation of the problem. The mathematical formulation describes the evolution of the fracture footprint $\mathcal{A}(t)$, bounded by the moving front $\mathcal{C}(t)$, along with the fracture aperture $w(x,y,t)$ and the fluid pressure $p(x,y,t)$.

        The rock is modeled as a linearly elastic medium with uniform elastic properties, characterized by Young's modulus $E$ and Poisson's ratio $\nu$. The toughness $K_{Ic}$ and leak-off coefficient $C_L$ are assumed to be spatially constant. However, the heterogeneity of these parameters can be incorporated into the algorithm without significant modifications. The minimum \textit{in-situ} compressive stress field $\sigma(y)$ acts normal to the fracture plane and is assumed to be piecewise constant across horizontal layers, varying only with depth. An incompressible Newtonian fluid of viscosity $\mu$ is injected at a point source located at the origin with volumetric flow rate $Q(t)$. The resulting fluid pressure $p(x,y,t)$ acts to separate the fracture faces, producing the aperture field $w(x,y,t)$. The fluid front is assumed to coincide with the fracture tip, as fluid lag is considered negligible under the typical \textit{in-situ} stress conditions found in field applications~\cite{garagash2000tip, lecampion2007implicit}. Fluid leak-off into the surrounding rock follows Carter's law~\cite{howard1957optimum}. Gravity effects are neglected. A detailed discussion of the typical assumptions underlying planar three-dimensional hydraulic fracture models can be found in~\cite{Peirce_Detournay_ILSA_2008, dontsov2017multiscale, detournay2016mechanics,lecampion2018numerical}.

        To simplify notation, the following scaled material and fluid parameters are introduced
        \begin{equation*}
            E' = \frac{E}{1-\nu^2}, \qquad \mu' = 12 \mu, \qquad K' = \left( \frac{32}{\pi} \right)^{1/2} K_{Ic}, \qquad \Cprime = 2C_L.
        \end{equation*}

        \subsubsection{Governing equations}

            The fracture aperture $w$ is related to the normal traction $T_n$ on the fracture surfaces through the hypersingular integral equation of linear elasticity derived from the displacement discontinuity method~\cite{crouch1983boundary, hills2013solution}
            \begin{equation}\label{eq:elasticity_equation}
                T_n(x, y, t) = \sigma(y) - \frac{E'}{8\pi}\int_{\mathcal{A}(t)}
                \frac{w(x', y', t)\,\dint x' \dint y'}{\big[(x'-x)^2+(y'-y)^2\big]^{3/2}},
            \end{equation}
            where $T_n$ denotes the normal component of the traction vector acting on the fracture surface, which is equal to the fluid pressure when the fracture is open and includes contact forces when the fracture closes. To account for potential closure, the fracture aperture is restricted by a minimum value
            \begin{equation}\label{eq:contact_condition}
                w(x, y, t) \geq \wmin(x, y, t).
            \end{equation}
            The minimum aperture $\wmin(x, y, t)$ accounts for physical mechanisms that prevent the fracture surfaces from closing completely, such as proppant placement or surface roughness. Its dependence on coordinates $(x, y)$ and time $t$ reflects the potential non-uniform distribution of these mechanisms. However, this study focuses on the case where $\wmin = 0$.

            Assuming that the fracturing fluid is incompressible and Newtonian, and that the flow regime within the fracture is laminar, the lubrication theory leads to the following governing equation for the fluid flow~\cite{dontsov2017multiscale}

            \begin{equation}\label{eq:lubrication_equation}
                \pd{w}{t} - \div\left(\frac{w^3}{\muprime}\nabla p\right) = Q(t)\delta(x,y) -\frac{\Cprime}{\sqrt{t\!-\!t_0(x,y)}},
            \end{equation}
            where $t_0(x,y)$ is the time at which the fracture front is located at the point $(x,y)$, and $\delta(x,y)$ is the Dirac delta function.

        \subsubsection{Initial and boundary conditions}
            The model requires an initial distribution of the fracture aperture $w(x,y,0)$ corresponding to a given initial fracture front $\mathcal{C}(0)$. These initial data are commonly obtained from solutions of one-dimensional hydraulic fracture models, such as Radial~\cite{Geertsma_KGD_1969}, PKN~\cite{Nordgren_PKN_1972}, or KGD~\cite{Khristianovic_KGD_1955} formulations.
            In practice, the precise choice of the initial fracture profile has no significant impact on the long-term evolution of the solution, provided that the initial fracture volume remains small compared to the total injected fluid volume~\cite{garagash2011multiscale, dontsov2015non}.

            At the moving fracture front $\mathcal{C}(t)$ shown in~\Cref{fig:planar3d_fracture_scheme}, two boundary conditions are imposed: (i) the fracture propagation criterion, and (ii) the zero fluid flux condition in the normal direction to the fracture front.

            According to linear elastic fracture mechanics, fracture propagation occurs when the stress intensity factor reaches the material's fracture toughness $K_{I} = K_{Ic}$, where $K_{Ic}$ is the critical stress intensity factor of the rock. The near-tip asymptotics of the fracture aperture follows the square-root behavior
            \begin{equation}\label{eq:asymptotic_aperture}
                \lim\limits_{s \to 0} \frac{w(s)}{s^{1/2}} = \frac{K'}{E'},
            \end{equation}
            where $s$ is the distance to the fracture front (\Cref{fig:planar3d_fracture_scheme}).

            Under propagation, the fracture aperture near the tip is determined by the toughness-scaled asymptotics~\eqref{eq:asymptotic_aperture}, while the normal velocity of the front satisfies $V > 0$.
            If the fracture is arrested, so that $V = 0$, the asymptotic relation~\eqref{eq:asymptotic_aperture} should be modified to include the actual stress intensity factor $K_{I}$ instead of the toughness parameter $K'$~\cite{dontsov2017multiscale}.

            The boundary condition at the fracture front $\mathcal{C}(t)$ is defined by the zero-flux condition
            \begin{equation}
                \bq \cdot \bn = 0,
            \end{equation}
            where $\bq$ is the fluid flux vector and $\bn$ is the outward unit normal to $\mathcal{C}(t)$.

    \subsection{Numerical scheme}\label{sec:Reference_planar_3D_ILSA_numerical_scheme}

\begin{figure}[H]
            \centering
            \includegraphics[width=0.7 \linewidth]{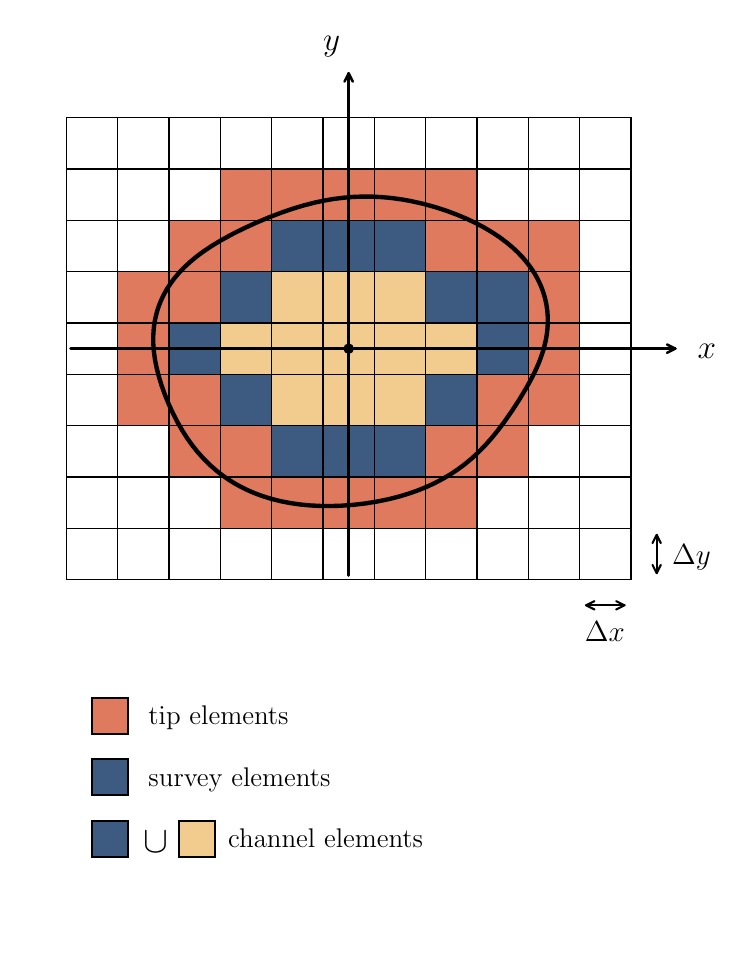}
            \caption{Discretization of the fracture domain using a fixed rectangular mesh. Elements are classified into channel, tip, and survey categories following~\cite{dontsov2017multiscale, Lecampion_Zia_Pyfrac_2020}. The black curve indicates the fracture front.}
            \label{fig:fracture_el_classification}
        \end{figure}

        The discretization of the fracture domain using a fixed rectangular mesh is illustrated in~\Cref{fig:fracture_el_classification}. Each cell $(i,j)$ is centered at $(x_i, y_j)$ and has dimensions $\Delta x \times \Delta y$, as indicated by the coordinate axes and the dimension markers. The fracture front is shown as a black curve. Following the classification in~\cite{dontsov2017multiscale, Lecampion_Zia_Pyfrac_2020}, mesh elements are divided into channel, tip, and survey categories. Channel elements are contained entirely within the fracture footprint and completely filled with fluid. Tip elements are intersected by the fracture front and are only partially filled, whereas survey elements form a subset of channel elements located adjacent to the tip elements. In addition, elements with active constraint, drawn from among the channel elements but tracked thereafter as a distinct category, are introduced to represent regions where the aperture equals the prescribed minimum value, indicating that the fracture is closed in these elements.

\begin{figure}[H]
            \centering
            \hspace*{1cm}
            \includegraphics[scale=0.8]{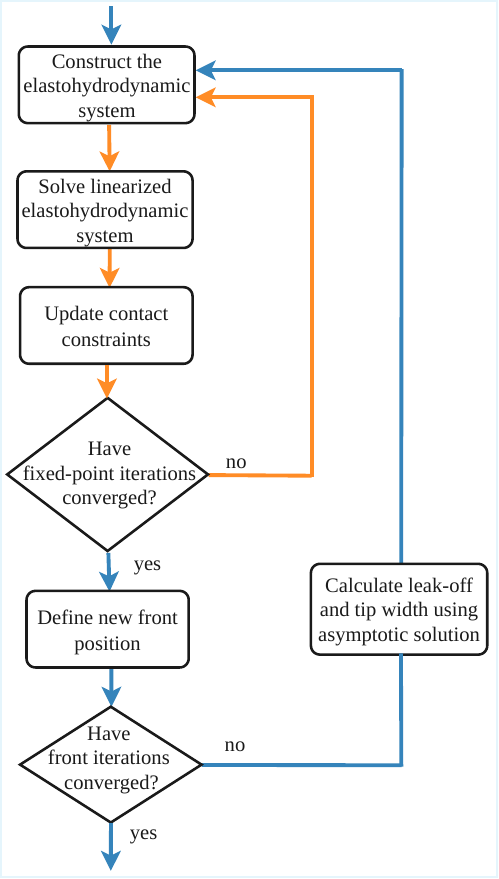}
            \caption{Flowchart of the reference planar 3D ILSA scheme. Orange arrows: fixed-point iteration loop for the nonlinear elastohydrodynamic system. Blue arrows: fracture front iteration loop.}
            \label{fig:planar3d_reference_algorithm_scheme}
        \end{figure}

        At each time step, fracture propagation is simulated by solving the nonlinear elastohydrodynamic system and updating the front position. \Cref{fig:planar3d_reference_algorithm_scheme} illustrates the overall algorithm of this process. The nonlinear elastohydrodynamic system is solved by the fixed-point iteration method. Each iteration of this method involves the solution of a linearized elastohydrodynamic system. To construct this system, the elasticity equation~\eqref{eq:elasticity_equation} is discretized with the displacement discontinuity method~\cite{crouch1983boundary}, and the lubrication equation~\eqref{eq:lubrication_equation} is discretized by the finite volume method with backward Euler time stepping. A piecewise constant approximation is used for the fracture aperture and fluid pressure within the elements. The resulting linearized system takes the following block form~\cite{valov2023implicit}:
        \begin{equation}\label{elastohydrodynamic_system}
            \begin{aligned}
                \left[\begin{array}{ccc}
                    \bbI^{C C}-\Delta t\bbL^{C C} \bbE^{C C} & -\Delta t\bbL^{C T} & -\Delta t\bbL^{C A} \\
                    -\Delta t\bbL^{T C} \bbE^{C C} & -\Delta t\bbL^{T T} & -\Delta t\bbL^{T A} \\
                    -\Delta t\bbL^{A C} \bbE^{C C} & -\Delta t\bbL^{A T} & -\Delta t\bbL^{A A}
                \end{array}\right]\left[\begin{array}{c}
                    \mathbf{w}^C \\
                    \mathbf{p}^T \\
                    {\mathbf{T}_n}^A
                \end{array}\right] =\left[\begin{array}{c}
                    \mathbf{R}^C\\
                    \mathbf{R}^T\\
                    \mathbf{R}^A
                \end{array}\right],
            \end{aligned}
        \end{equation}
        where $\bbI$ is the identity matrix, $\bbL$ is the fluid flux matrix, $\bbE$ is the elasticity matrix, and $\Delta t$ is the time step size. The exact form of $\bbL$ and $\bbE$ can be found in~\cite{valov2023implicit}. The vectors $\mathbf{w}$, $\mathbf{p}$, and $\mathbf{T}_n$ represent the fracture aperture, fluid pressure, and normal traction within the elements. Superscripts $C, T, A$ denote channel, tip, and active-constraint sub-blocks, respectively. The right-hand side vector $\mathbf{R}$ accounts for fluid injection and leak-off, among other components detailed in~\cite{valov2023implicit}.

        Depending on cell type, the primary unknown of the elastohydrodynamic system is one of the fracture aperture $w$, the fluid pressure $p$, or the normal traction $T_n$. In channel elements, the fracture aperture serves as the primary unknown, with pressure derived from the elasticity equation. However, near the fracture front, direct discretization can lead to inaccuracies due to steep solution gradients. Therefore, to enhance solution accuracy, the aperture in the tip elements is calculated using asymptotic relations~\cite{garagash2011multiscale, dontsov2015non}. Following~\cite{dontsov2017multiscale}, the asymptotic solution is integrated over the tip elements to capture the near-tip behavior in a weak sense, ensuring that the prescribed volume is consistent with the asymptotic solution rather than enforcing the exact aperture profile. Consequently, in tip elements, the aperture is prescribed by the asymptotic solution, and the pressure becomes the primary unknown to be determined. Similarly, for elements with active constraint, the aperture is prescribed to be equal to the minimum aperture, and therefore the normal traction is obtained as the primary unknown.

        At each fixed-point iteration, the set of elements with active constraint is updated. Contact conditions are handled directly within the fixed-point iterations, without employing a separate active-set iteration loop as in~\cite{Lecampion_Zia_Pyfrac_2020}. An element $i$ enters the contact set at iteration $k$ if the aperture falls to or below the minimum admissible value
        \begin{equation}\label{eq:contact_condition_min_width}
            w^{i}_{k} \leq w_{\mathrm{min}}^{i},
        \end{equation}
        where $w^{i}_{k}$ denotes the aperture of element $i$ at iteration $k$, and $w_{\mathrm{min}}^{i}$ is the minimum aperture for element $i$. At each front update, the contact set is discarded, and all active-constraint elements are reclassified as channel elements.

        The iterative procedure for solving the elastohydrodynamic system is shown in \Cref{fig:planar3d_reference_algorithm_scheme}. The fixed-point iterations continue until the maximum relative change in the aperture of the channel and active-constraint elements falls below a prescribed tolerance:
        \begin{equation}\label{lubrication_exit_condition}
            \frac{
                \| \mathbf{w}_{k+1} - \mathbf{w}_{k} \|_{\infty}
            }{
                \| \mathbf{w}_{k} \|_{\infty}
            } < \varepsilon_{\text{EHD}},
        \end{equation}
        where $\mathbf{w}_{k}$ is the vector of these apertures at iteration $k$, and $\|\cdot\|_{\infty}$ denotes the maximum norm. The tolerance $\varepsilon_{\text{EHD}} > 0$ is a parameter controlling the accuracy of the nonlinear elastohydrodynamic solution. If this criterion is not satisfied, the elastohydrodynamic system is reassembled and the process is repeated. This inner loop is represented by the orange arrow in~\Cref{fig:planar3d_reference_algorithm_scheme}. Once convergence is achieved, the algorithm proceeds to update the fracture front.

        The new front position is defined by inverting the near-tip asymptotic relations~\cite{dontsov2015non} at survey elements to compute shortest distances to the front. These distances serve as initial data for solving the eikonal equation~\cite{dontsov2017multiscale} by the fast marching method~\cite{malladi1996level, sethian1996fast}. The solution to this equation yields a signed-distance function, which enables approximation of the front as a piecewise-linear curve. A detailed description of the algorithm can be found in~\cite{dontsov2017multiscale}.

        The updated front is compared against its previous position, and convergence is achieved if
        \begin{equation}\label{front_convergence_criteria}
            \frac{
                \| \mathbf{s}_{k+1} - \mathbf{s}_{k} \|_{\infty}
            }{
                \| \mathbf{s}_{k} \|_{\infty}
            } < \varepsilon_{\text{front}},
        \end{equation}
        where $\mathbf{s}_{k}$ is the vector of distances between the center of survey elements and the fracture front at iteration $k$. The tolerance $\varepsilon_{\text{front}}$ is a parameter controlling the accuracy of fracture front tracking. If the condition~\eqref{front_convergence_criteria} is not satisfied, the algorithm returns to the start of the time step loop following the blue arrow in~\Cref{fig:planar3d_reference_algorithm_scheme}. This path involves recalculating the apertures in the tip elements using the near-tip asymptotic solution and updating the leak-off distribution on the new geometry. The process then restarts with the construction of the elastohydrodynamic system. If the condition~\eqref{front_convergence_criteria} is satisfied, the time step is completed.

        Within a time step the set of tip elements is rebuilt at each front iteration. As the front advances, a cell that was intersected by the front may end up entirely behind it. Such a cell retains the tip classification and is reclassified as channel only at the beginning of the next time step. Since elements with active constraint are drawn from among the channel elements, the union of the channel and active-constraint categories remains unchanged within the time step: an element may move between these two categories as the contact set evolves, but no element enters or leaves their union while the front iterations proceed.

\section{Performance-enhancing approaches}\label{sec:Performance_enhancing_approaches}

    \subsection{Unified planar 3D ILSA scheme}

        This section introduces a computational approach for simulating hydraulic fracture propagation, referred to as the unified planar 3D ILSA scheme. The main idea of the method is to merge the loops responsible for resolving the elastohydrodynamic system nonlinearity and tracking the fracture front into a single iterative loop. For clarity, we present two formulations of this algorithm: a \textit{basic unified scheme} that captures the essence of the method, and an \textit{optimized unified scheme} that incorporates additional efficiency strategies.

\begin{figure}[H]
                \centering
                \includegraphics[width=0.99\textwidth]{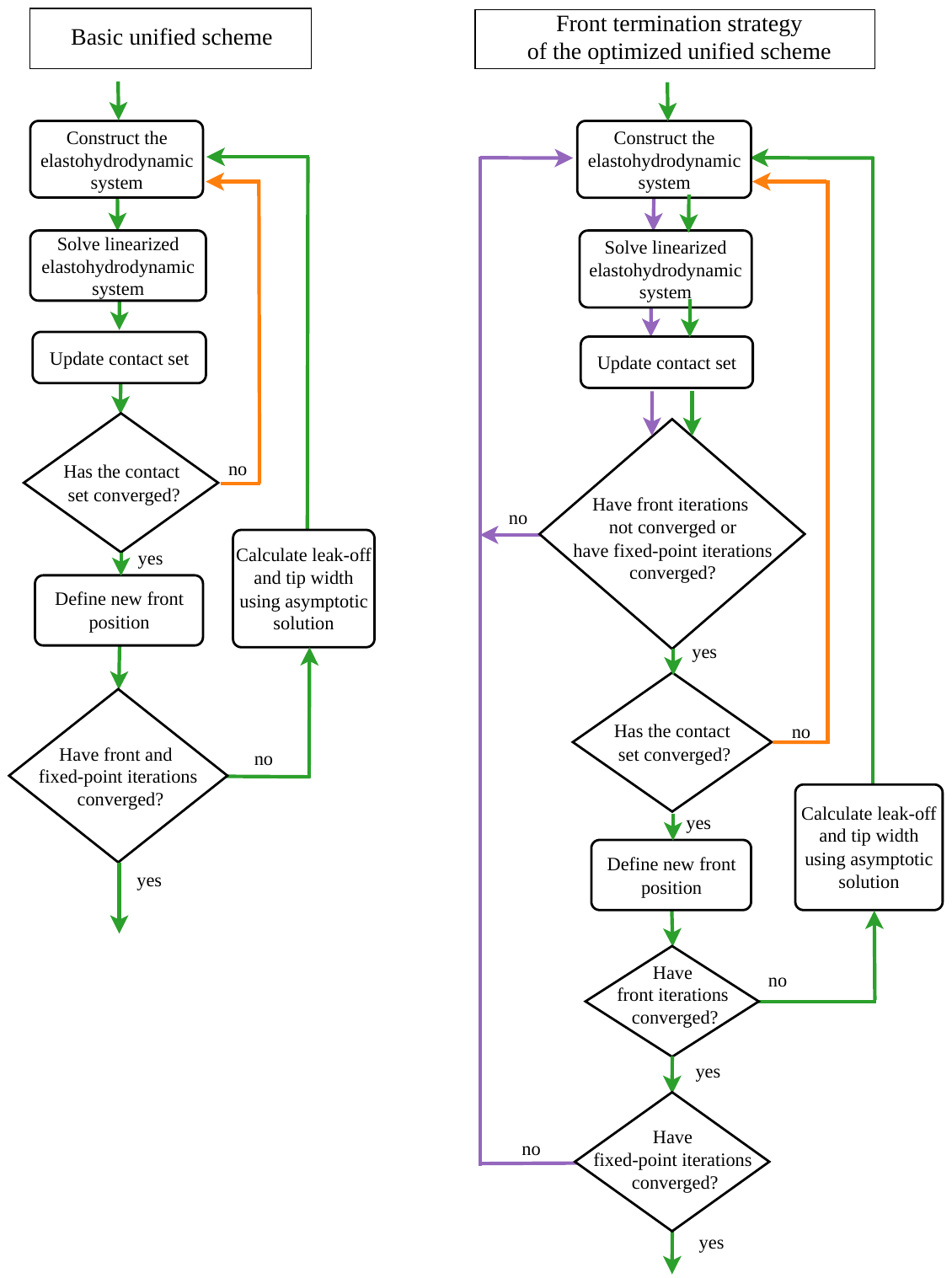}
                \caption{Flowcharts of the unified planar 3D ILSA schemes. Left panel: basic unified scheme. Right panel: front termination strategy of the optimized unified scheme.}
                \label{fig:planar3d_unified}
            \end{figure}

        \paragraph{Basic unified scheme}
            A schematic representation of the scheme is shown in the left panel of~\Cref{fig:planar3d_unified}. Unlike the reference planar 3D ILSA scheme, which updates the front position only after the elastohydrodynamic solver has converged, the unified scheme incorporates the front iterations into a single iterative loop alongside the fixed-point iterations of the elastohydrodynamic solver. A key advantage of this scheme is that the location of the fracture front can be updated prior to the full convergence of the elastohydrodynamic solver. Consequently, early fracture front updates prevent unnecessary fixed-point iterations that attempt to converge to an elastohydrodynamic system solution with an outdated front position.

            Merging the two loops also affects the treatment of the contact set. In the reference scheme the contact set is discarded at each front update, which occurs once the elastohydrodynamic solver converges. In the unified scheme the front is updated within the iterative loop, so applying the same rule would discard the contact set at nearly every iteration. Let $A_{k}$ denote the contact set at iteration $k$. The front update is therefore permitted only at the iterations for which
            \begin{equation}\label{eq:contact_set_convergence}
                A_{k+1} = A_{k},
            \end{equation}
            that is, once the contact set has stopped changing and no new element has
            been added to it at the current iteration. When this condition holds, the
            front is updated and the contact set is discarded together with the update,
            as in the reference scheme. When it does not hold, the front update is
            skipped and the contact set is carried over to the next iteration.

            Another feature of the unified scheme is that elements may enter and leave the tip category and the size of the linear system~\eqref{elastohydrodynamic_system} may change across iterations. Convergence is therefore monitored using a fixed-size vector of apertures of the channel and active-constraint elements, whose classification, as noted in \Cref{sec:Reference_planar_3D_ILSA_numerical_scheme}, is not altered by the front update. This vector is ordered by mesh cell rather than by the ordering of the linear system, so each entry stays attached to its cell when an element changes category, and successive iterations are always compared in the fixed order.

        \paragraph{Optimized unified scheme}

            Although the basic scheme demonstrates the unification principle, it may perform redundant front iterations. The optimized unified scheme introduces two strategies to reduce such redundant operations: delaying the initiation of front updates until the elastohydrodynamic solution reaches sufficient accuracy, and terminating front updates once the front tracking algorithm has converged, keeping it fixed for the remaining fixed-point iterations of the elastohydrodynamic solver.

            The first strategy concerns the initiation of front updates. Front updates are activated
            once the relative change in the fracture aperture between iterations falls to the order of unity,
            \begin{equation}
                \frac{\| \mathbf{w}_{k} - \mathbf{w}_{k-1} \|_{\infty}}{\| \mathbf{w}_{k} \|_{\infty}} \leq 1,
            \end{equation}
            where $\mathbf{w}_{k}$ denotes the fracture aperture vector at iteration $k$ and $\|\cdot\|_{\infty}$ denotes the maximum norm. The choice of the threshold is motivated by numerical experiments, which demonstrate that this value is sufficient and that there is no advantage in choosing a smaller value.

            The second strategy concerns the termination of the fracture front updates and is illustrated in the right panel of~\Cref{fig:planar3d_unified}. Once the fracture front has converged, it is no longer updated at each subsequent iteration. The algorithm then continues iterating only the nonlinear elastohydrodynamic system solution while keeping the front fixed. Once the elastohydrodynamic solver has converged, a new front position is computed and compared with the previously fixed position to verify front convergence. If the front convergence condition~\eqref{front_convergence_criteria} remains satisfied, the time step is completed. Otherwise, the coupled iterations of the fracture front and the elastohydrodynamic system are resumed until convergence is achieved.

            The optimized unified scheme serves as the final computational algorithm. Performance evaluations reported in the subsequent sections correspond to this particular formulation.

    \subsection{Matrix splitting for the elastohydrodynamic system}

        Numerical solutions of coupled differential equations often benefit from a matrix splitting approach, which decomposes complex system matrices into components that can be solved more efficiently. A general framework for such an approach is discussed in~\cite{nuca2024splitting}, where the system matrix is split into a sparse implicit part and a dense explicit component. The implicit component is handled at the current iteration, while the explicit part is evaluated using values from the previous iteration. This approach reduces the computational cost by replacing a dense, computationally expensive system matrix with a sparse one, allowing the system to be solved faster. A similar matrix splitting strategy has already been successfully applied to accelerate magma dike propagation simulations in quasi-2-D settings, thereby confirming its practical utility~\cite{abdullin2026quasi}.

        In this section, we describe the application of this methodology for accelerating the solution of the linearized elastohydrodynamic system~\eqref{elastohydrodynamic_system}. This system is solved repeatedly within several nested iteration loops: a time-stepping loop, an iteration loop for front tracking, and a fixed-point iteration loop for solving the nonlinear elastohydrodynamic system. Since the solution of the linearized elastohydrodynamic system dominates the computational cost of the algorithm, optimizing it may significantly improve simulation performance.

        The linearized elastohydrodynamic system exhibits a block structure comprising the elasticity and fluid flux matrices. The elasticity matrix $\bbE$ represents the non-local elastic response, with each column containing the elastic influence coefficients that relate the pressure over the entire grid to a unit displacement discontinuity in a single cell, so that the matrix is dense. To improve computational efficiency, we decompose the elasticity matrix $\bbE$ into implicit and explicit components based on the range of elastic interactions, as follows:
        \begin{equation}
            \label{eq:E_decompose}
            \bbE = \bbE_{\text{impl}} + \bbE_{\text{expl}},
        \end{equation}
        where the implicit part $\bbE_{\text{impl}}$ captures short-range interactions within a prescribed stencil, and the explicit component $\bbE_{\text{expl}}$ accounts for long-range interactions across the remaining domain.

\begin{figure}[H]
            \centering
            \begin{subfigure}[b]{0.48\textwidth}
                \centering
                \includegraphics[width=\textwidth]{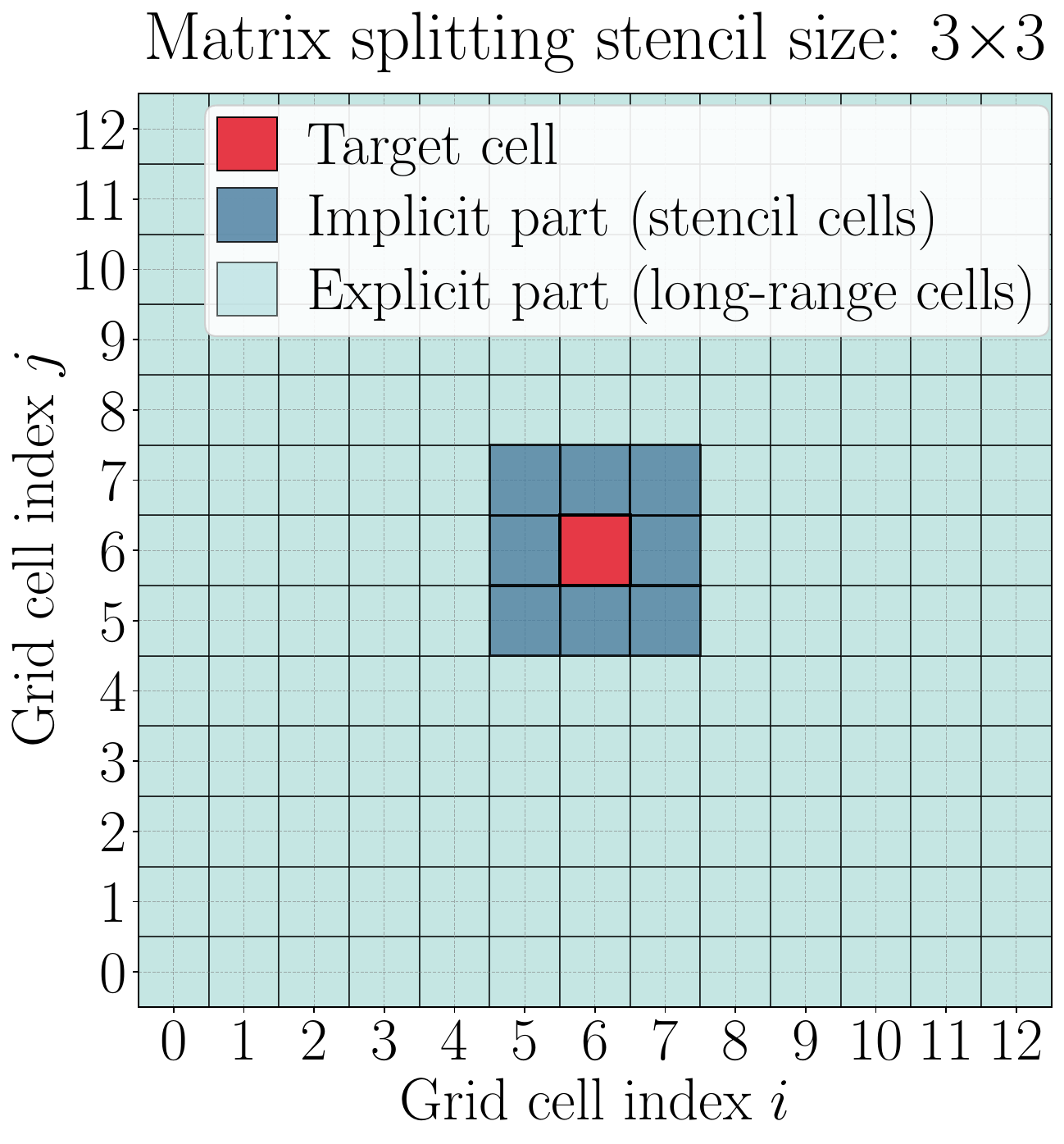}
            \end{subfigure}
            \hfill
            \begin{subfigure}[b]{0.48\textwidth}
                \centering
                \includegraphics[width=\textwidth]{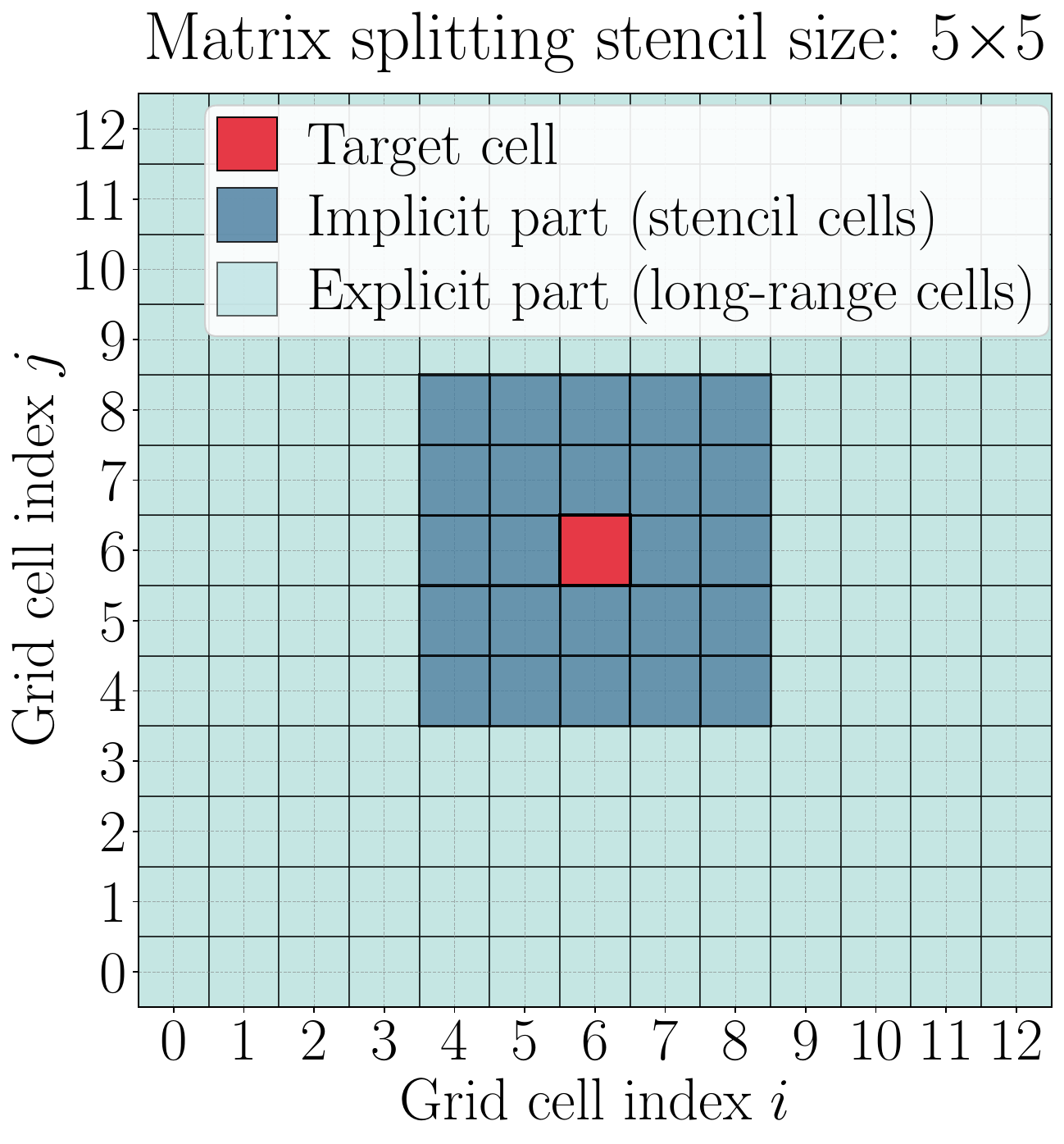}
            \end{subfigure}
            \caption{Stencils for decomposing the elasticity matrix into the sparse implicit part and the dense explicit part for $3 \times 3$ and $5 \times 5$ stencils. The red cell is the target cell for which elastic influence coefficients are being partitioned.}
            \label{fig:stencils}
        \end{figure}

        \Cref{fig:stencils} illustrates this decomposition for $3 \times 3$ and $5 \times 5$ stencils. For a given target cell (red), the implicit part $\bbE_{\text{impl}}$ includes only coefficients associated with its immediate neighbors within the stencil (blue). The contributions of the remaining cells (light blue) correspond to the explicit part $\bbE_{\text{expl}}$.

        Decomposition~\eqref{eq:E_decompose} enables the following approximation for the elasticity operator applied to the solution vector:
        \begin{equation}
            \label{eq:approx_operator_E}
                \bbE^{C C} \mathbf{w}^C_{k} \approx \bbE^{C C}_{\text{impl}} \mathbf{w}^C_{k} + \bbE^{C C}_{\text{expl}} \mathbf{w}^C_{k-1}.
        \end{equation}
        This approximation makes it possible to construct the modified elastohydrodynamic system by keeping $\bbE_{\text{impl}} \mathbf{w}^C_{k}$ on the left-hand side and moving $\bbE_{\text{expl}} \mathbf{w}^C_{k-1}$ to the right-hand side as follows:
        \begin{equation}
            \label{eq:modified_elastohydrodynamic_system}
            \begin{gathered}
                \begin{bmatrix}
                    \bbI^{C C} -\Delta t\bbL^{C C} \bbE^{C C}_{\text{impl}} & -\Delta t\bbL^{C T} & -\Delta t\bbL^{C A} \\
                                -\Delta t\bbL^{T C} \bbE^{C C}_{\text{impl}} & -\Delta t\bbL^{T T} & -\Delta t\bbL^{T A} \\
                                -\Delta t\bbL^{A C} \bbE^{C C}_{\text{impl}} & -\Delta t\bbL^{A T} & -\Delta t\bbL^{A A}
                \end{bmatrix}
                \begin{bmatrix}
                    \mathbf{w}^C_{k} \\
                    \mathbf{p}^T_{k} \\
                    {\mathbf{T}_n}^A_{k}
                \end{bmatrix} = \\
                \begin{bmatrix}
                    \mathbf{R}^C + \Delta t\bbL^{C C} \bbE^{C C}_{\text{expl}} \mathbf{w}^C_{k-1} \\
                    \mathbf{R}^T + \Delta t\bbL^{T C} \bbE^{C C}_{\text{expl}} \mathbf{w}^C_{k-1} \\
                    \mathbf{R}^A + \Delta t\bbL^{A C} \bbE^{C C}_{\text{expl}} \mathbf{w}^C_{k-1}
                \end{bmatrix}.
            \end{gathered}
        \end{equation}
        This transformation yields a sparse system matrix, enabling the use of efficient sparse linear solvers. However, such an approximation may require additional fixed-point iterations to maintain accuracy and ensure convergence.

        The choice of the stencil for the elasticity matrix decomposition is flexible. Different stencil sizes can be used, for example, $3 \times 3$, $5 \times 5$, $7 \times 7$, and $9 \times 9$. A larger implicit stencil accounts for elastic interactions over a greater range, potentially improving the approximation and reducing the number of additional iterations required for solving the nonlinear elastohydrodynamic system. However, a larger stencil reduces the sparsity of the system matrix, increasing the cost of each linear solve. Therefore, the stencil size should be chosen to balance the cost per iteration against the number of iterations required for convergence. The impact of different stencil sizes is investigated in~\Cref{sec:operator_splitting_results}.

    \subsection{Anderson acceleration}

        The nonlinear problem at each time step can be formulated as a fixed-point problem
        \begin{equation}
            \mathbf{x} = \mathbf{f}(\mathbf{x}),
            \label{eq:fixed-point_equation}
        \end{equation}
        which is solved via the iterative scheme
        \begin{equation}
            \mathbf{x}_{k+1} = \mathbf{f}(\mathbf{x}_{k}),
        \end{equation}
        where $\mathbf{x}_{k}$ denotes the vector of unknowns at iteration $k$, and $\mathbf{f}$ is a nonlinear fixed-point mapping. Anderson acceleration~\cite{anderson1965iterative} is a technique designed to improve the convergence rate of such iterative procedures. The method has proven to be efficient for linearly convergent processes~\cite{evans2020proof} and is used to accelerate convergence in many problems~\cite{he2022solve}.

        The basic concept of the method is to construct an improved iteration as an affine combination of the $(m+1)$ most recent iterates, with the weights chosen to minimize the norm of the corresponding combination of the fixed-point residuals $\mathbf{r}(\mathbf{x}_{i}) = \mathbf{f}(\mathbf{x}_{i}) - \mathbf{x}_{i}$. A detailed description of the method is given in~\cite{anderson1965iterative}. Throughout this study the memory parameter is set to $m = 3$, as numerical experiments indicate that larger values do not yield a consistent improvement.

        Two aspects of the present implementation are specific to the planar 3D ILSA algorithm. The first is that the acceleration is applied to the apertures of the channel and active-constraint elements. As noted in \Cref{sec:Reference_planar_3D_ILSA_numerical_scheme}, these two categories together form a set of elements whose union does not change under the reclassification of individual elements, so the corresponding aperture vector keeps a fixed size and a fixed correspondence to mesh elements throughout the iterations.

        The second is the strategy used to reset the accumulated history, which is informative only as long as the underlying fixed-point mapping does not change appreciably between the iterations. The history is therefore discarded when an element enters the contact set, since the aperture of that element is obtained from the solution of the linearized system~\eqref{elastohydrodynamic_system} while it belongs to the channel category and is set to the prescribed minimum value once it becomes active-constraint. The history is also discarded once the elastohydrodynamic solver has converged~\eqref{lubrication_exit_condition}, as the iterations that follow belong to a different fixed-point process.

        In the unified scheme the mapping is additionally modified by the front updates. However, unlike the discontinuous change of category discussed above, a front update does not signal that the mapping has changed enough to make the accumulated history uninformative, so the history is retained across the front updates and discarded only when an element enters the contact set or the elastohydrodynamic solver converges. This choice is also a practical one, as the front is updated at nearly every iteration, and discarding the history at each update would leave almost no accumulated history to accelerate the next iteration.

\section{Results}\label{sec:results}

    \subsection{Benchmark setup and performance metrics}

        To assess the computational efficiency and accuracy of the proposed planar 3D ILSA acceleration approaches, five representative benchmark cases are considered. These cases are designed to cover a broad range of geometrical and mechanical configurations, from idealized to industry-relevant scenarios.

\begin{figure}[H]
            \centering
            \includegraphics[width=0.99\textwidth]{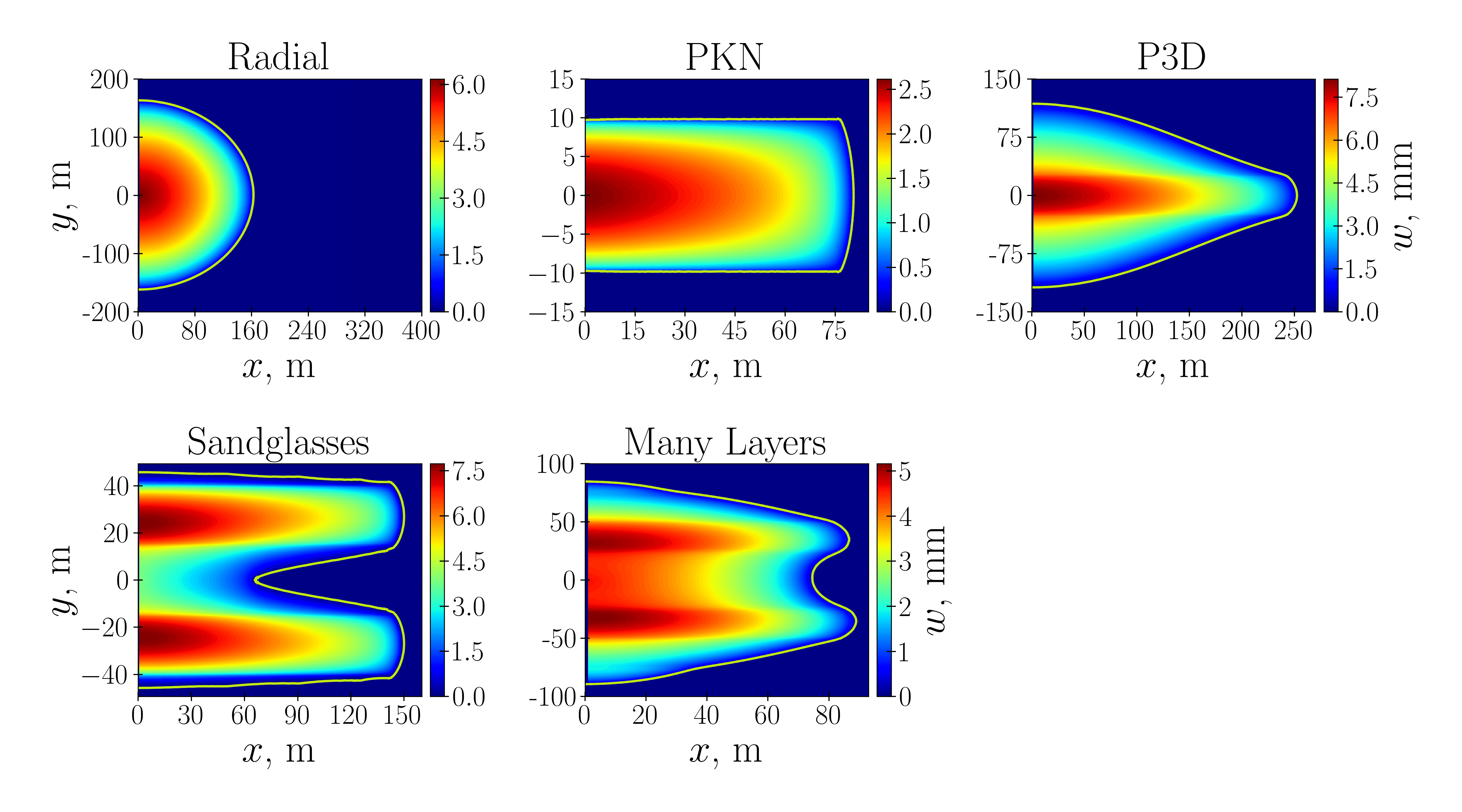}
            \caption{Fracture geometries used in the benchmark cases. The aperture distribution is visualized using a colormap. The solid yellow line indicates the fracture front.}
            \label{fig:cases}
        \end{figure}

\begin{figure}[H]
            \centering
            \includegraphics[width=0.99\textwidth]{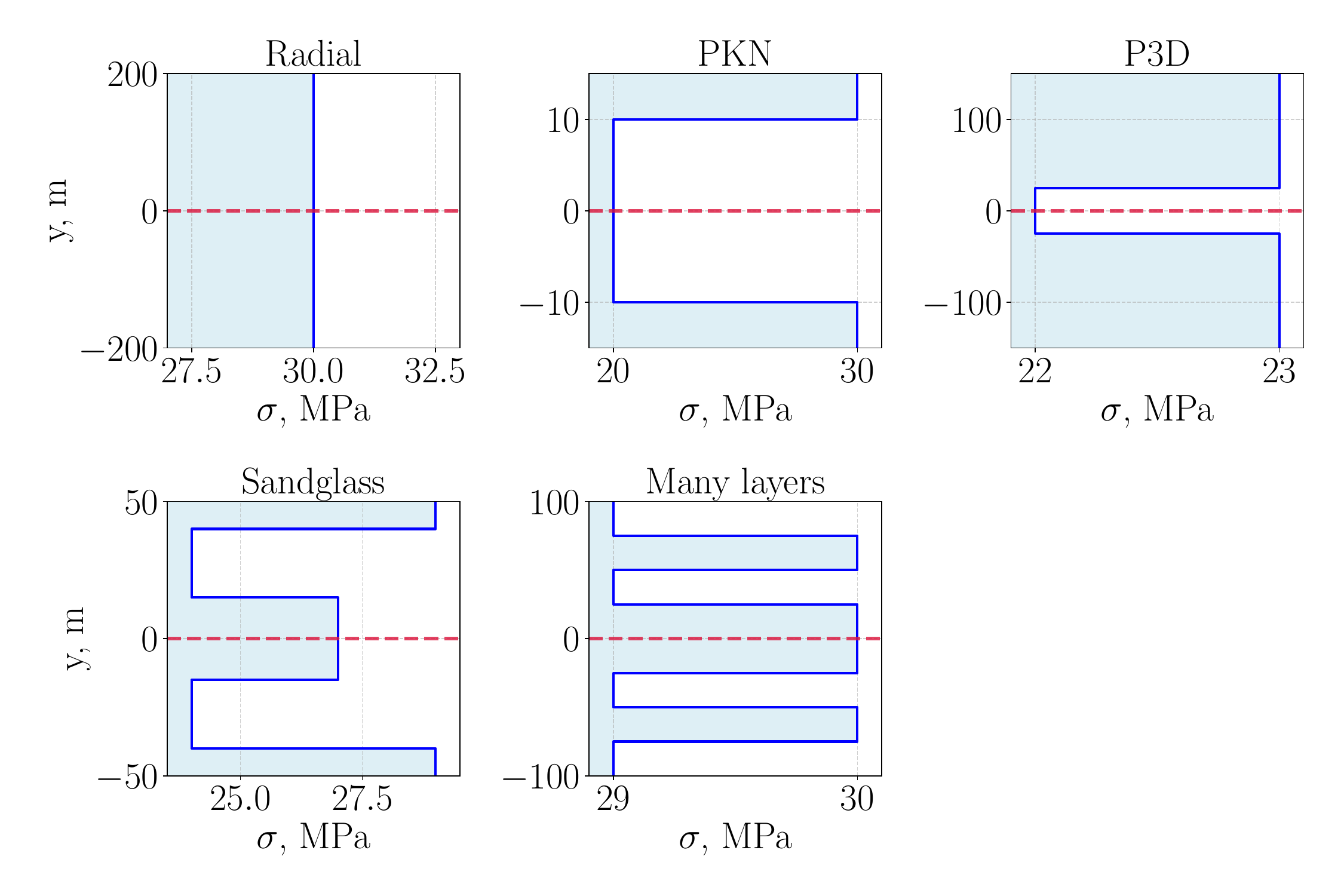}
            \caption{Minimum compressive stress profiles $\sigma(y)$ used in the benchmark cases against the vertical depth. The red dashed line indicates the fluid injection point.}
            \label{fig:cases_stresses}
        \end{figure}

        The first two benchmark cases correspond to classical radial and PKN fractures with well-studied behavior. The third case represents a pseudo-3D fracture configuration that accounts for fracture height growth. The fourth case considers a ``sandglass''-shaped fracture geometry with a non-convex fracture front caused by initiation in a high-stress layer. Finally, the fifth case corresponds to a multilayered formation and provides the most realistic setting by incorporating field-representative stress profiles. \Cref{fig:cases} illustrates the characteristic fracture footprints for the described cases. Input parameters and material properties employed in the simulations are summarized in~\Cref{tab:benchmark_cases}. The corresponding minimum compressive stress profiles $\sigma(y)$ for each configuration are shown in~\Cref{fig:cases_stresses}.

        \begin{table}[h!]
            \caption{Simulation parameters for the benchmark cases.}
            \centering
            \begin{tabular}{lccccc}
            \toprule
            \textbf{Parameter} & \textbf{Radial} & \textbf{PKN} & \textbf{P3D} & \textbf{Sandglass} & \textbf{Many Layers} \\
            \midrule
            \textbf{Injection time}, $10^{3}\,\text{s}$ & $5$ & $5$ & $5$ & $0.9$ & $1$ \\
            \textbf{Flow rate}, $10^{-3}\,\text{m}^{3}/\text{s}$ & $90$ & $1$ & $90$ & $100$ & $90$ \\
            \textbf{Young's modulus}, GPa & $33$ & $1$ & $33$ & $20$ & $33$ \\
            \textbf{Poisson's ratio} & $0.3$ & $0.3$ & $0.3$ & $0.3$ & $0.3$ \\
            \textbf{Toughness}, MPa$\cdot\text{m}^{1/2}$ & $3$ & $0.001$ & $3$ & $0.001$ & $3$ \\
            \textbf{Leak-off}, $10^{-5}\,\text{m}/\text{s}^{1/2}$ & $1$ & $0$ & $1$ & $0$ & $1$ \\
            \textbf{Viscosity}, Pa$\cdot$s & $0.1$ & $0.01$ & $0.1$ & $0.1$ & $0.1$ \\
            \bottomrule
            \end{tabular}
            \label{tab:benchmark_cases}
        \end{table}

        The numerical scheme described in~\Cref{sec:Reference_planar_3D_ILSA_numerical_scheme} is implemented in C++ following~\cite{valov2023implicit} for the toughness-corrected tip asymptotic solution. A fixed time step of $\Delta t = 10~\text{s}$ is employed. The convergence tolerances for the fracture front iterations $\varepsilon_{\text{front}}$ and the nonlinear elastohydrodynamic solver $\varepsilon_{\text{EHD}}$ are set to $10^{-6}$. The nonlinear elastohydrodynamic system is solved by fixed-point iteration. At each fixed-point iteration, the linearized elastohydrodynamic system~\eqref{elastohydrodynamic_system} is assembled, preconditioned using the approach described in~\cite{peirce2006localized}, and solved by the BiCGStab iterative method~\cite{saad2003iterative}. When the matrix splitting technique is employed, the linearized system takes the modified form~\eqref{eq:modified_elastohydrodynamic_system} and is solved instead by the direct sparse solver SparseLU~\cite{demmel1999supernodal}, without preconditioning. If a converged solution is not obtained within the prescribed tolerances for a given time step, the step size is halved and the solution is recomputed. The fixed, rather than adaptive, time step is used throughout the study in order to isolate the effect of the proposed acceleration techniques from the time-stepping strategy and thereby ensure a consistent comparison across all schemes and benchmark cases. In practice, adaptive time-stepping is commonly used and provides an additional, independent source of acceleration~\cite{Lecampion_Zia_Pyfrac_2020, chen2020explicit}.

        To characterize the spatial resolution used in the benchmark study, the individual computational meshes employed for each case are summarized in \Cref{tab:mesh_all}. For every mesh, the cell size is reported as $\Delta x \times \Delta y$.

        \begin{table}[H]
            \caption{Cell sizes $\Delta x \times \Delta y$ (m) of the computational meshes used for the Radial, PKN, P3D, Sandglass, and Many Layers benchmark cases, ordered from coarsest to finest refinement level.}
            \centering
            \small
            \setlength{\tabcolsep}{6pt}
            \renewcommand{\arraystretch}{1.15}
            \resizebox{\textwidth}{!}{%
            \begin{tabular}{l c c c c c}
            \toprule
            \textbf{Case} & \multicolumn{5}{c}{$\Delta x \times \Delta y$, m} \\
            \midrule
            \multirow{2}{*}{Radial}
                & $20.00\times17.50$
                & $11.54\times11.67$
                & $7.69\times7.78$
                & $5.88\times5.83$
                & $4.69\times4.67$ \\
                & $3.90\times3.89$
                & $3.49\times3.50$
                & $2.80\times2.80$
                & $2.26\times2.24$
                & \\
            \midrule
            \multirow{3}{*}{PKN}
                & $3.00\times3.00$
                & $2.50\times2.50$
                & $1.25\times1.25$
                & $1.00\times1.00$
                & $0.83\times0.83$ \\
                & $0.77\times0.77$
                & $0.71\times0.71$
                & $0.67\times0.67$
                & $0.59\times0.59$
                & $0.53\times0.53$ \\
                & $0.47\times0.47$
                & \\
            \midrule
            \multirow{2}{*}{P3D}
                & $16.67\times16.67$
                & $8.33\times8.33$
                & $5.56\times5.56$
                & $4.17\times4.17$
                & $3.33\times3.33$ \\
                & $3.13\times3.13$
                & $2.78\times2.78$
                & $2.50\times2.50$
                & $2.25\times2.25$
                & \\
            \midrule
            \multirow{2}{*}{Sandglass}
                & $10.02\times10.00$
                & $5.01\times5.00$
                & $3.01\times3.33$
                & $2.00\times2.00$
                & $1.67\times1.67$ \\
                & $1.25\times1.25$
                & & & & \\
            \midrule
            \multirow{2}{*}{Many Layers}
                & $10.00\times10.00$
                & $5.00\times5.00$
                & $3.33\times3.33$
                & $2.50\times2.50$
                & $2.22\times2.22$ \\
                & $2.00\times2.00$
                & $1.43\times1.43$
                & $1.25\times1.25$
                & & \\
            \bottomrule
            \end{tabular}%
            }
            \label{tab:mesh_all}
        \end{table}

        To quantitatively assess the efficiency of the proposed acceleration techniques, we introduce the acceleration metric as
        \begin{equation}
            \label{eq:acceleration}
            \mathrm{Acceleration} = \frac{P_{\text{mod}} - P_{\text{ref}}}{P_{\text{ref}}} \times 100\%,
        \end{equation}
        where $P_{\text{ref}}$ denotes the computational performance of the reference planar 3D ILSA scheme, and $P_{\text{mod}}$ denotes the computational performance of the corresponding modified algorithm under identical simulation conditions. The computational performance $P$ is defined as
        \begin{equation}
            P = \frac{1}{T_{\text{run}}},
        \end{equation}
        where $T_{\text{run}}$ is the total simulation time. For instance, an acceleration of 100\% corresponds to a twofold increase in performance, meaning the modified algorithm completes the simulation in half the time of the reference, whereas a value of $-50\%$ indicates a performance degradation by a factor of two, which means the modified algorithm requires twice as much time as the reference.

\begin{figure}[H]
            \centering
            \includegraphics[width=0.99\textwidth]{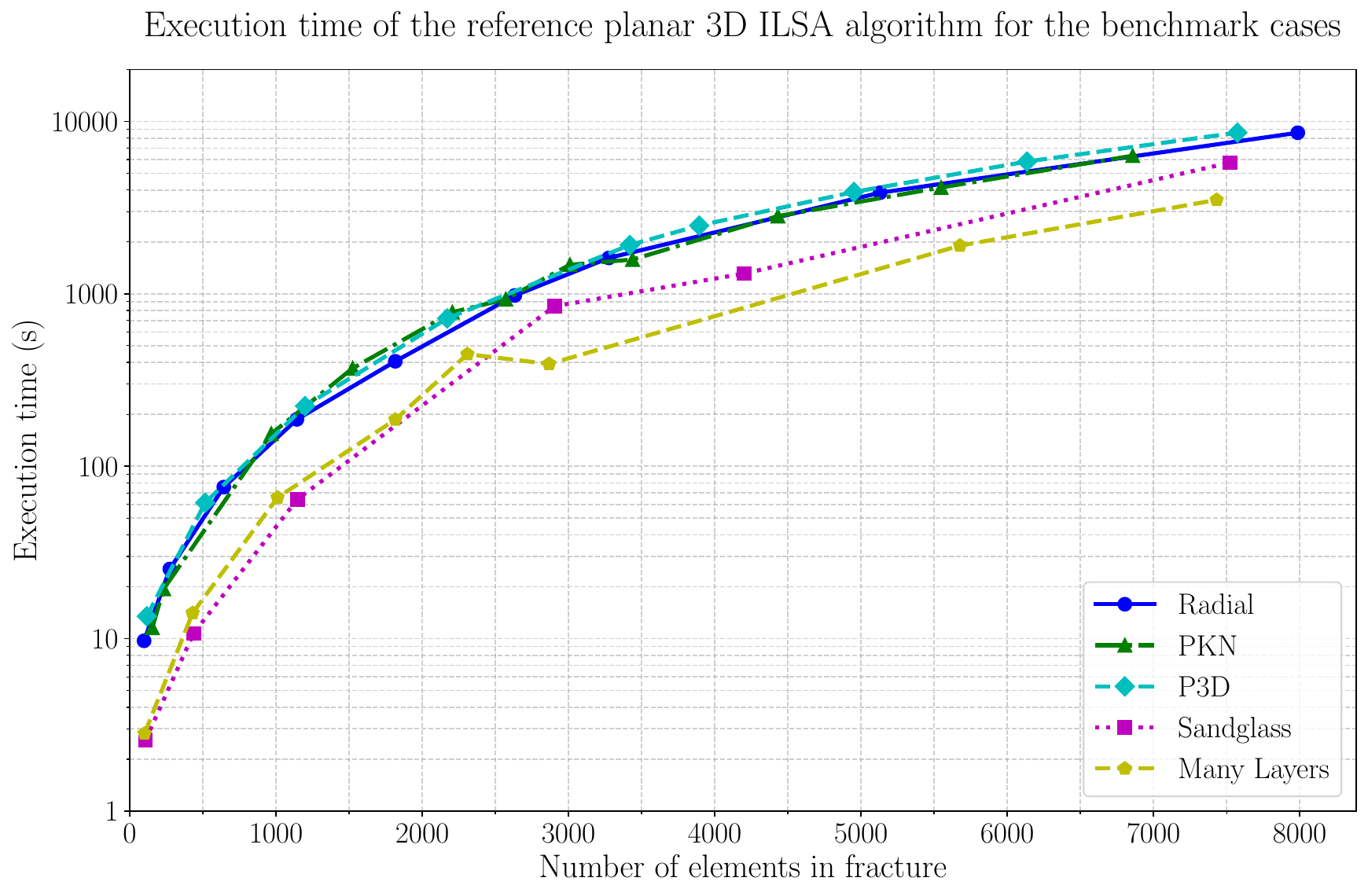}
            \caption{Execution time of the reference planar 3D ILSA scheme against the number of fracture elements at simulation completion for the benchmark cases.}
            \label{fig:reference_execution_times}
        \end{figure}

        To establish a performance baseline, the reference scheme is evaluated for the benchmark cases using multiple computational meshes with varying resolution. The corresponding execution times are reported in~\Cref{fig:reference_execution_times}.
        This figure presents the execution time against the number of fracture elements at simulation completion for the benchmark cases, with the element count ranging from 98 to 7987 depending on the case. These reference execution times provide the baseline against which the modified algorithms are compared in the sections below.

    \subsection{Plan of calculations}

\begin{figure}[H]
            \centering
            \includegraphics[width=0.99\linewidth]{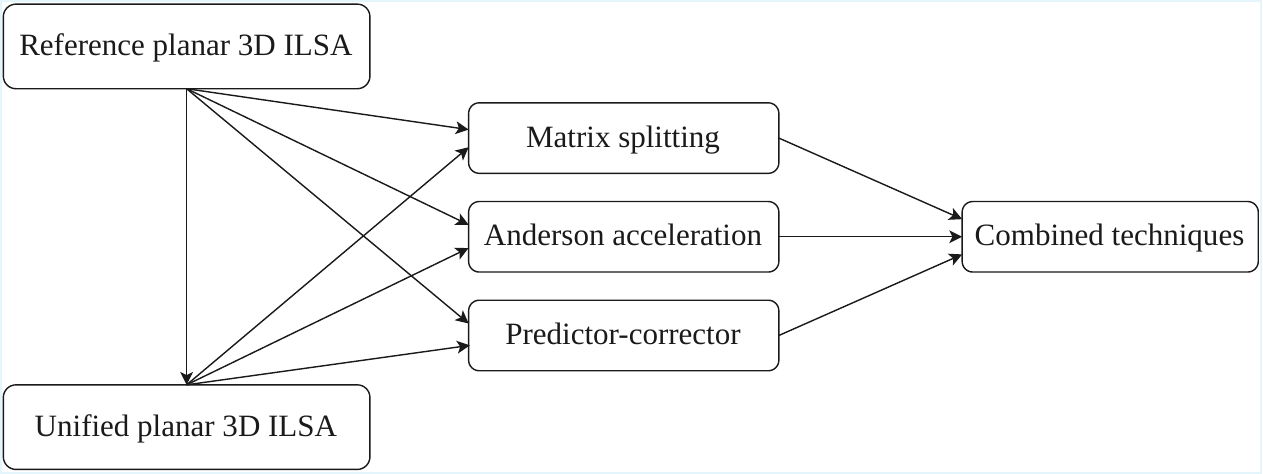}
            \caption{Schematic of the calculation plan. Three acceleration techniques, namely the matrix splitting, Anderson acceleration, and the predictor--corrector scheme, are applied in a modular manner to both the reference planar 3D ILSA scheme and the unified planar 3D ILSA scheme, individually and in combination.}
            \label{fig:plan_of_calculations}
        \end{figure}

        Both the reference and the unified planar 3D ILSA schemes serve as the basis for performance assessment. Three acceleration techniques are applied to each: the matrix splitting, Anderson acceleration, and the predictor--corrector scheme adopted from~\cite{zia2019explicit}. The predictor--corrector scheme is included to examine its interaction with the proposed methods. All three techniques are applied in a modular manner, allowing us to assess the individual effect of each modification. Their combined impact on convergence behavior and computational performance is then evaluated by applying them together. The combinations of algorithms considered in this study and their relationships are summarized in~\Cref{fig:plan_of_calculations}.

        The results are presented in three stages. First, the acceleration techniques are applied to the reference scheme. Second, a comparison between the reference and the unified schemes is performed to highlight the differences introduced by the unified formulation. Finally, the same acceleration techniques are applied to the unified scheme.

        The accuracy of each method is verified against the reference scheme to ensure that the results remain consistent. The discrepancy in the fracture aperture is quantified by the relative $L_2$ and $L_\infty$ error norms, defined in Section~S1 of the Supplementary Material, with error histories for every variant and case reported in Section~S2. The relative $L_2$ error does not exceed 1\% for any case, and the $L_\infty$ error stays below 3\% except for the Sandglass case, where it peaks at up to 9.5\% in a single element adjacent to the fracture front, reflecting the sensitivity of the near-tip treatment rather than a loss of accuracy of the acceleration techniques.

    \subsection{Acceleration techniques applied to the reference planar 3D ILSA scheme}

        \subsubsection{Effect of the matrix splitting on the reference planar 3D ILSA scheme}\label{sec:operator_splitting_results}

            The results are presented for the $3 \times 3$ elastic response stencil. The effect of varying the stencil size is examined separately at the end of the section.

\begin{figure}[H]
                \centering
                \includegraphics[width=0.99\textwidth]{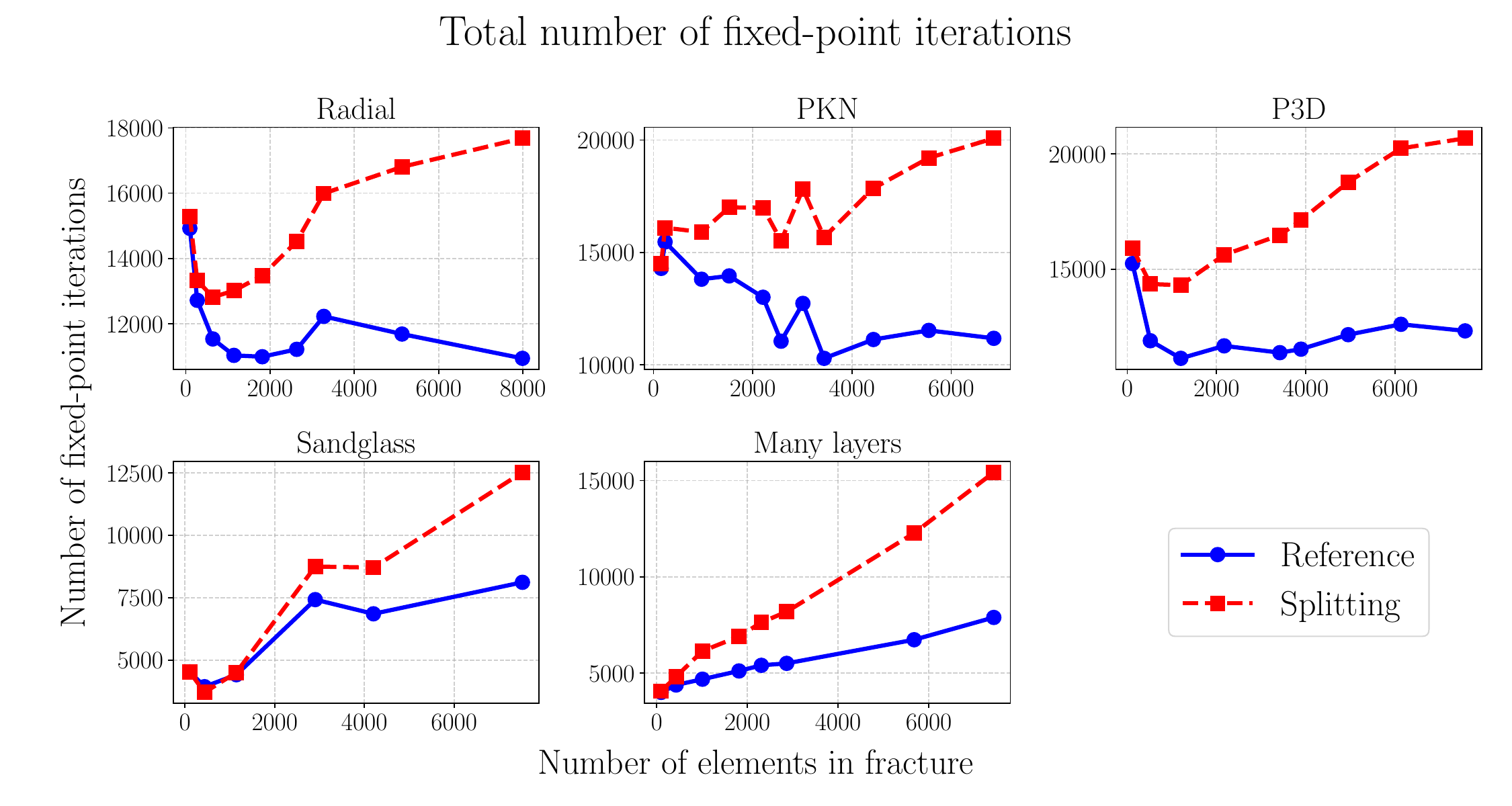}
                \caption{Total number of fixed-point iterations for the reference planar 3D ILSA scheme and the reference scheme with the matrix splitting against the number of fracture elements for the benchmark cases.}
                \label{fig:iterations_nonlinear_split_in_orig}
            \end{figure}

            The fixed-point iteration count, accumulated over all time steps and front iterations, is shown in~\Cref{fig:iterations_nonlinear_split_in_orig}. For most cases and mesh resolutions, the matrix splitting increases the number of fixed-point iterations, indicating a slower convergence rate compared to the reference scheme. The front iteration count remains almost unchanged (see Fig.~S21 in the Supplementary Material), confirming that the modification has a negligible impact on the front tracking process.

\begin{figure}[H]
                \centering
                \includegraphics[width=0.99\textwidth]{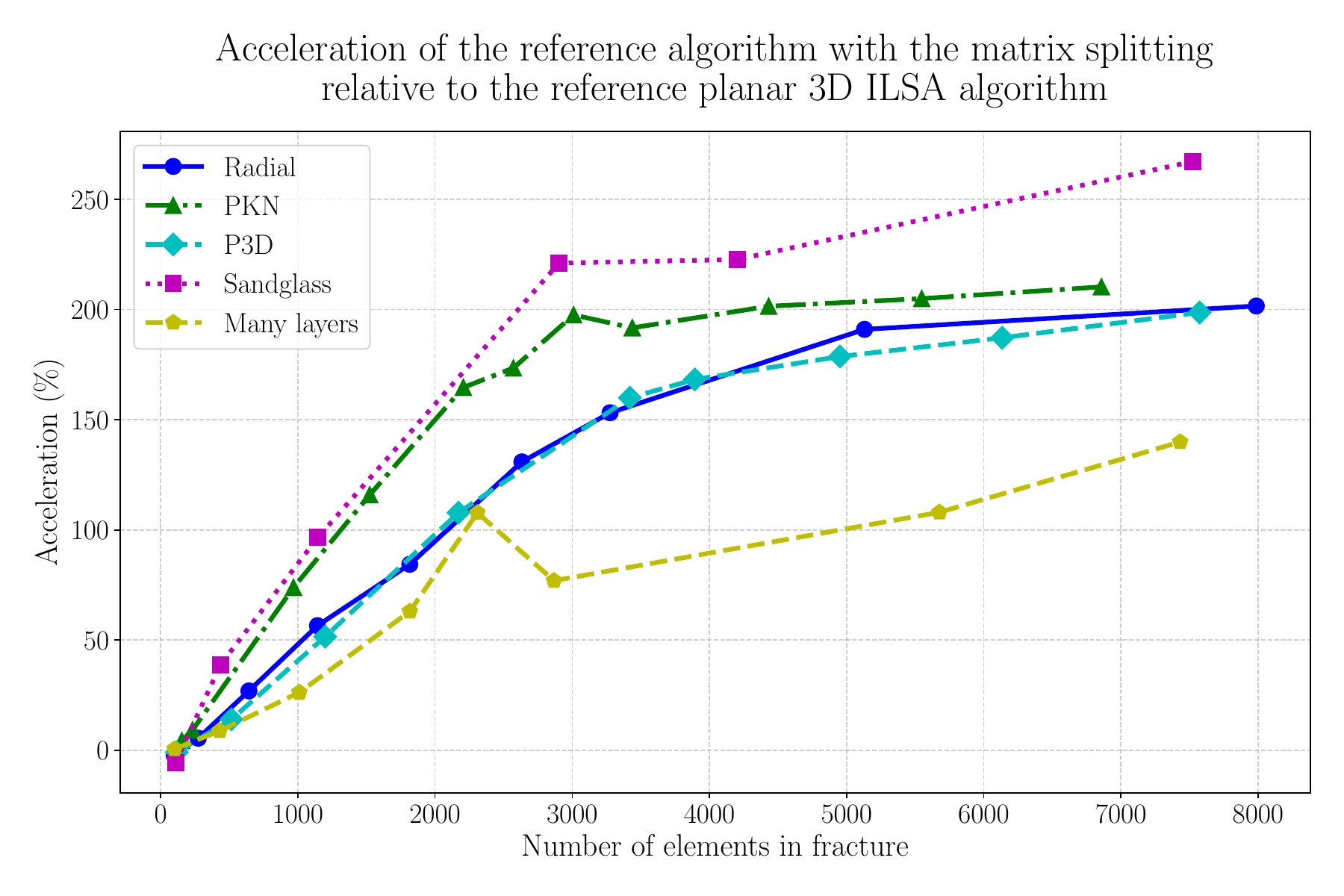}
                \caption{Acceleration resulting from the application of the matrix splitting with the $3 \times 3$ stencil to the reference planar 3D ILSA scheme against the number of fracture elements for the benchmark cases. Acceleration is computed according to~\eqref{eq:acceleration}.}
                \label{fig:split_acceleration}
            \end{figure}

            A consistent speed-up is obtained for all benchmark cases except on the coarsest meshes, where the effect is marginal and may be slightly negative, as reported in~\Cref{fig:split_acceleration}. At the finest resolutions the acceleration reaches approximately 200\% for the Radial, PKN, and P3D cases, 267\% for the Sandglass case, and 140\% for the Many Layers case. The sparse structure of the split system allows each linear solve to be performed more efficiently than with the original dense matrix. This reduction in per-iteration cost more than compensates for the additional fixed-point iterations. The acceleration, however, does not increase monotonically with mesh refinement: a fixed-size stencil covers a smaller physical region as the mesh is refined, capturing a diminishing fraction of the global elastic interaction. This increases the number of fixed-point iterations required for convergence, causing the growth of the acceleration to slow down at fine resolutions.

\begin{figure}[H]
                \centering
                \includegraphics[width=0.99\textwidth]{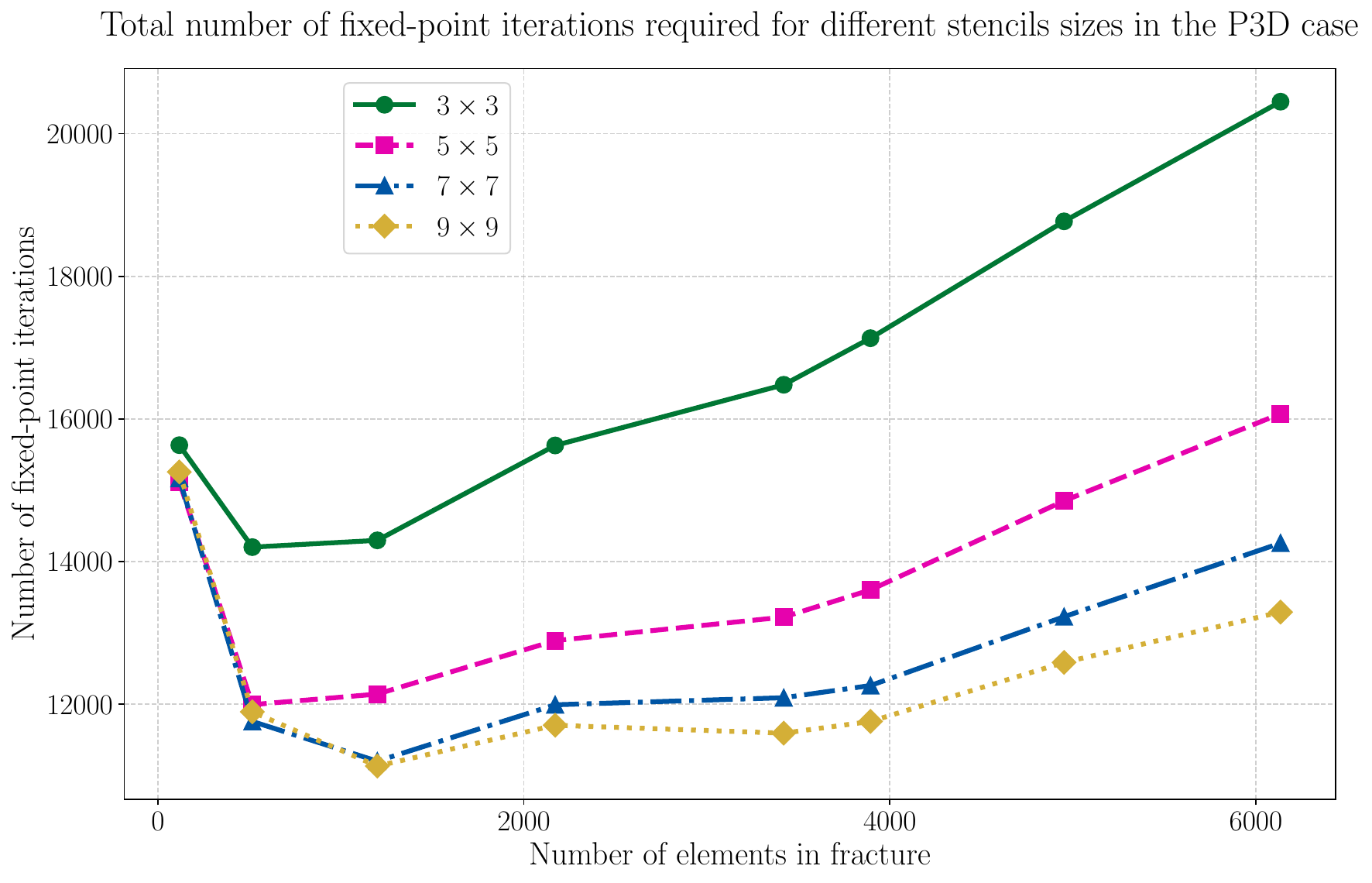}
                \caption{Total number of fixed-point iterations for the P3D case against the number of fracture elements, shown for stencil sizes $3 \times 3$, $5 \times 5$, $7 \times 7$, and $9 \times 9$.}
                \label{fig:stencil_nonlin_iter}
            \end{figure}

\begin{figure}[H]
                \centering
                \includegraphics[width=0.99\textwidth]{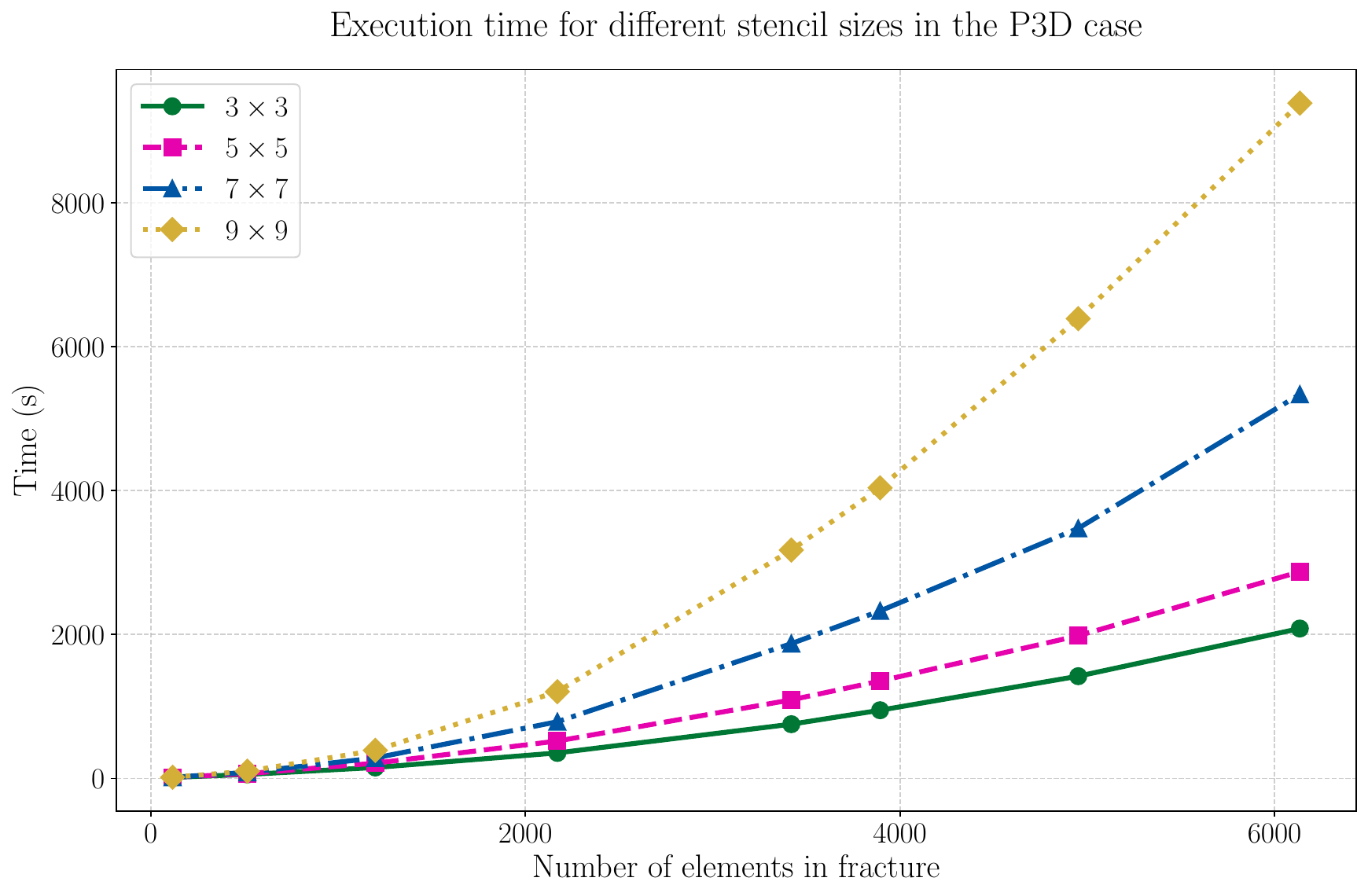}
                \caption{Execution time against the number of fracture elements for the P3D case, shown for stencil sizes $3 \times 3$, $5 \times 5$, $7 \times 7$, and $9 \times 9$.}
                \label{fig:stencil_time}
            \end{figure}

            The influence of the stencil size is illustrated for the P3D case. \Cref{fig:stencil_nonlin_iter} shows that smaller stencils require more iterations, with the $3 \times 3$ configuration yielding the highest count across all mesh resolutions. The corresponding execution times are given in~\Cref{fig:stencil_time}. In contrast to the iteration counts, larger stencils lead to longer runtimes due to the increased density of the system matrix and the higher cost of each linear solve. Numerical experiments indicate that the $3 \times 3$ stencil offers the best balance for the considered mesh resolutions.

        \subsubsection{Effect of Anderson acceleration on the reference planar 3D ILSA scheme}\label{sec:anderson_acceleration_results}

\begin{figure}[H]
                \centering
                \includegraphics[width=0.99\textwidth]{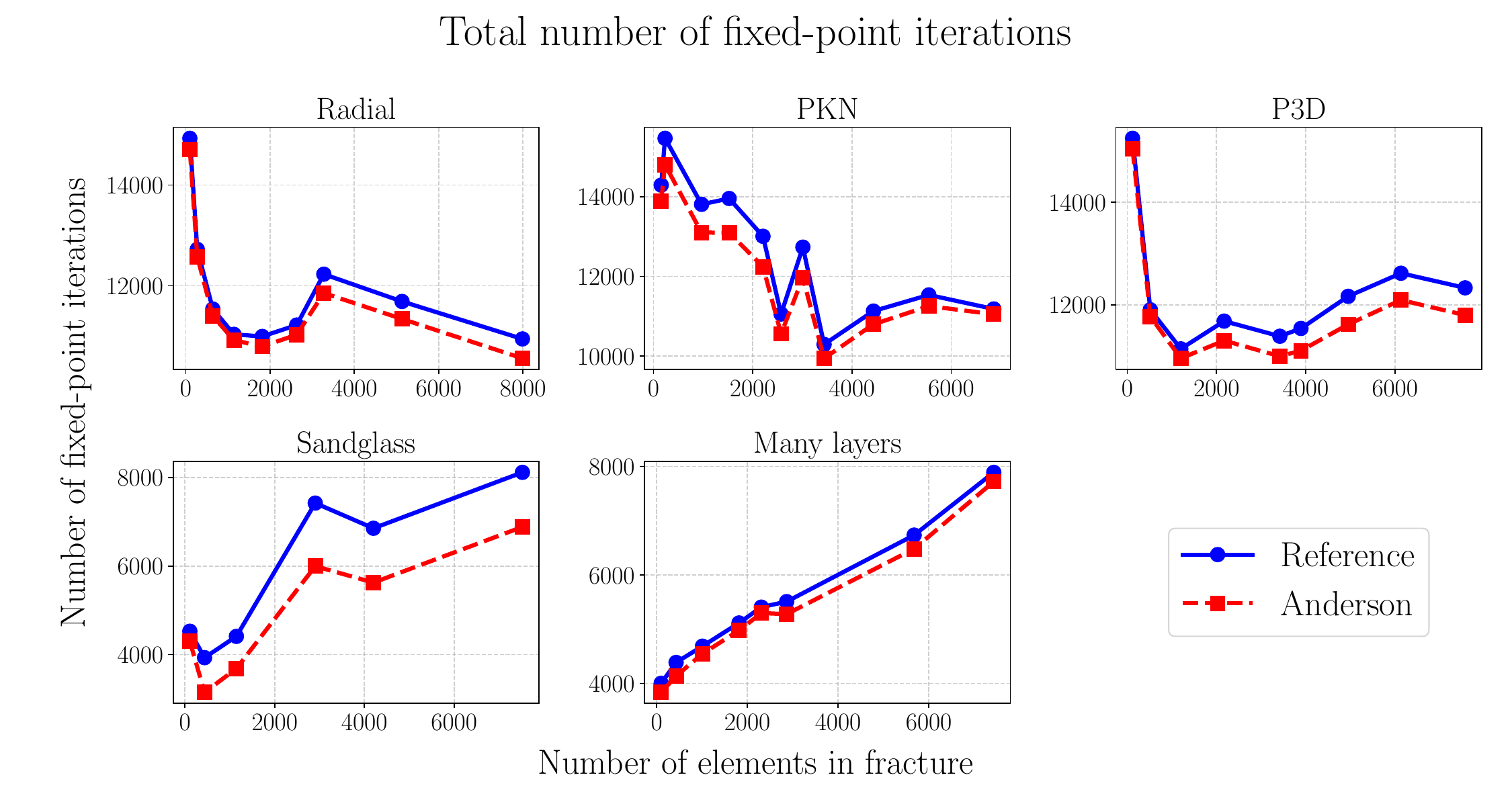}
                \caption{Total fixed-point iteration count of the reference planar 3D ILSA scheme and the reference scheme with Anderson acceleration against the number of fracture elements for the benchmark cases.}
                \label{fig:iterations_nonlinear_anderson_in_orig}
            \end{figure}

            The total number of fixed-point iterations for the reference scheme and the reference scheme with Anderson acceleration is shown in~\Cref{fig:iterations_nonlinear_anderson_in_orig}. Anderson acceleration reduces the iteration count in all cases: the reduction remains modest, within about 6\% for the Radial, PKN, P3D, and Many Layers cases, and is markedly larger, reaching nearly 20\%, for the Sandglass case. This behavior can be explained by the baseline convergence characteristics. For all cases except Sandglass, the nonlinear elastohydrodynamic solver in the reference scheme converges in approximately 3--4 iterations, leaving little scope for improvement given the solver tolerance $\varepsilon_{\text{EHD}}$ of $10^{-6}$. In contrast, the Sandglass case, with an average of 5 fixed-point iterations, provides greater potential for acceleration. The tolerance of $10^{-6}$ is commonly used and generally considered sufficient for practical applications. Additional tests with a stricter tolerance (e.g., $10^{-10}$), which requires more iterations, indicate that Anderson acceleration becomes more effective under such conditions. However, such a stringent tolerance is rarely used in practice, and these results are therefore beyond the scope of the present analysis. The front iteration count remains almost unchanged when Anderson acceleration is introduced (see Fig.~S22 in the Supplementary Material), since the method only influences the convergence of the nonlinear elastohydrodynamic solver and does not modify the front iteration process.

\begin{figure}[H]
                \centering
                \includegraphics[width=0.99\textwidth]{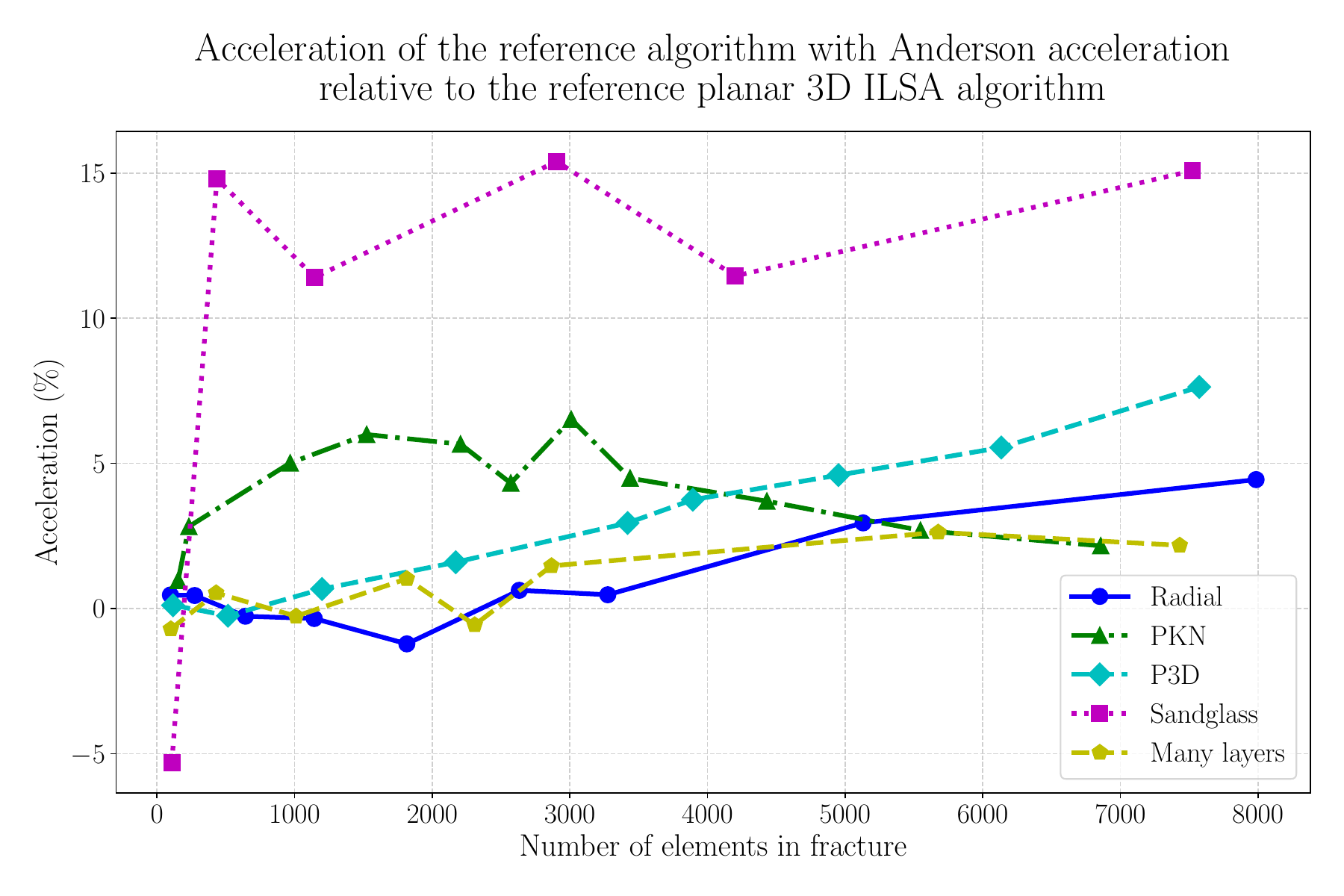}
                \caption{Acceleration resulting from the application of Anderson acceleration to the reference planar 3D ILSA scheme against the number of fracture elements for the benchmark cases.}
                \label{fig:anderson_acceleration}
            \end{figure}

            The acceleration across the benchmark cases is presented in~\Cref{fig:anderson_acceleration}. For the Radial, PKN, P3D, and Many Layers cases, the effect is modest, with gains of up to 8\% of the reference, and consistent with the minimal change in iteration count. For the Sandglass case, the acceleration is more pronounced and stays within 11\%--15\% for all meshes above 400 elements, with a slight slowdown of about 5\% observed only for the coarsest mesh. The computational overhead of Anderson acceleration is typically below 0.1\% of the total time, so that the observed effect stems solely from the change in the iteration count.

        \subsubsection{Effect of the predictor--corrector scheme on the reference planar 3D ILSA scheme}

            To enhance the efficiency of the fracture front tracking, the predictor--corrector scheme~\cite{zia2019explicit} is employed. This approach uses local front velocities from the previous time step to explicitly advance the fracture front and obtain an improved trial front position, which serves as the initial guess for the implicit corrector iterations.

\begin{figure}[H]
                \centering
                \includegraphics[width=0.99\textwidth]{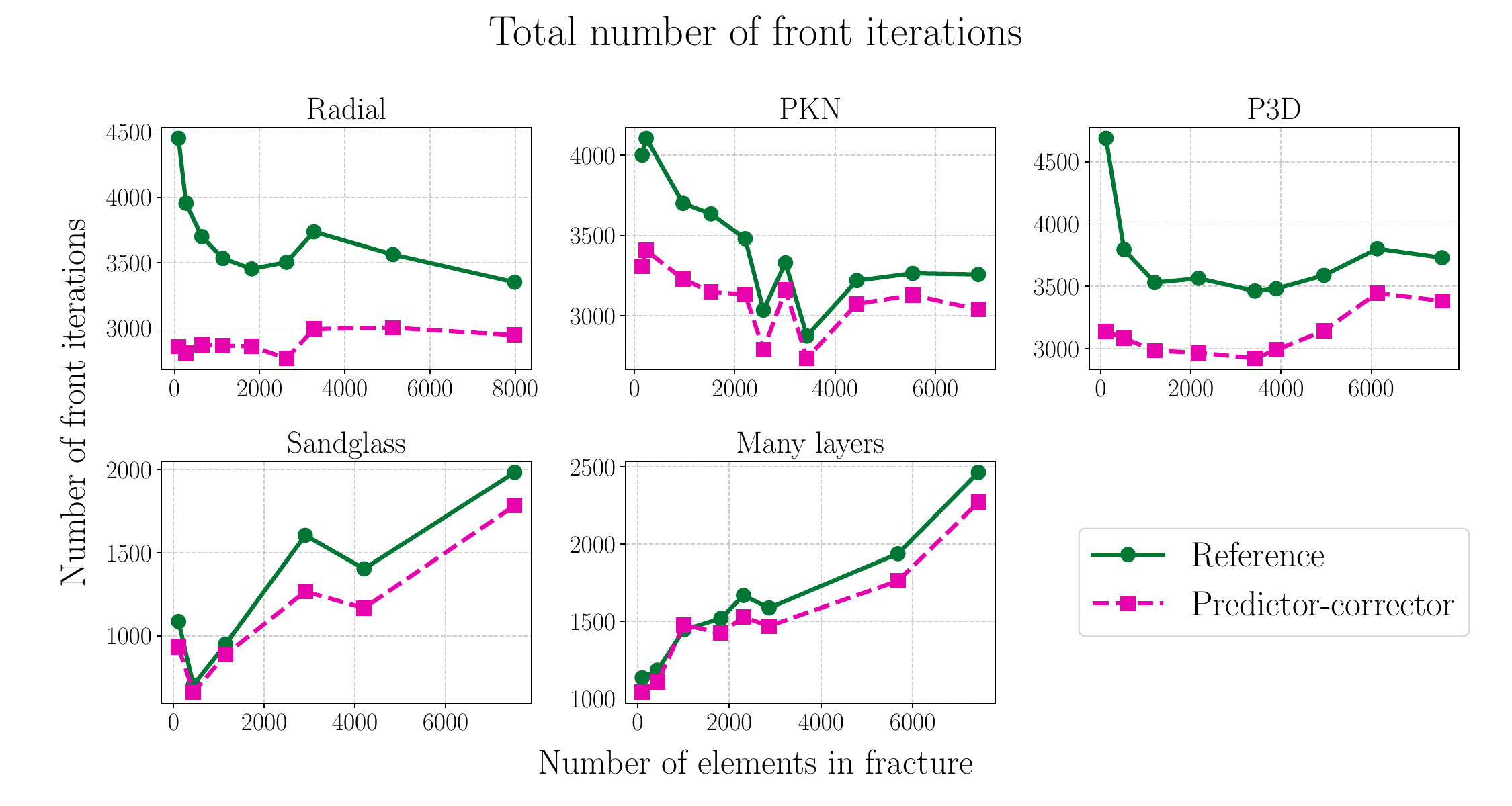}
                \caption{Total front iteration count of the reference planar 3D ILSA scheme and the reference scheme with the predictor--corrector method against the number of fracture elements for the benchmark cases.}
                \label{fig:iterations_front_predict_in_orig}
            \end{figure}

            The total number of front iterations is shown in~\Cref{fig:iterations_front_predict_in_orig}. The predictor--corrector scheme reduces the front iteration count by an average of 13\%, with the reduction reaching 36\% for the Radial case at the coarsest mesh. This improvement is primarily attributed to the accuracy of the initial guess provided by the predictor step: by using the front velocity from the previous time step, the algorithm starts the implicit iterative process closer to the final solution at each time step.

\begin{figure}[H]
                \centering
                \includegraphics[width=0.99\textwidth]{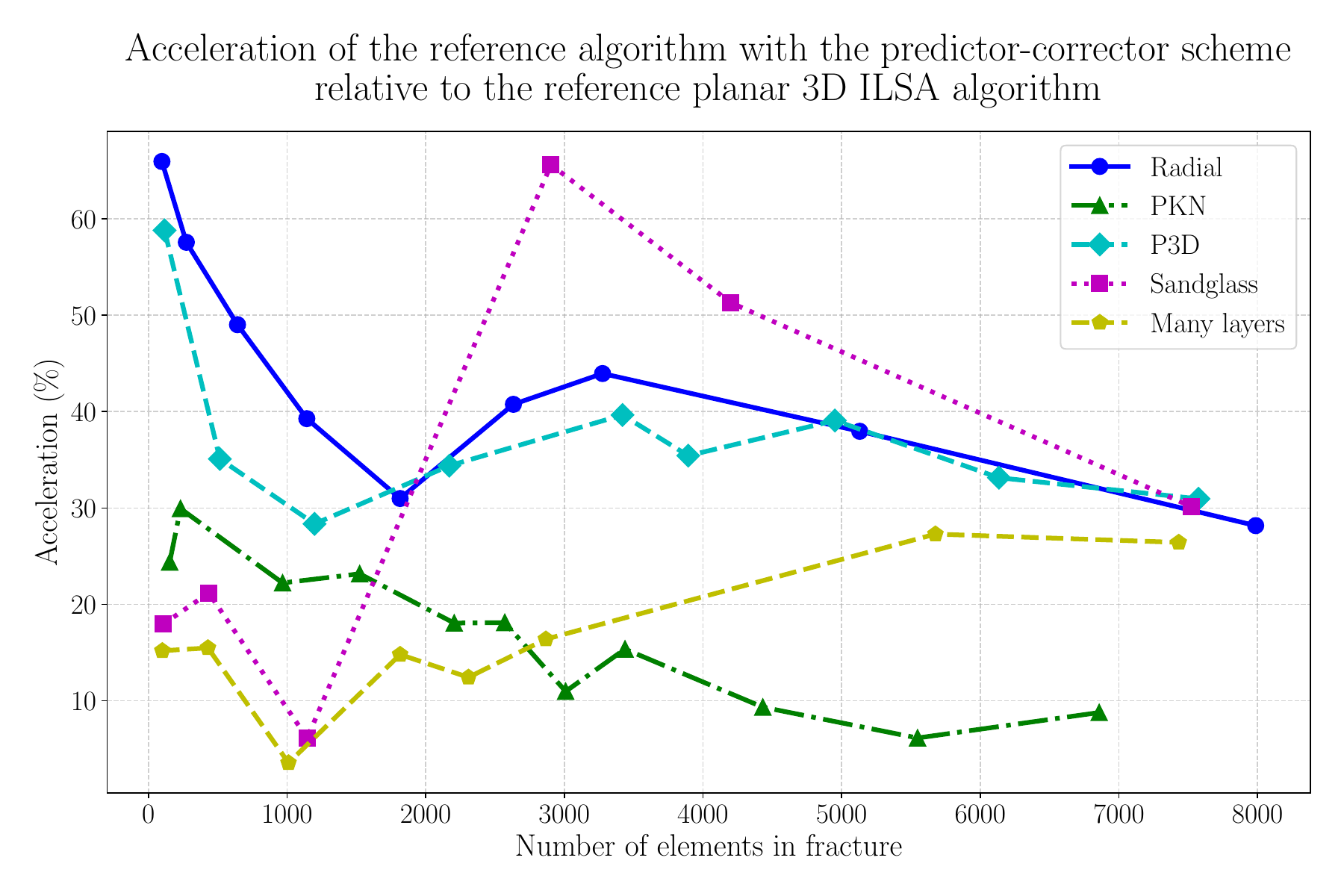}
                \caption{Acceleration resulting from the application of the predictor--corrector scheme to the reference planar 3D ILSA scheme against the number of fracture elements for the benchmark cases.}
                \label{fig:predict_acceleration}
            \end{figure}

            \Cref{fig:predict_acceleration} reports the corresponding acceleration. For the Radial, PKN, and P3D cases the acceleration is largest at the coarsest meshes, ranging from 24\% to 66\% there, and decreases with mesh refinement to 9\%--31\% at the finest meshes. For the Sandglass case the acceleration peaks at approximately 66\% at an intermediate mesh resolution, whereas for the Many Layers case it stays near 15\% over most of the range and grows to approximately 27\% at the two finest meshes. The acceleration stems from the reduced number of front iterations and the associated reduction in elastohydrodynamic solver calls, since each front iteration involves a solution of the elastohydrodynamic system.

            These results are within the range reported in~\cite{zia2019explicit}. In that study, the predictor--corrector scheme is evaluated across several test configurations, including a penny-shaped fracture in a homogeneous permeable medium, accelerating and decelerating fracture front cases with stress heterogeneities, and a numerical simulation of a three-layer laboratory experiment. For the three-layer case, CPU times of 492~s for the predictor--corrector scheme against 619~s for the implicit scheme are reported, corresponding to a reduction in computational time of approximately 20\%. Using the acceleration metric defined in~\eqref{eq:acceleration}, this corresponds to an acceleration of approximately 26\%, which falls within the range of values observed in the present benchmarks. The study~\cite{zia2019explicit} concludes that the predictor--corrector approach reduces computational time by 25\%--50\% relative to the implicit scheme, corresponding to 33\%--100\% acceleration by the metric used in the present study. The results presented in this section, with accelerations of 4\%--66\% depending on mesh resolution, are consistent with the reported range. The differences may be attributed to variations in problem configuration or solver implementation details.

        \subsubsection{Combined effect of the matrix splitting, Anderson acceleration, and the predictor--corrector scheme on the reference planar 3D ILSA scheme}

\begin{figure}[H]
                \centering
                \includegraphics[width=0.99\textwidth]{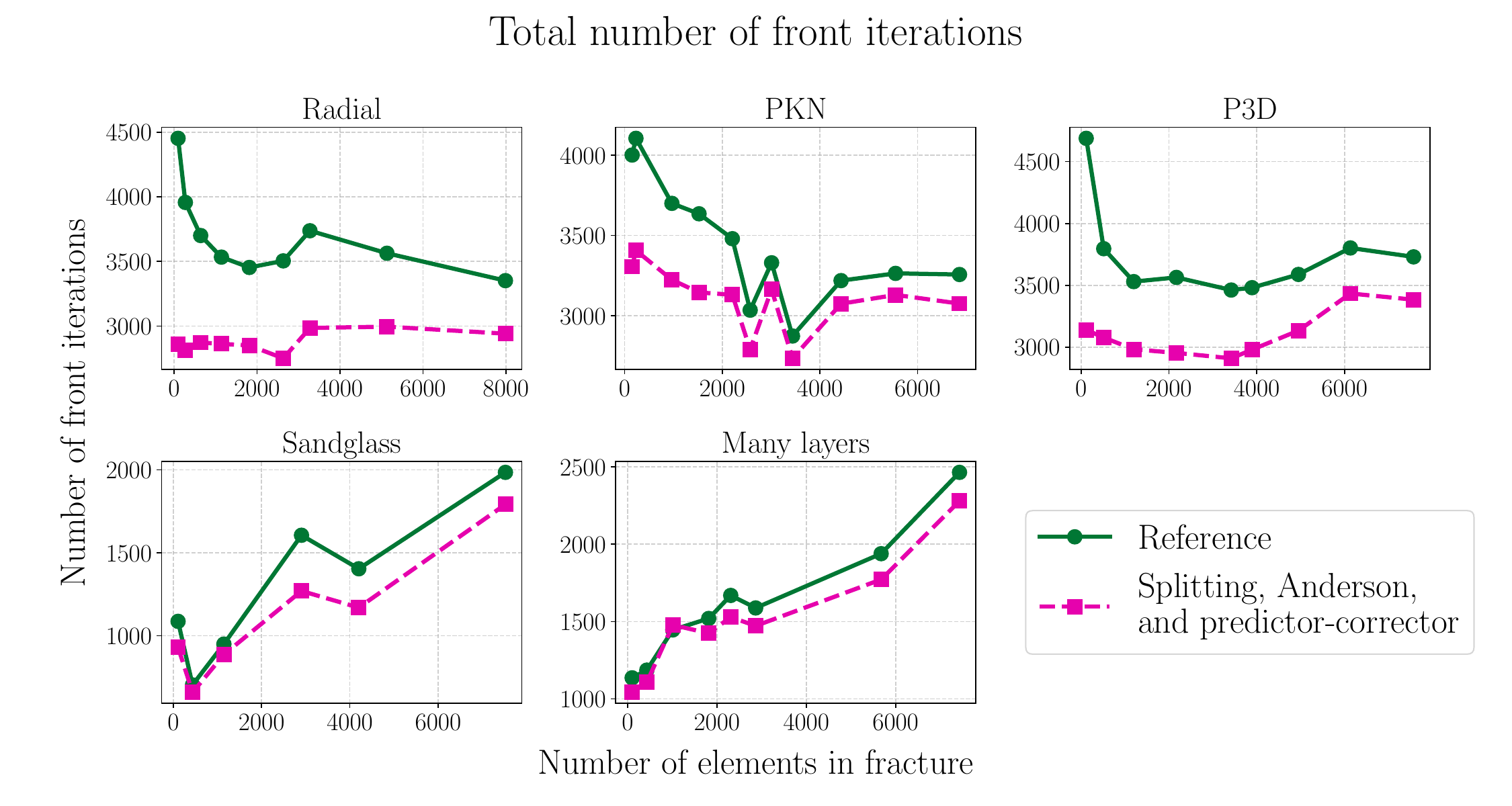}
                \caption{Total number of front iterations for the reference planar 3D ILSA scheme and the reference scheme with the matrix splitting, Anderson acceleration, and the predictor--corrector scheme against the number of fracture elements for the benchmark cases.}
                \label{fig:iterations_front_anderson_split_predict_in_orig}
            \end{figure}

            The total number of front iterations for the reference planar 3D ILSA scheme and the reference scheme with all three acceleration techniques is reported
            in~\Cref{fig:iterations_front_anderson_split_predict_in_orig}. A noticeable reduction in the front iteration count is observed, driven primarily by the
            predictor--corrector scheme, since the matrix splitting and Anderson acceleration have almost no influence on front convergence.

\begin{figure}[H]
                \centering
                \includegraphics[width=0.99\textwidth]{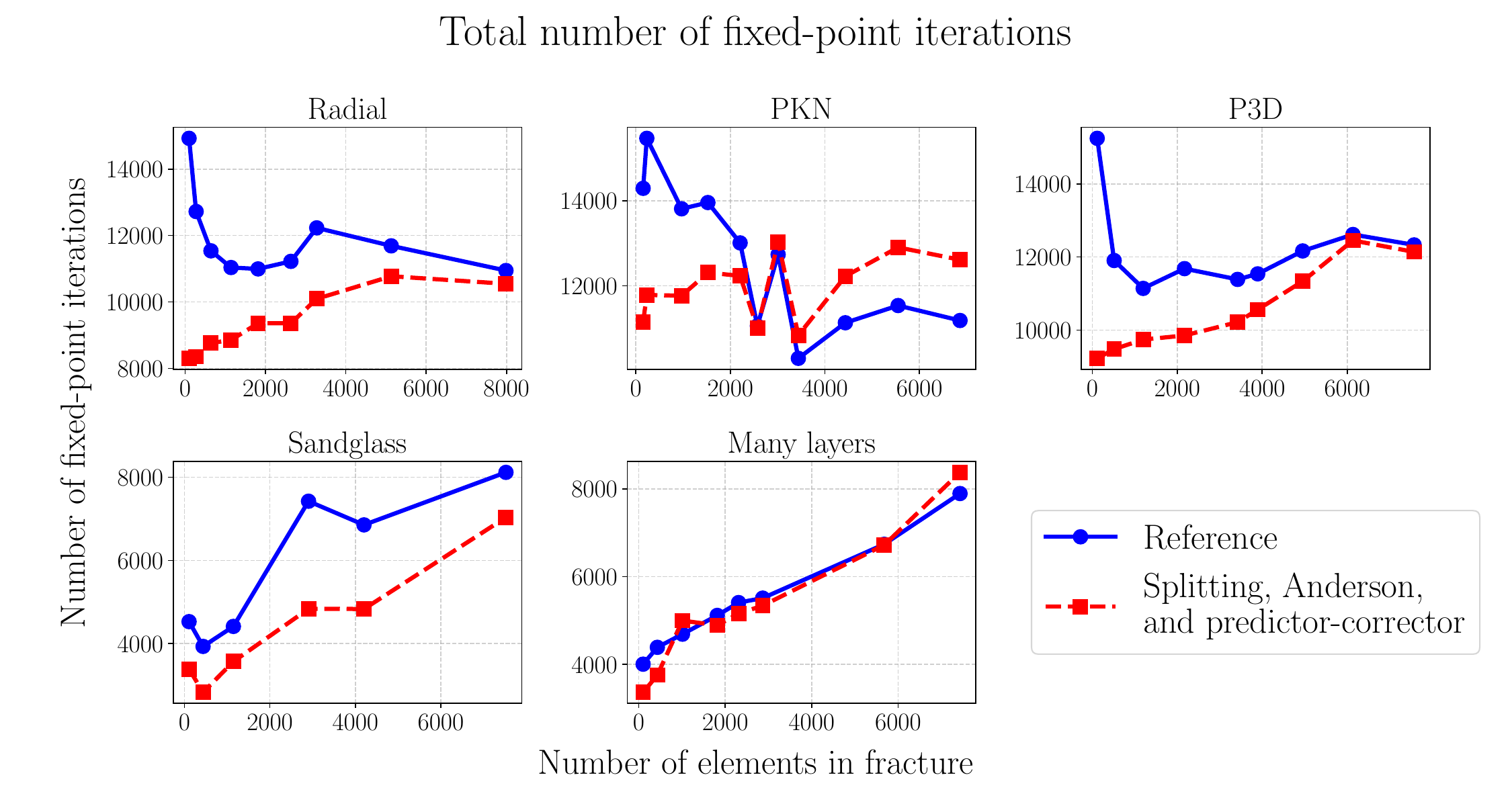}
                \caption{Total number of fixed-point iterations for the reference planar 3D ILSA scheme and the reference scheme with the matrix splitting, Anderson acceleration, and the predictor--corrector scheme against the number of fracture elements for the benchmark cases.}
                \label{fig:iterations_nonlinear_anderson_split_predict_in_orig}
            \end{figure}

            The fixed-point iteration count is shown in~\Cref{fig:iterations_nonlinear_anderson_split_predict_in_orig}. The combined effect of the three techniques produces a resolution-dependent result: the iteration count falls markedly below the reference at coarse grids, and the difference shrinks as the mesh is refined, turning into a slight excess of up to 13\% for the PKN case and up to 6\% for the Many Layers case at the finest grids. This pattern reflects the opposing contributions of the matrix splitting, which increases the iteration count, and Anderson acceleration together with the predictor--corrector scheme, which reduce it. At coarse resolutions the iteration-reducing techniques fully compensate for the splitting-induced overhead, while at finer resolutions the compensation is smaller and, for the PKN and Many Layers cases, no longer suffices to offset it completely.

\begin{figure}[H]
                \centering
                \includegraphics[width=0.99\textwidth]{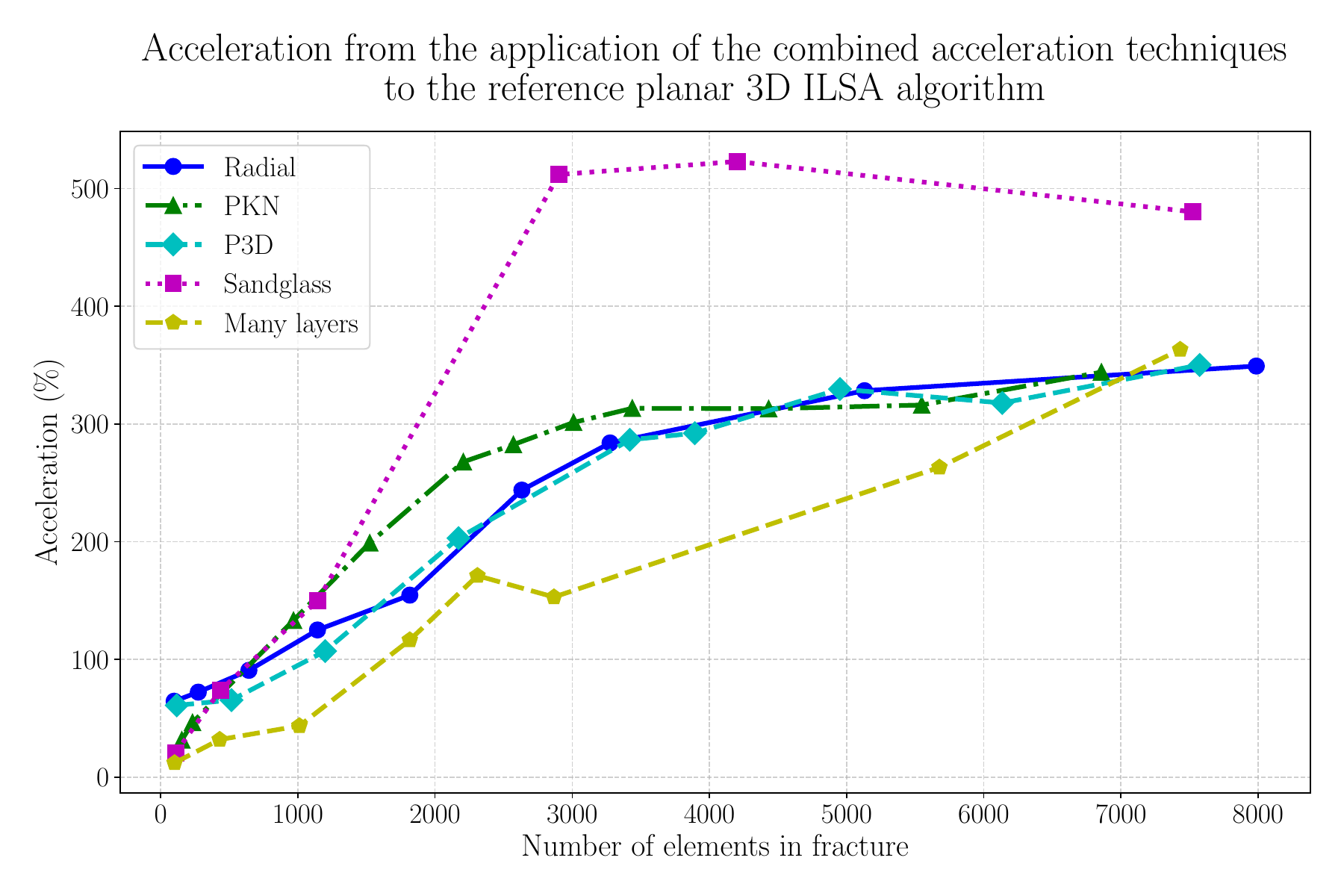}
                \caption{Acceleration resulting from the combined application of the matrix splitting, Anderson acceleration, and the predictor--corrector scheme to the reference planar 3D ILSA scheme against the number of fracture elements for the benchmark cases.}
                \label{fig:anderson_split_predict_acceleration}
            \end{figure}

            The resulting acceleration is presented in~\Cref{fig:anderson_split_predict_acceleration}. The combined application of the three techniques yields a substantial speed-up across all benchmark cases. The techniques act in a complementary fashion. The matrix splitting substantially reduces the per-iteration cost by transforming the system matrix into a sparse one. The per-iteration cost reduction, however, comes with an increased number of fixed-point iterations. The elevated iteration count amplifies the benefit of Anderson acceleration, which improves convergence and reduces the splitting-induced overhead. A further reduction in computational cost is provided by the predictor--corrector scheme through a decrease in the number of front iterations. The acceleration grows with mesh refinement, from 12\%--65\% at the coarsest meshes to 344\%--363\% at the finest ones for the Radial, PKN, P3D, and Many Layers cases. The Sandglass case benefits the most, with the acceleration reaching 523\% at an intermediate resolution and 480\% at the finest mesh. Across all considered cases, the acceleration growth slows down or slightly decreases on fine meshes. This behavior is consistent with the trend described in~\Cref{sec:operator_splitting_results}, where the locality of the matrix splitting stencil leads to the increase in the fixed-point iteration count for finer meshes, causing the growth of the acceleration to slow down or slightly decline.

    \subsection{Performance comparison of the unified and reference planar 3D ILSA schemes}

        The unified planar 3D ILSA scheme introduces a different iteration structure, which requires clarification before a performance comparison can be made. The iterations of the unified scheme are referred to as unified iterations to emphasize their distinction from the iterations of the reference scheme. In the reference scheme, fixed-point iterations are performed at a fixed fracture front, whereas unified iterations simultaneously update the front position and the elastohydrodynamic system solution. Despite the difference, each iteration in both schemes corresponds to one solve of the linearized elastohydrodynamic system, making the total iteration count a meaningful basis for comparison.

\begin{figure}[H]
            \centering
            \includegraphics[width=0.99\textwidth]{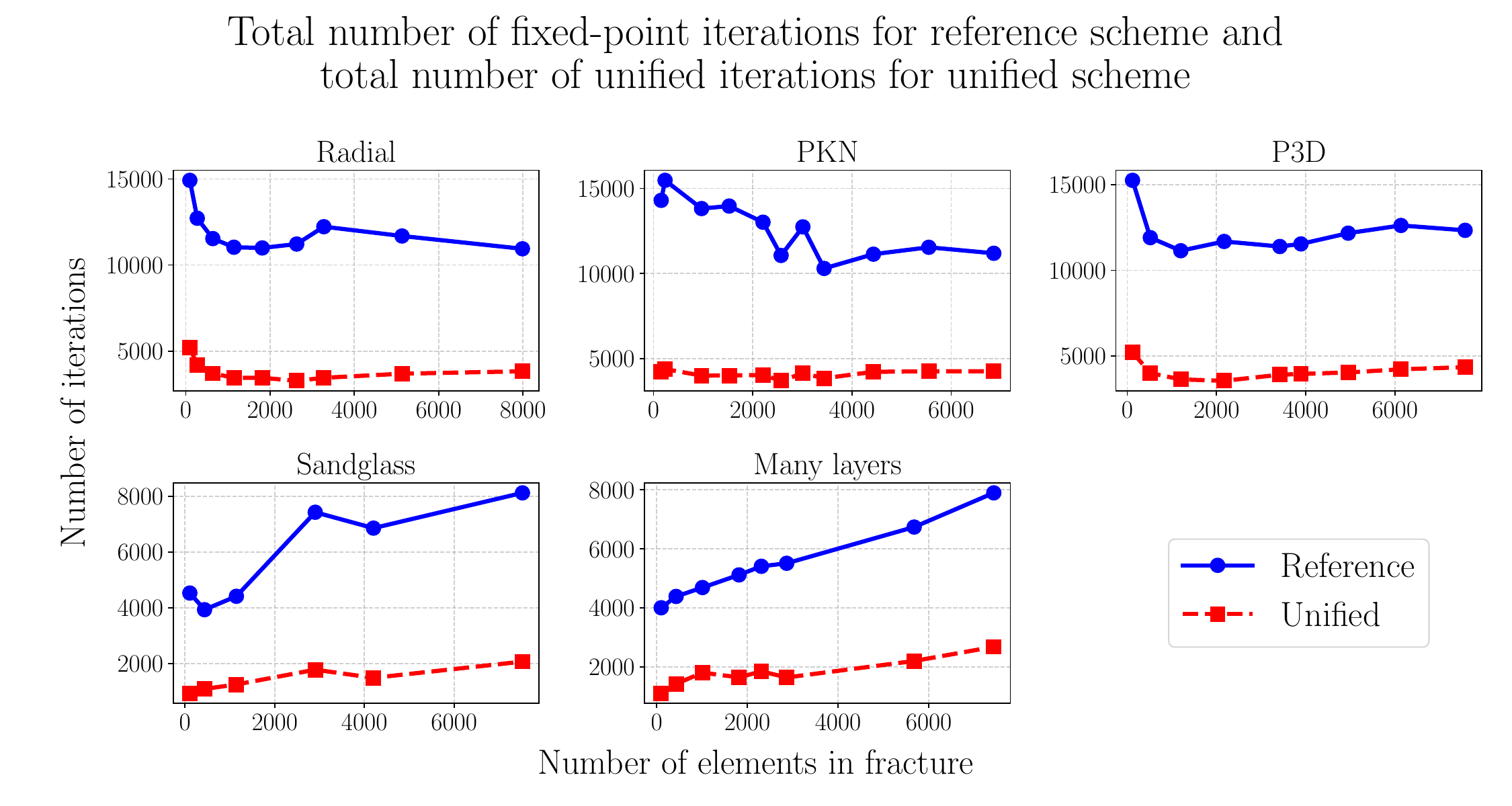}
            \caption{Total number of unified and fixed-point iterations against the number of fracture elements for the unified and the reference planar 3D ILSA schemes for the benchmark cases.}
            \label{fig:iterations_nonlinear_unified}
        \end{figure}

\begin{figure}[H]
            \centering
            \includegraphics[width=0.99\textwidth]{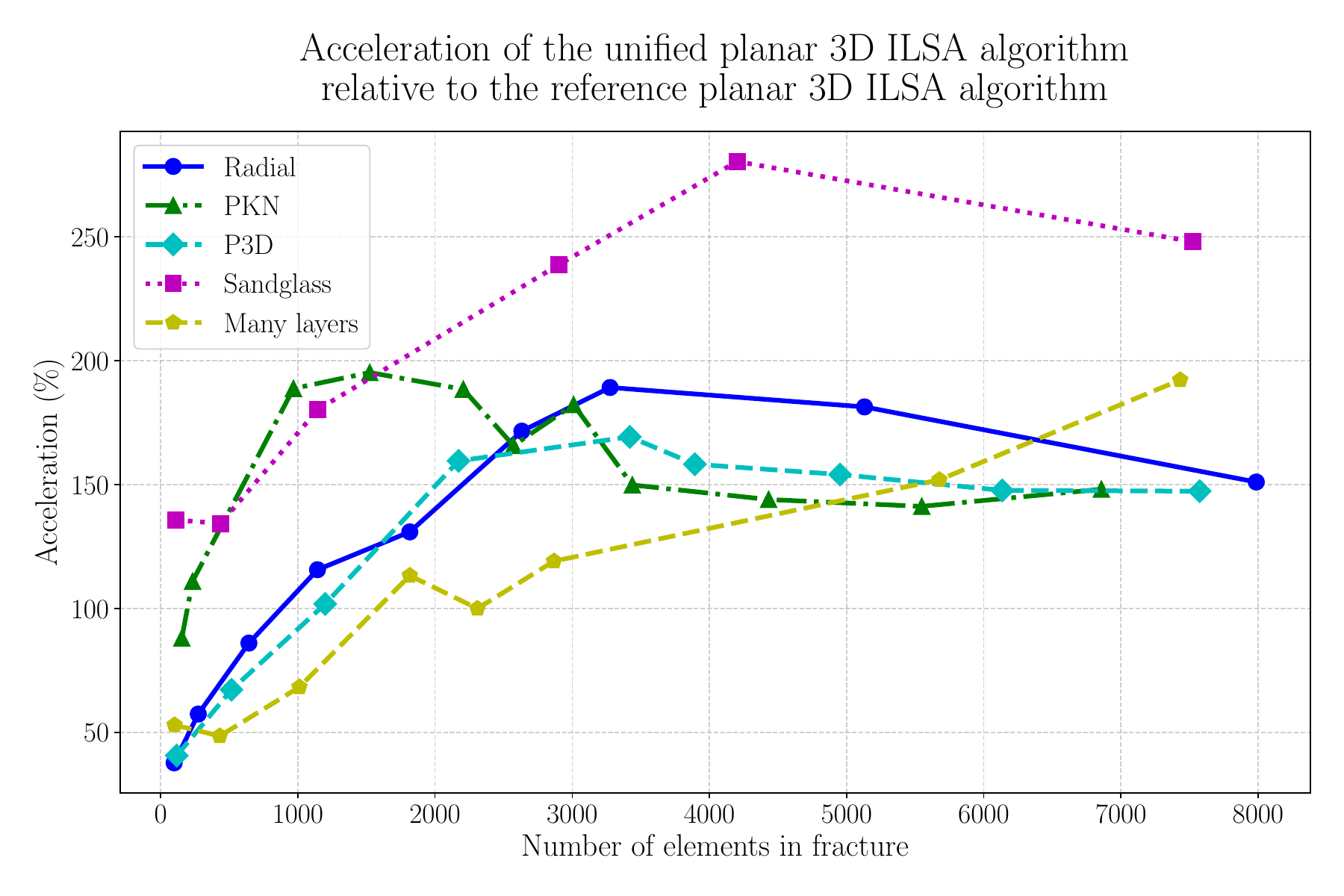}
            \caption{Acceleration of the unified planar 3D ILSA scheme relative to the reference scheme against the number of fracture elements for the benchmark cases.}
            \label{fig:unified_acceleration}
        \end{figure}

        The total number of iterations for both schemes is reported in~\Cref{fig:iterations_nonlinear_unified}. The unified scheme achieves a reduction
        of approximately 62\%--72\% for the Radial, PKN, P3D, and Many Layers cases, and reaches 79\% for the Sandglass case. The acceleration values are presented in~\Cref{fig:unified_acceleration}. For the Radial, PKN, P3D, and Many Layers cases, the acceleration ranges between 38\% and 195\% depending on the mesh resolution. The Sandglass case stands out, with acceleration reaching 280\%. This scenario highlights the efficiency of the unified scheme for cases with complex fracture geometries. In the reference scheme, resolving such a configuration requires a large number of front iterations, each involving full convergence of the nonlinear elastohydrodynamic solver. In contrast, the unified scheme adapts the front position dynamically during the nonlinear solution process. This approach eliminates redundant calculations and significantly reduces the computational cost.

    \subsection{Acceleration techniques applied to the unified planar 3D ILSA scheme}

        With the performance gain of the unified planar 3D ILSA scheme over the reference scheme having been established, this section examines whether the acceleration techniques can further reduce the computational cost of the unified scheme.

        \subsubsection{Effect of the matrix splitting on the unified planar 3D ILSA scheme}

\begin{figure}[H]
                \centering
                \includegraphics[width=0.99\textwidth]{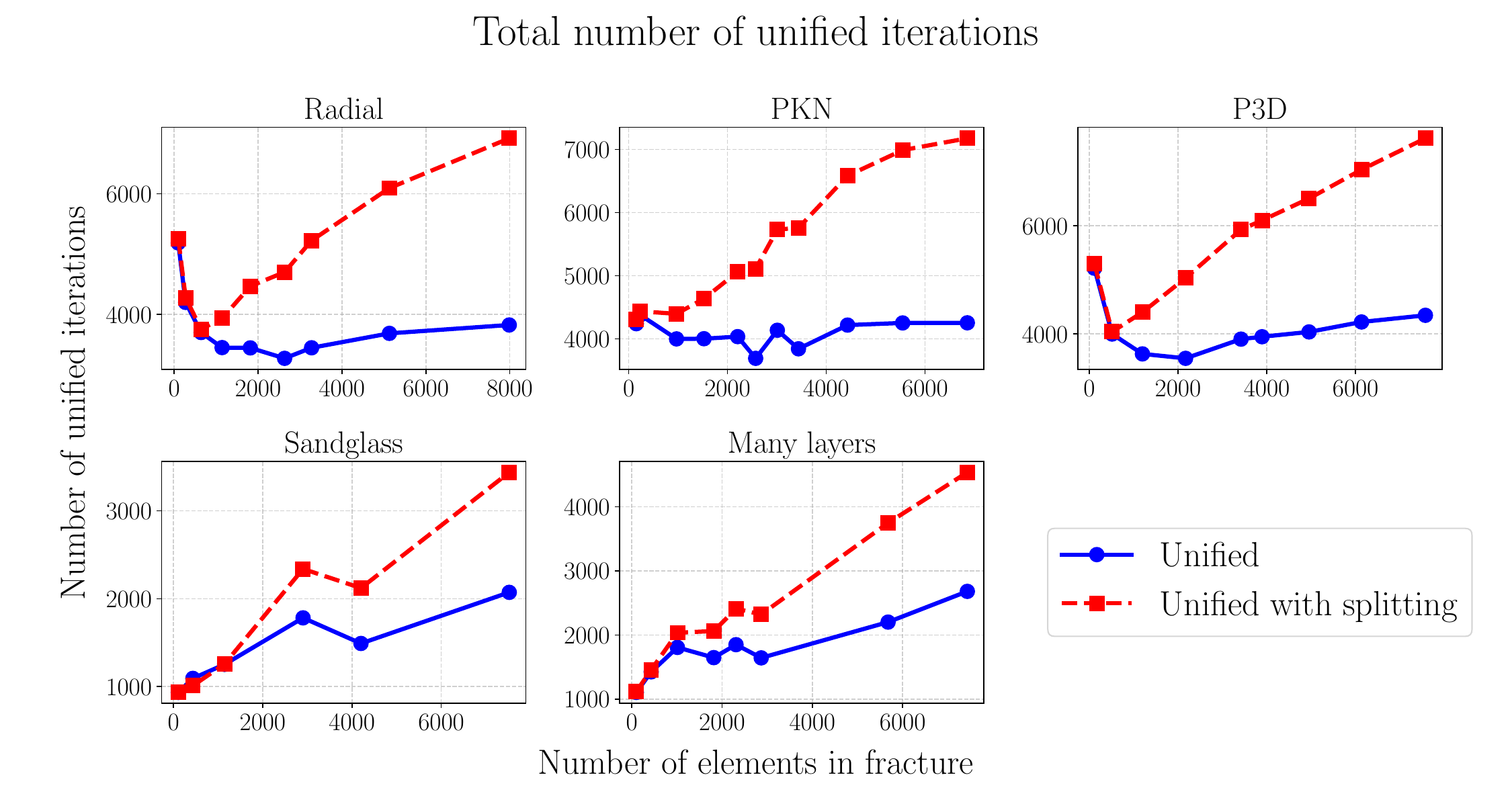}
                \caption{Total number of unified iterations against the number of fracture elements for the unified scheme and the unified scheme with the matrix splitting for the benchmark cases.}
                \label{fig:iterations_nonlinear_unified_split}
            \end{figure}

            The total number of unified iterations for the unified planar 3D ILSA scheme with and without the matrix splitting is presented in~\Cref{fig:iterations_nonlinear_unified_split}. In almost all benchmark cases, the matrix splitting raises the unified iteration count, which is consistent with the behavior observed for the reference scheme in~\Cref{sec:operator_splitting_results}. This increase stems from the approximation of the original dense system matrix by a sparse one: although each iteration becomes less expensive, more iterations are required to reach convergence.

\begin{figure}[H]
                \centering
                \includegraphics[width=0.99\textwidth]{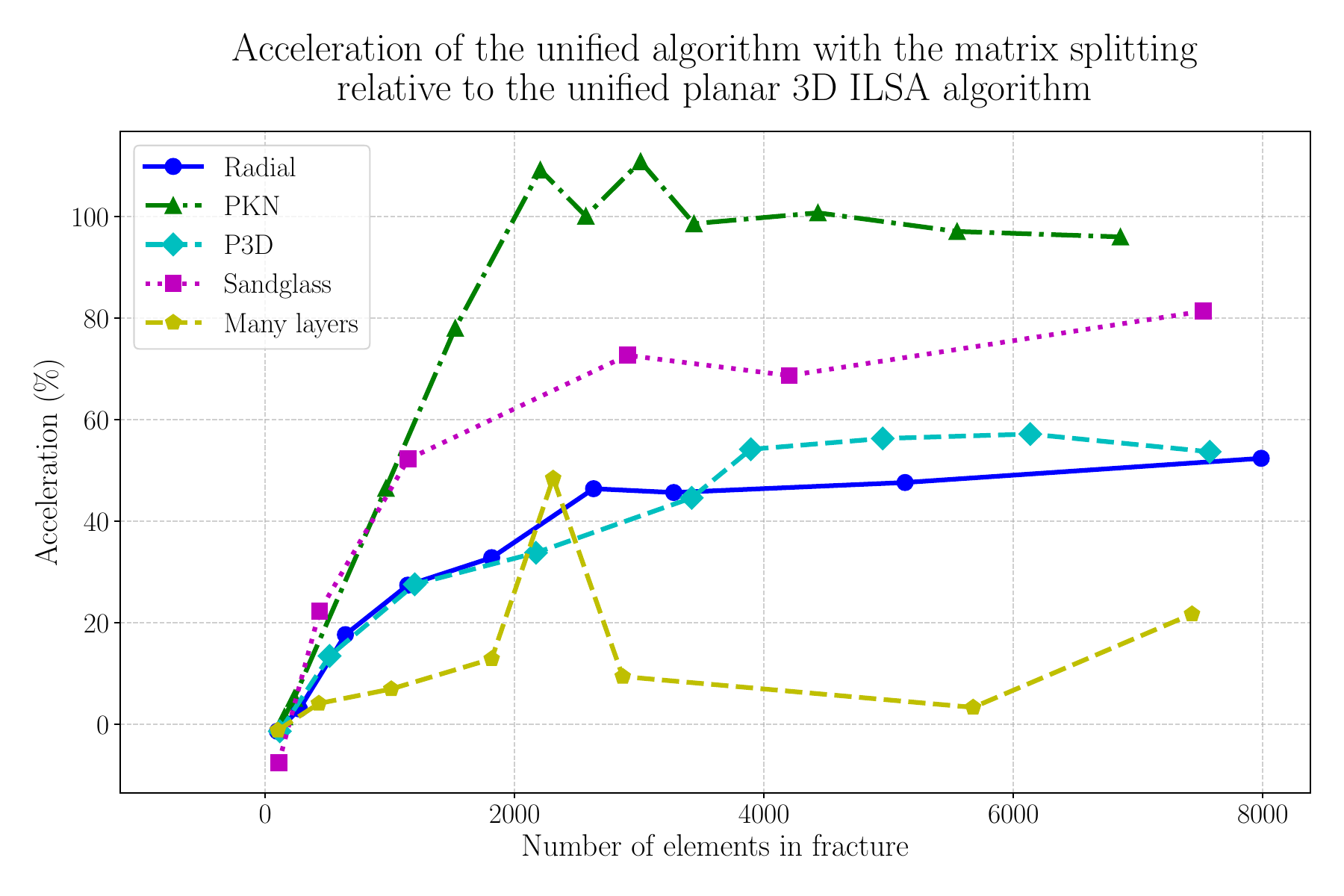}
                \caption{Acceleration resulting from the application of the matrix splitting to the unified planar 3D ILSA scheme against the number of fracture elements for the benchmark cases.}
                \label{fig:unified_split_acceleration}
            \end{figure}

            The acceleration achieved by the matrix splitting is shown in~\Cref{fig:unified_split_acceleration}. In contrast to the reference scheme,
             where the acceleration reaches about 200\%, the gain obtained for the unified scheme is smaller and strongly case-dependent. It grows with mesh refinement for the Radial and Sandglass cases, reaching 52\% and 81\% at the finest meshes, and levels off near 100\% and 55\% for the PKN and P3D cases. The Many Layers case gains the least: the acceleration stays below 13\% at most resolutions and does not exceed 49\%.

        \subsubsection{Effect of Anderson acceleration on the unified planar 3D ILSA scheme}

\begin{figure}[H]
                \centering
                \includegraphics[width=0.99\textwidth]{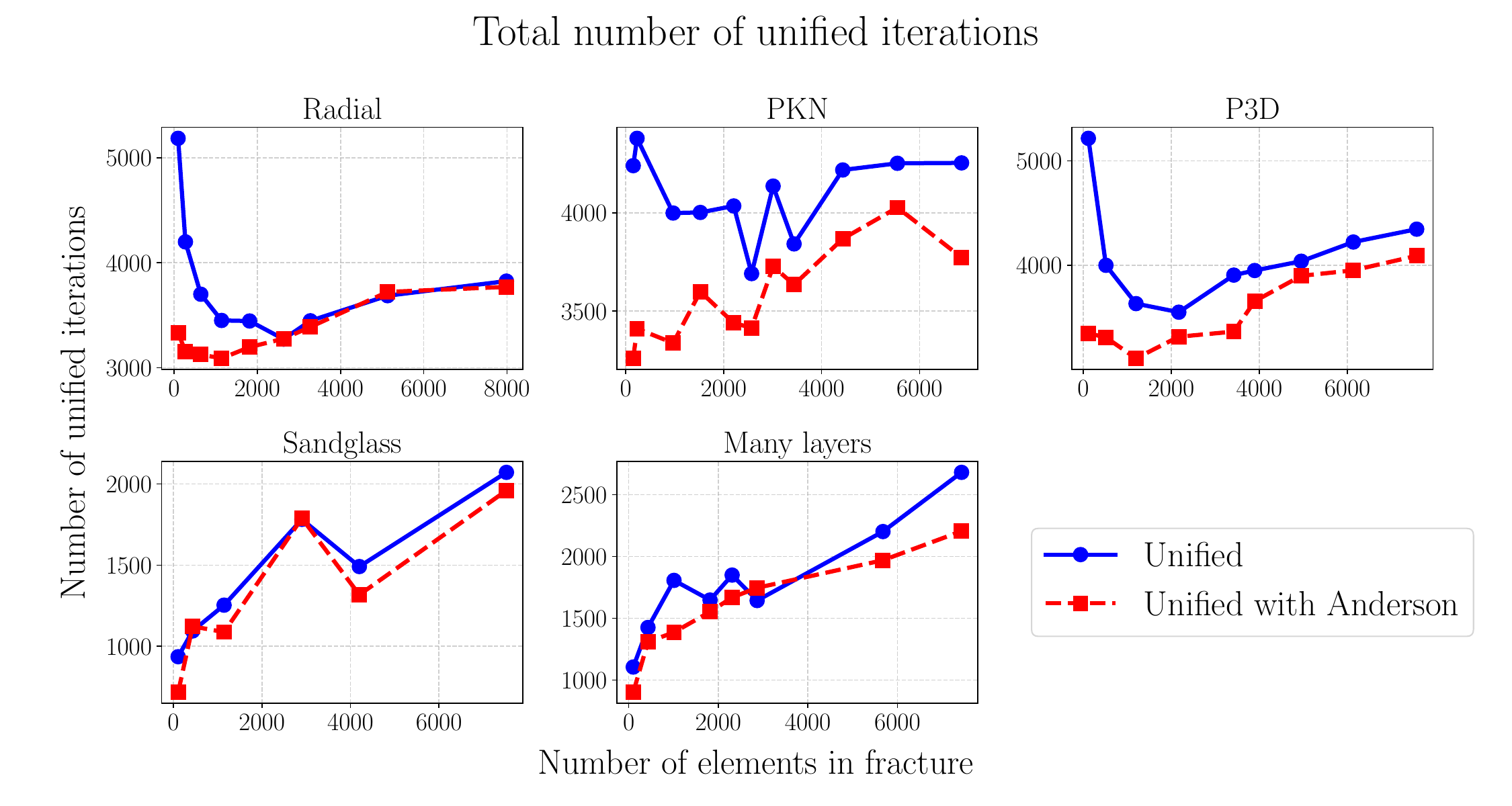}
                \caption{Total number of unified iterations against the number of fracture elements for the unified planar 3D ILSA scheme and the unified scheme with Anderson acceleration for the benchmark cases.}
                \label{fig:iterations_nonlinear_unified_anderson}
            \end{figure}

            \Cref{fig:iterations_nonlinear_unified_anderson} shows the total number of unified iterations for the unified planar 3D ILSA scheme with and without Anderson acceleration. Anderson acceleration reduces the iteration count for nearly all benchmark cases and mesh resolutions. For the Radial, PKN, and P3D cases the reduction is largest at the coarsest mesh, reaching up to 36\%, and diminishes with mesh refinement. For the Sandglass and Many Layers cases the reduction remains irregular across resolutions, varying up to 24\%.

\begin{figure}[H]
                \centering
                \includegraphics[width=0.99\textwidth]{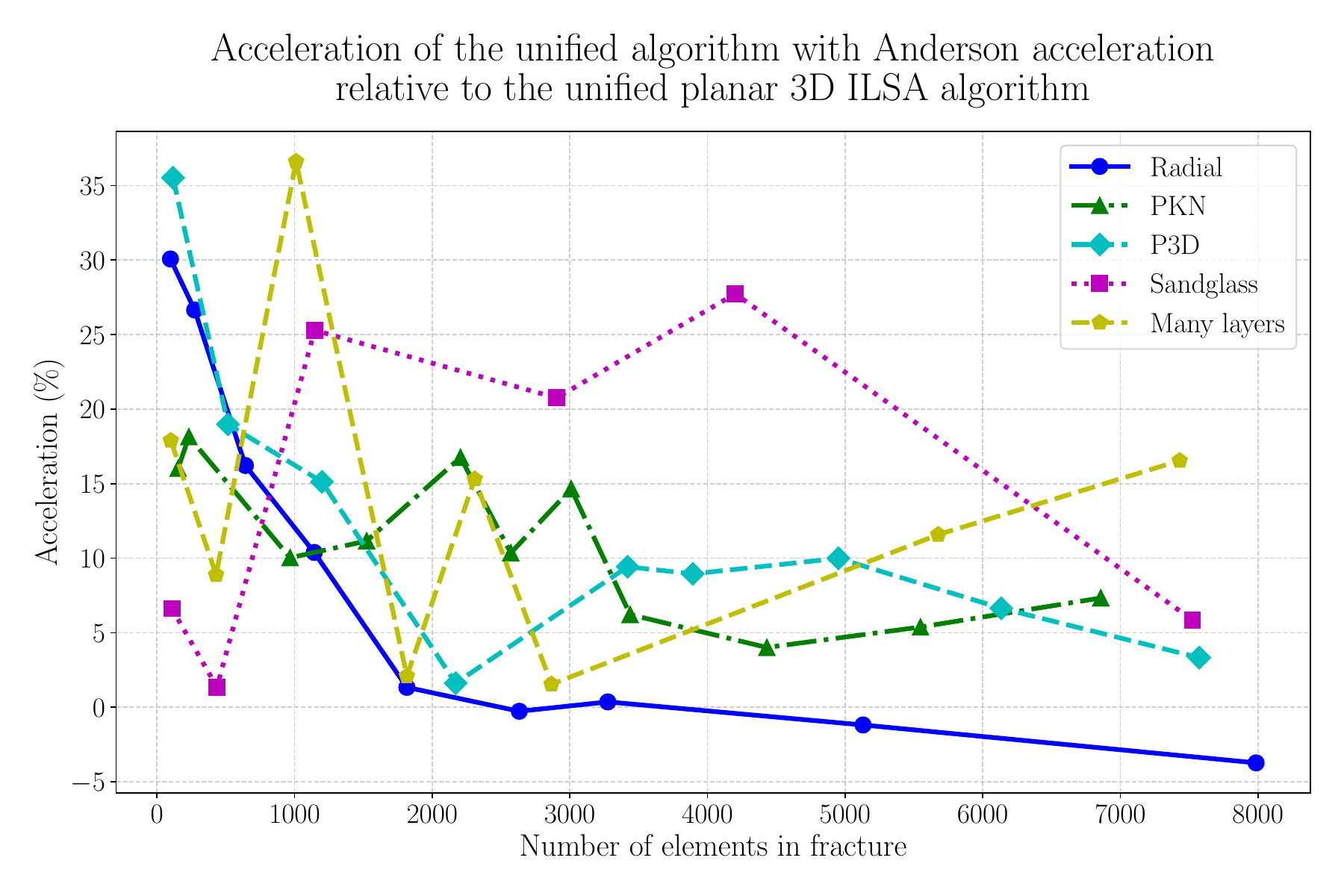}
                \caption{Acceleration resulting from the application of Anderson acceleration to the unified planar 3D ILSA scheme against the number of fracture elements for the benchmark cases.}
                \label{fig:unified_anderson_acceleration}
            \end{figure}

            The resulting acceleration is reported in~\Cref{fig:unified_anderson_acceleration}. The acceleration follows the reduction in the iteration count. It attains 30\%--36\% at the coarsest meshes for the Radial and P3D cases and decreases with mesh refinement. For the PKN case the gain stays within 4\%--18\%, while for the Sandglass and Many Layers cases it reaches up to 28\% and 37\% at intermediate resolutions. This behavior contrasts with the results for the reference planar 3D ILSA scheme (see~\Cref{sec:anderson_acceleration_results}), where Anderson acceleration provides less benefit. The difference is rooted in the algorithmic structure of the two formulations. In the reference scheme, the elastohydrodynamic system is solved at a fixed fracture front and typically converges in a few iterations, leaving little scope for improvement. In the unified scheme, the elastohydrodynamic system solution and the front iterations are coupled within a single iterative loop, resulting in a higher baseline iteration count and a residual that reflects both sources of nonlinearity. The higher number of baseline iterations provides Anderson acceleration with greater scope to reduce their count.

        \subsubsection{Effect of the predictor--corrector scheme on the unified planar 3D ILSA scheme}

\begin{figure}[H]
                \centering
                \includegraphics[width=0.99\textwidth]{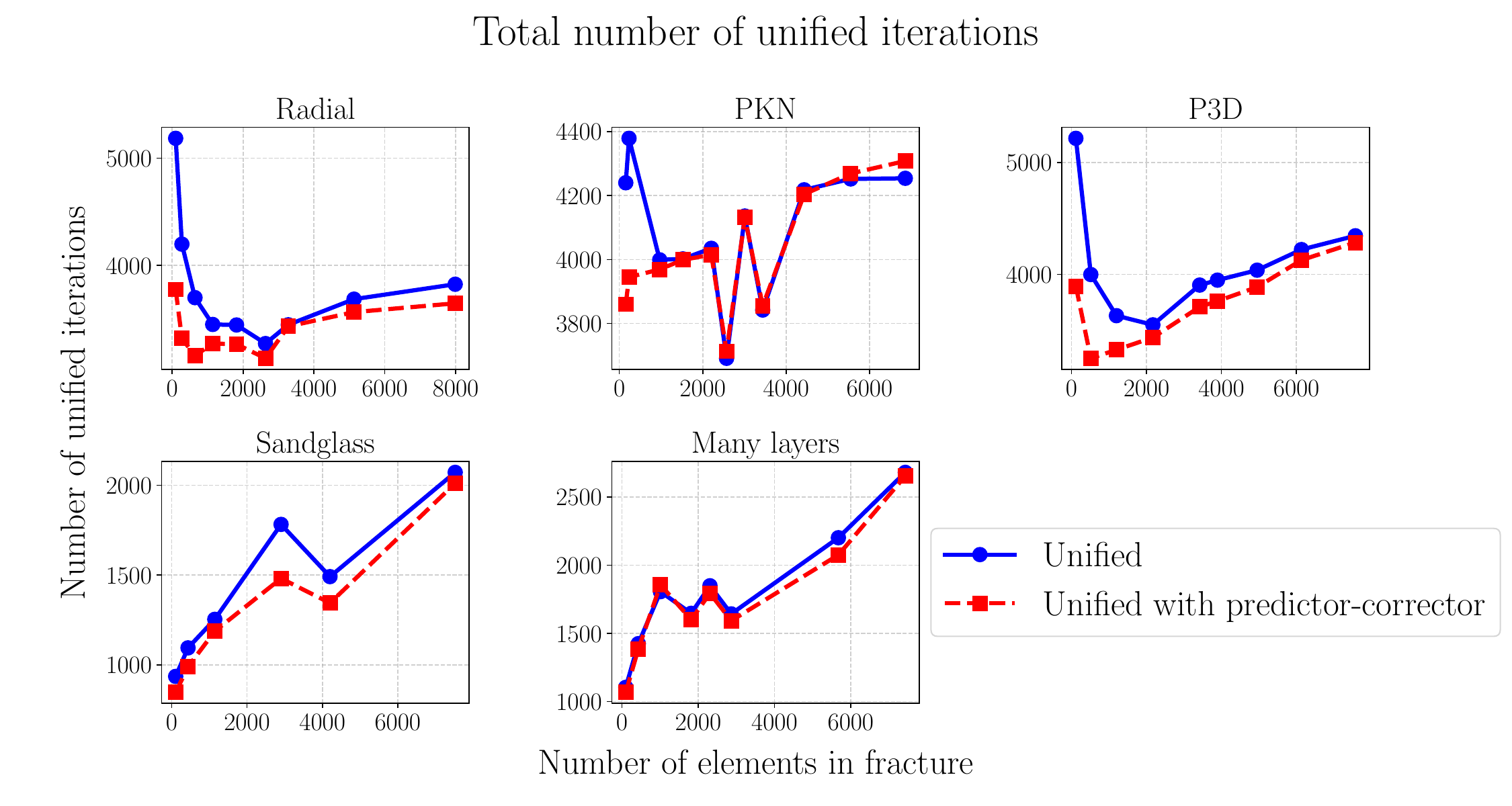}
                \caption{Total number of unified iterations against the number of fracture elements for the unified planar 3D ILSA scheme and the unified scheme with the predictor--corrector method for the benchmark cases.}
                \label{fig:iterations_nonlinear_unified_predictor}
            \end{figure}

            The total number of unified iterations for the unified planar 3D ILSA scheme with and without the predictor--corrector method is shown in~\Cref{fig:iterations_nonlinear_unified_predictor}. For most benchmark cases, the predictor--corrector technique reduces the iteration count by roughly 6\% on average. The reduction is largest at the coarsest meshes for the Radial and P3D cases, where it reaches approximately 27\% and 25\%, and diminishes with mesh refinement.

\begin{figure}[H]
                \centering
                \includegraphics[width=0.99\textwidth]{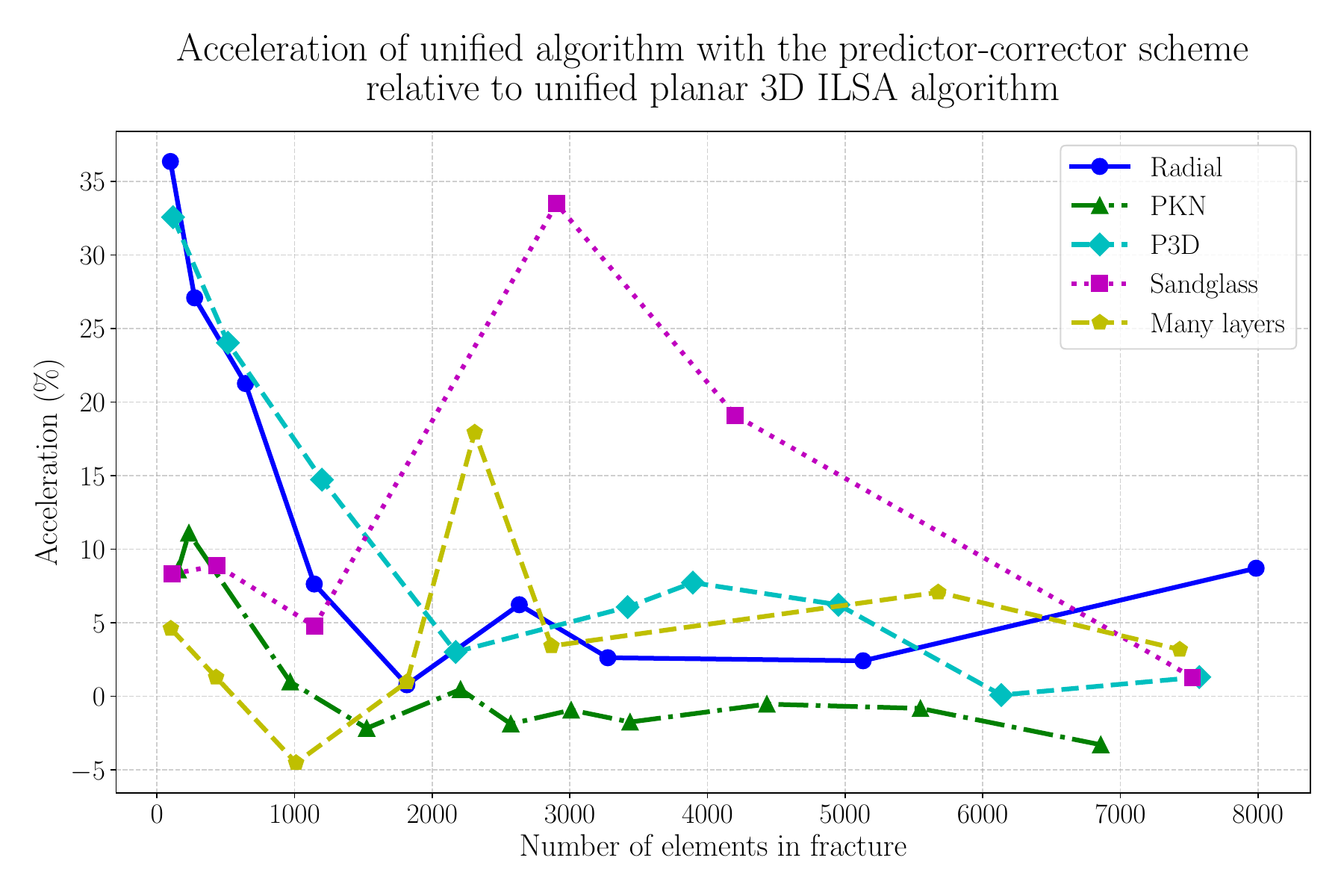}
                \caption{Acceleration resulting from the application of the predictor--corrector scheme to the unified planar 3D ILSA scheme against the number of fracture elements for the benchmark cases.}
                \label{fig:unified_predictor_acceleration}
            \end{figure}

            The corresponding acceleration is reported in~\Cref{fig:unified_predictor_acceleration}. The overall performance impact is modest, with most cases varying within 5\%. Exceptions include the Many Layers case, which reaches 18\% at an intermediate mesh, the Sandglass case, which reaches 34\%, and the Radial and P3D cases, which show a speed-up of about 33\%--36\% at the coarsest resolutions that diminishes as the mesh is refined. The slight slowdown observed for the PKN case at fine meshes is consistent with its marginal increase in iteration count.

            The limited benefit of the predictor--corrector scheme follows from the structure of the unified algorithm. In the reference scheme, the fracture front remains fixed throughout each solve of the nonlinear elastohydrodynamic system. Consequently, an inaccurate initial front approximation directly increases the number of required elastohydrodynamic system solves. In the unified scheme, the front is updated within the same iterative loop as the elastohydrodynamic system solution, meaning that the front position is corrected during convergence regardless of the initial guess. This reduces the sensitivity of the unified scheme to the quality of the initial front estimate and thus diminishes the benefit provided by the predictor--corrector scheme.

        \subsubsection{Combined effect of the matrix splitting, Anderson acceleration, and the predictor--corrector scheme on the unified planar 3D ILSA scheme}

\begin{figure}[H]
                \centering
                \includegraphics[width=0.99\textwidth]{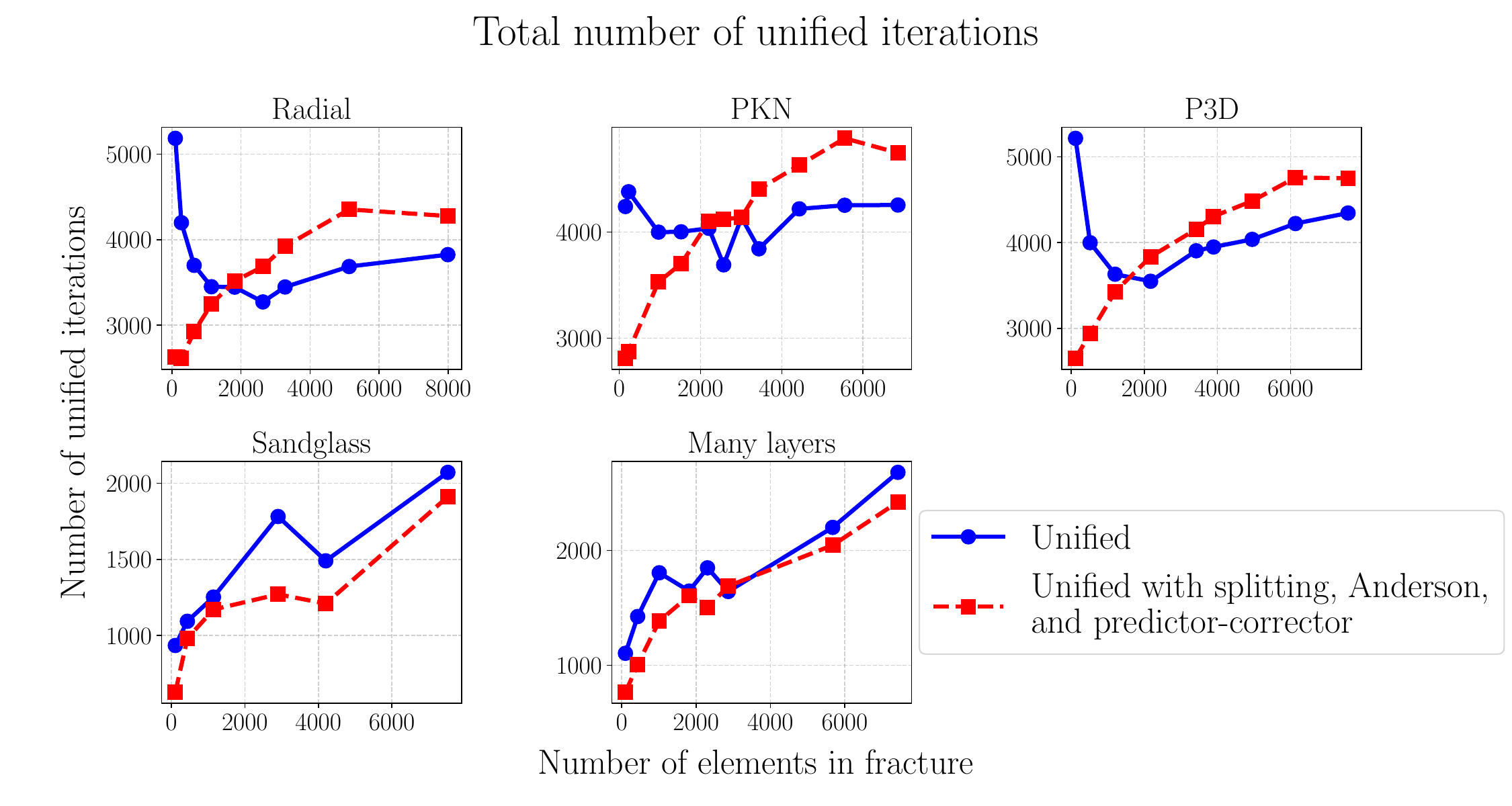}
                \caption{Total number of unified iterations for the unified planar 3D ILSA scheme and the unified scheme with the matrix splitting, Anderson acceleration, and the predictor--corrector scheme against the number of fracture elements for the benchmark cases.}
                \label{fig:iterations_nonlinear_unified_predictor_anderson_split}
            \end{figure}

            The total number of unified iterations for the unified planar 3D ILSA scheme and the unified scheme with the matrix splitting, Anderson acceleration, and the predictor--corrector scheme is presented in~\Cref{fig:iterations_nonlinear_unified_predictor_anderson_split}. For the Radial, PKN, and P3D cases, the effect on the iteration count varies with the mesh resolution: coarse grids yield fewer iterations than the plain unified scheme, while fine grids exhibit a higher count. This behavior reflects the competition between the increased iteration count introduced by the matrix splitting and the reductions provided by Anderson acceleration and the predictor--corrector scheme. Complex cases such as Sandglass and Many Layers, however, show a decrease across almost all mesh resolutions, as their high baseline iteration counts leave greater scope for the convergence-enhancing techniques to outweigh the splitting overhead.

\begin{figure}[H]
                \centering
                \includegraphics[width=0.99\textwidth]{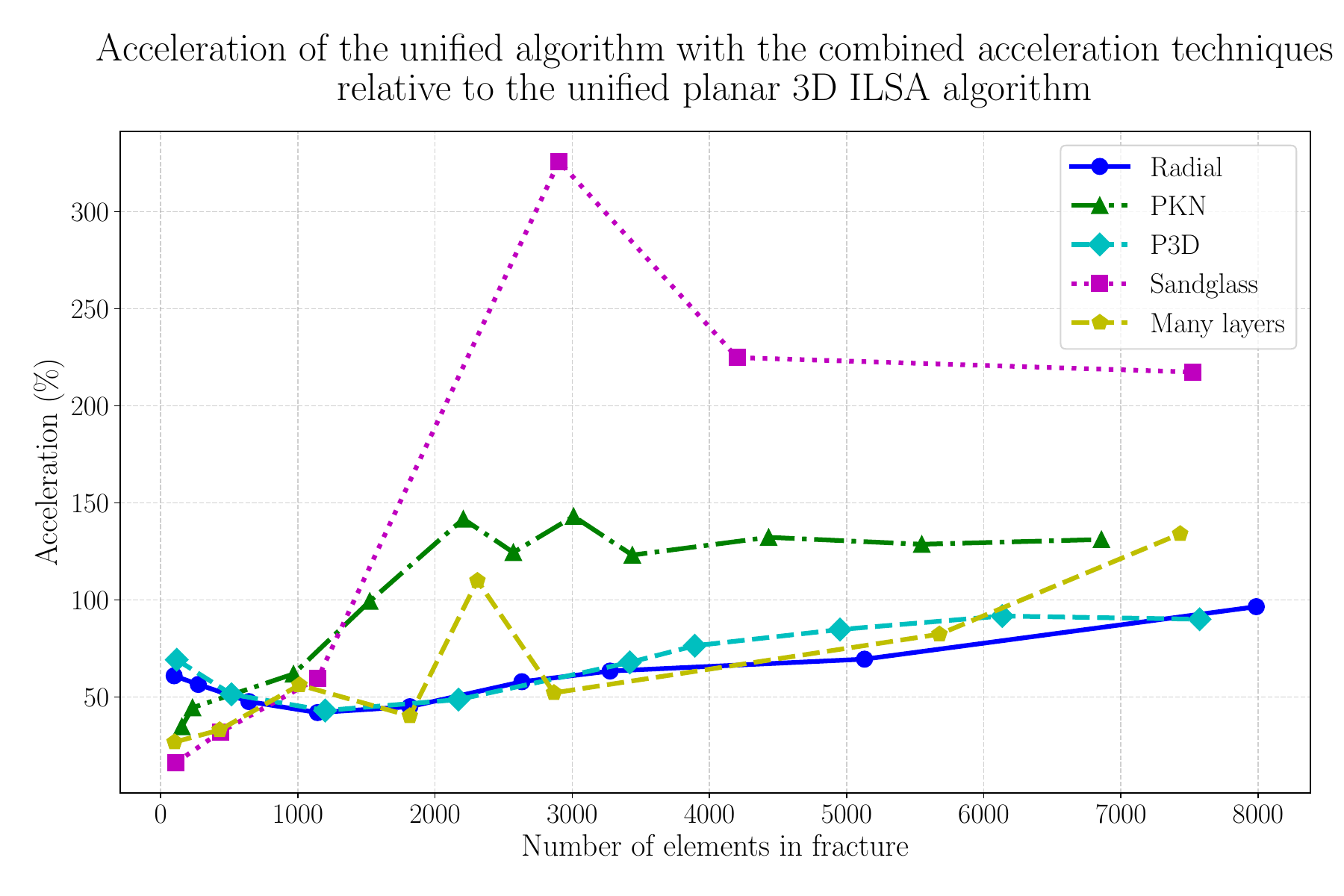}
                \caption{Acceleration for the unified planar 3D ILSA scheme with the matrix splitting, Anderson acceleration, and the predictor--corrector scheme relative to the unified scheme against the number of fracture elements for the benchmark cases.}
                \label{fig:unified_predictor_anderson_split_acceleration}
            \end{figure}

            The resulting acceleration is shown in~\Cref{fig:unified_predictor_anderson_split_acceleration}, reaching up to 143\% relative to the plain unified scheme for the Radial, PKN, P3D, and Many Layers cases, and up to 326\% for the Sandglass case. This gain combines the reduced per-iteration cost from the sparse system matrix with the lower iteration counts from Anderson acceleration and the predictor--corrector scheme.

\begin{figure}[H]
                \centering
                \includegraphics[width=0.99\textwidth]{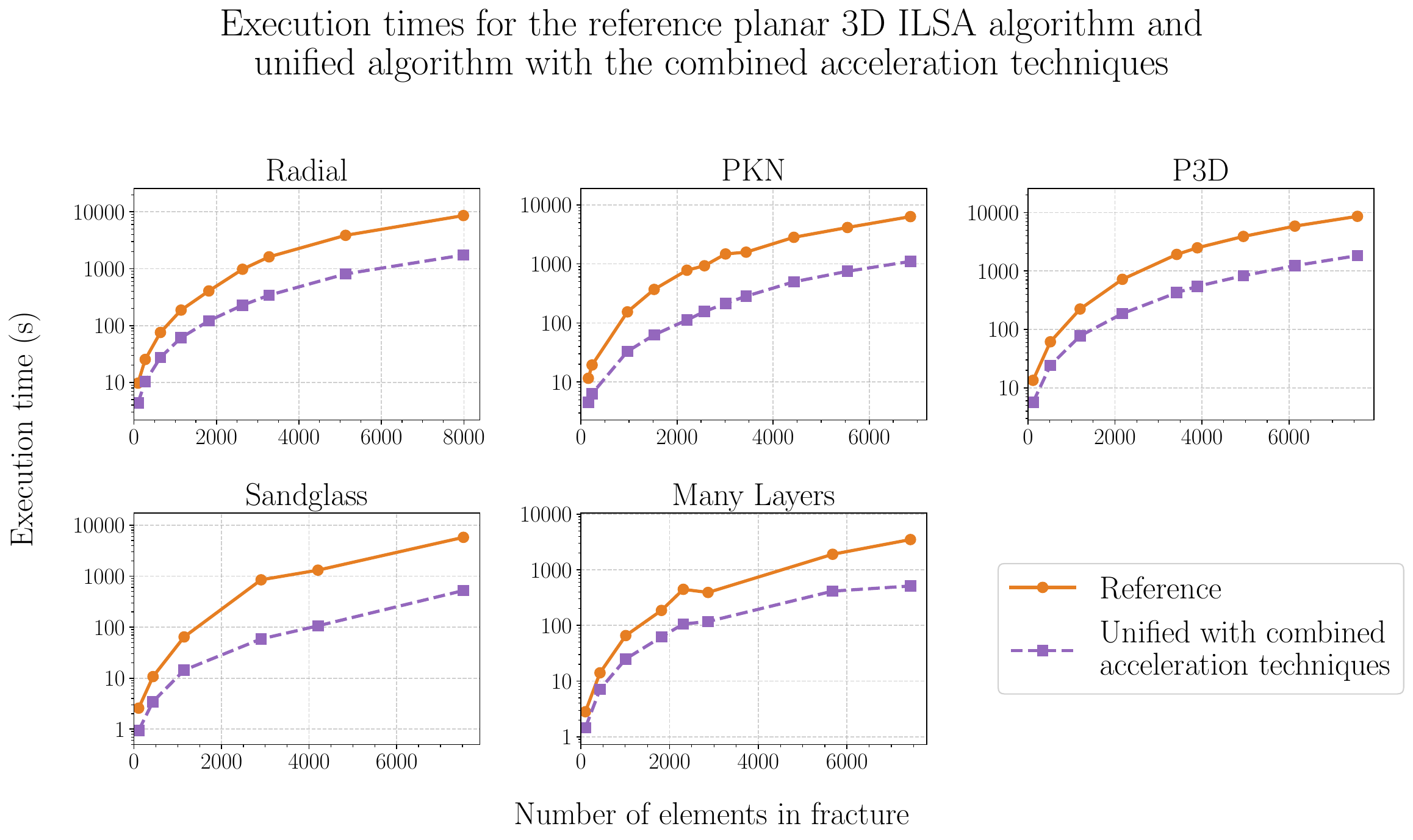}
                \caption{Execution time of the reference planar 3D ILSA scheme and the unified planar 3D ILSA scheme with the matrix splitting, Anderson acceleration, and the predictor--corrector scheme against the number of fracture elements for the benchmark cases.}
                \label{fig:final_execution_times}
            \end{figure}

\begin{figure}[H]
                \centering
                \includegraphics[width=0.99\textwidth]{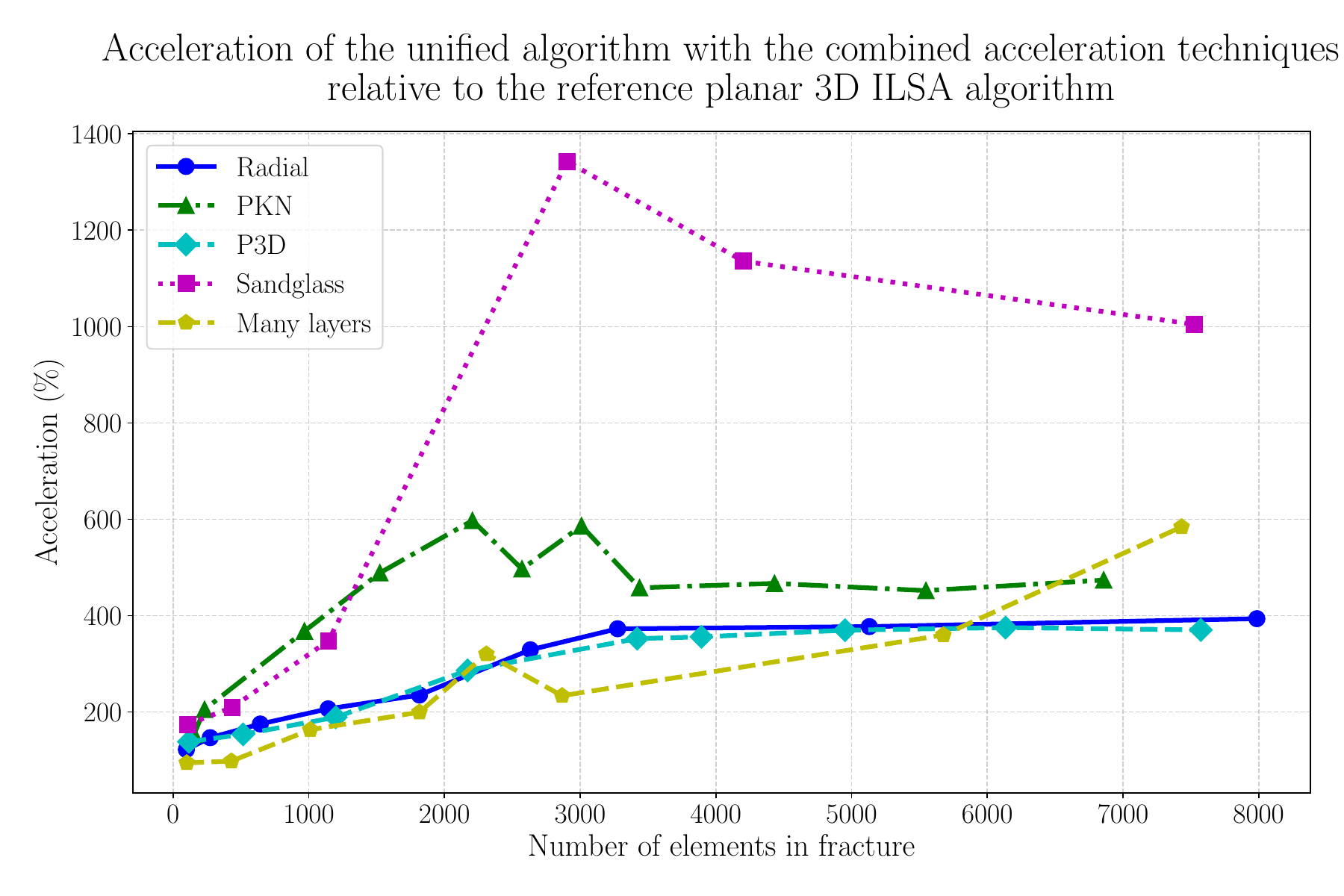}
                \caption{Acceleration of the unified planar 3D ILSA scheme with the matrix splitting, Anderson acceleration, and the predictor--corrector scheme relative to the reference planar 3D ILSA scheme against the number of fracture elements for the benchmark cases.}
                \label{fig:unified_predictor_anderson_split_acceleration_reference}
            \end{figure}

            \Cref{fig:final_execution_times} compares the execution times of the reference planar 3D ILSA scheme and the unified scheme with all acceleration techniques applied across all benchmark cases. The overall performance gain relative to the reference scheme is summarized in~\Cref{fig:unified_predictor_anderson_split_acceleration_reference}. The improvement is substantial across all configurations: the Sandglass case reaches an acceleration of up to 1340\%, while the remaining cases achieve an average acceleration of approximately 320\%. These results confirm that the unified scheme, when combined with all three acceleration techniques, represents a significant advancement over the reference implementation.

    \subsection{Behavior of the schemes during fracture recession and closure}\label{sec:closure_results}

         The benchmark cases considered above cover the propagation stage only. Once the fracture front arrests, the algorithms operate in a different regime, and their relative performance is not predicted by the results obtained during propagation. This stage is therefore examined separately in this section.

         Closure is treated by the contact algorithm introduced in~\Cref{sec:Reference_planar_3D_ILSA_numerical_scheme}. The fracture front remains at the arrested position, and an element whose aperture reaches the prescribed minimum value $\wmin$ is transferred to the active-constraint category. Recession is thus represented by the growth of the contact set on a fixed grid rather than by an explicit motion of the front. A similar treatment underlies the contact-based fixed-grid scheme examined in~\cite{talebkeikhah2025numerical}, where it is compared with the implicit moving-mesh algorithm that tracks the receding front using the recession near-tip asymptote. That comparison shows that the two formulations give close results for reasonable spatial discretizations. The differences are concentrated at the onset of recession, which the contact-based scheme predicts with a delay, and near full closure, where the number of open elements becomes small and the resolution of the fixed grid deteriorates. These conclusions support the use of the contact algorithm adopted here as a basis for the performance study presented below, whose subject is the computational cost of the closure stage rather than the closure physics itself.

        \subsubsection{Setup of the closure benchmarks}

             Two of the benchmark cases of~\Cref{tab:benchmark_cases}, P3D and Many Layers, are considered. The pumping schedule of each case is extended by a shut-in stage with zero injection rate, during which the fracture deflates and closes. To reach closure within a reasonable simulated time, the leak-off coefficient is increased by one order of magnitude with respect to~\Cref{tab:benchmark_cases}, and the injection stage of the Many Layers case is extended from $1 \times 10^{3}$ to $2 \times 10^{3}$~s. All remaining material and fluid parameters, as well as the stress profiles shown in~\Cref{fig:cases_stresses}, are unchanged. The results are presented for a single mesh resolution for each case. The modified parameters and the mesh description are collected in~\Cref{tab:closure_cases}.

            \begin{table}[H]
                \caption{Parameters of the closure benchmarks, including those that differ from~\Cref{tab:benchmark_cases}, together with the mesh description, the number of fracture elements of the arrested footprint, and the resulting front arrest time. All other parameters are those of the corresponding case in~\Cref{tab:benchmark_cases}.}
                \centering
                \begin{tabular}{lcc}
                \toprule
                \textbf{Parameter} & \textbf{P3D} & \textbf{Many Layers} \\
                \midrule
                \textbf{Injection time}, $10^{3}\,\text{s}$ & $5$ & $2$ \\
                \textbf{Shut-in time}, $10^{3}\,\text{s}$ & $1.5$ & $0.75$ \\
                \textbf{Leak-off}, $10^{-5}\,\text{m}/\text{s}^{1/2}$ & $10$ & $10$ \\
                \textbf{Cell size} $\Delta x \times \Delta y$, m & $3.13 \times 3.13$& $2.22 \times 2.22$ \\
                \textbf{Number of fracture elements} $N$ & $875$ & $1074$ \\
                \textbf{Front arrest time}, $10^{3}\,\text{s}$ & $5.33$ & $2.12$ \\
                \bottomrule
                \end{tabular}
                \label{tab:closure_cases}
            \end{table}

             To isolate the effect of each acceleration technique on the closure phase itself, every variant employs the reference planar 3D ILSA scheme throughout propagation, and the corresponding technique is switched on only from the arrest of the front onward. All quantities reported in this section, both iteration counts and execution times, are accumulated over the closure phase only, that is, from the arrest of the fracture front to the end of the simulation. The acceleration is evaluated according to~\eqref{eq:acceleration} with the execution time of the closure phase used in place of the total simulation time, and the reference planar 3D ILSA scheme is taken as the baseline for all variants, including those built on the unified scheme. In the present setup the closure phase accounts for 6\% of the total runtime of the reference scheme for the P3D case and for 14\% for the Many Layers case.

        \subsubsection{Iteration counts during closure}

\begin{figure}[H]
                \centering
                \includegraphics[width=0.99\textwidth]{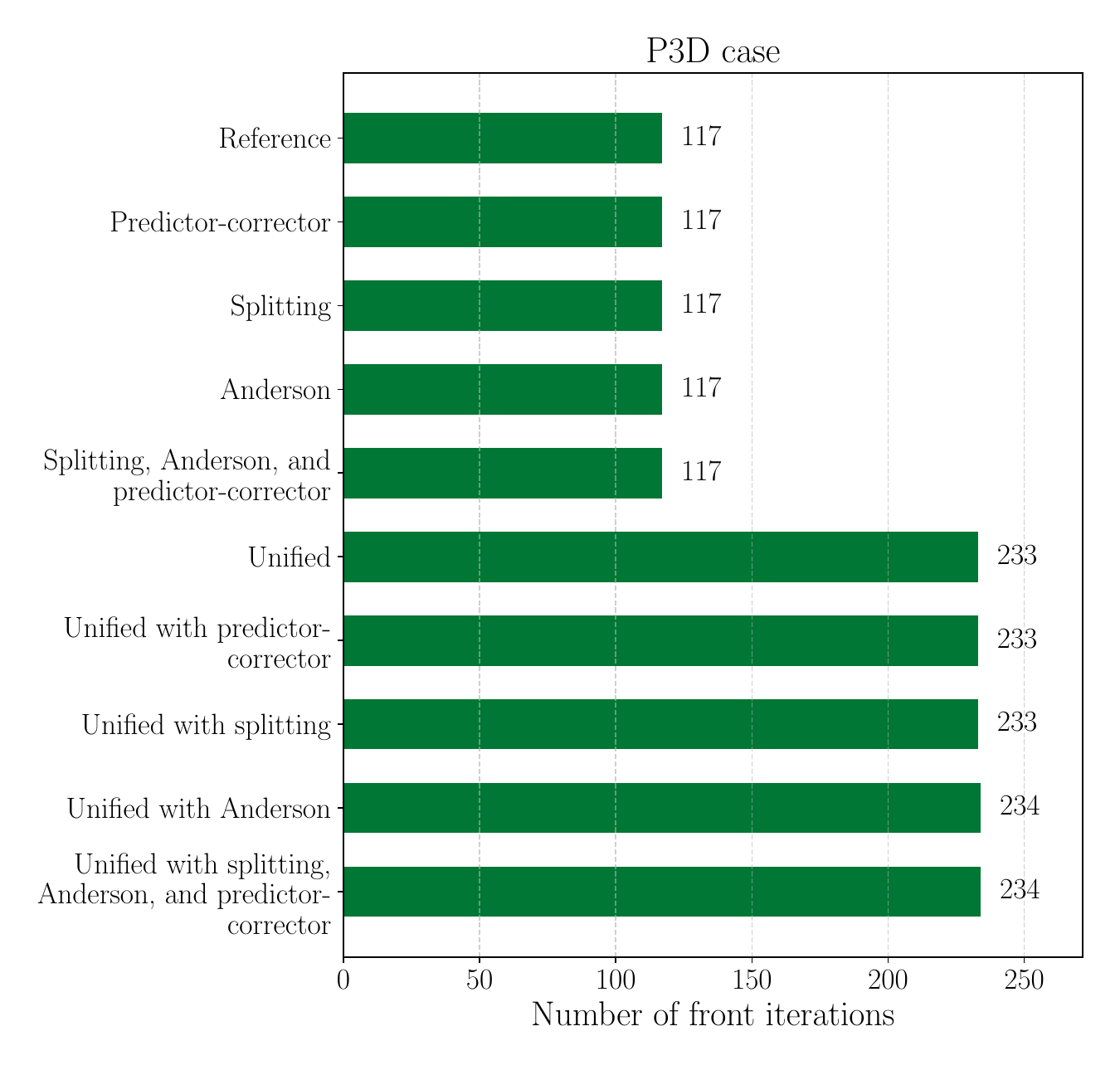}
                \caption{Total number of front iterations accumulated over the closure phase for the P3D case.}
                \label{fig:closure_front_iterations_p3d}
            \end{figure}

\begin{figure}[H]
                \centering
                \includegraphics[width=0.99\textwidth]{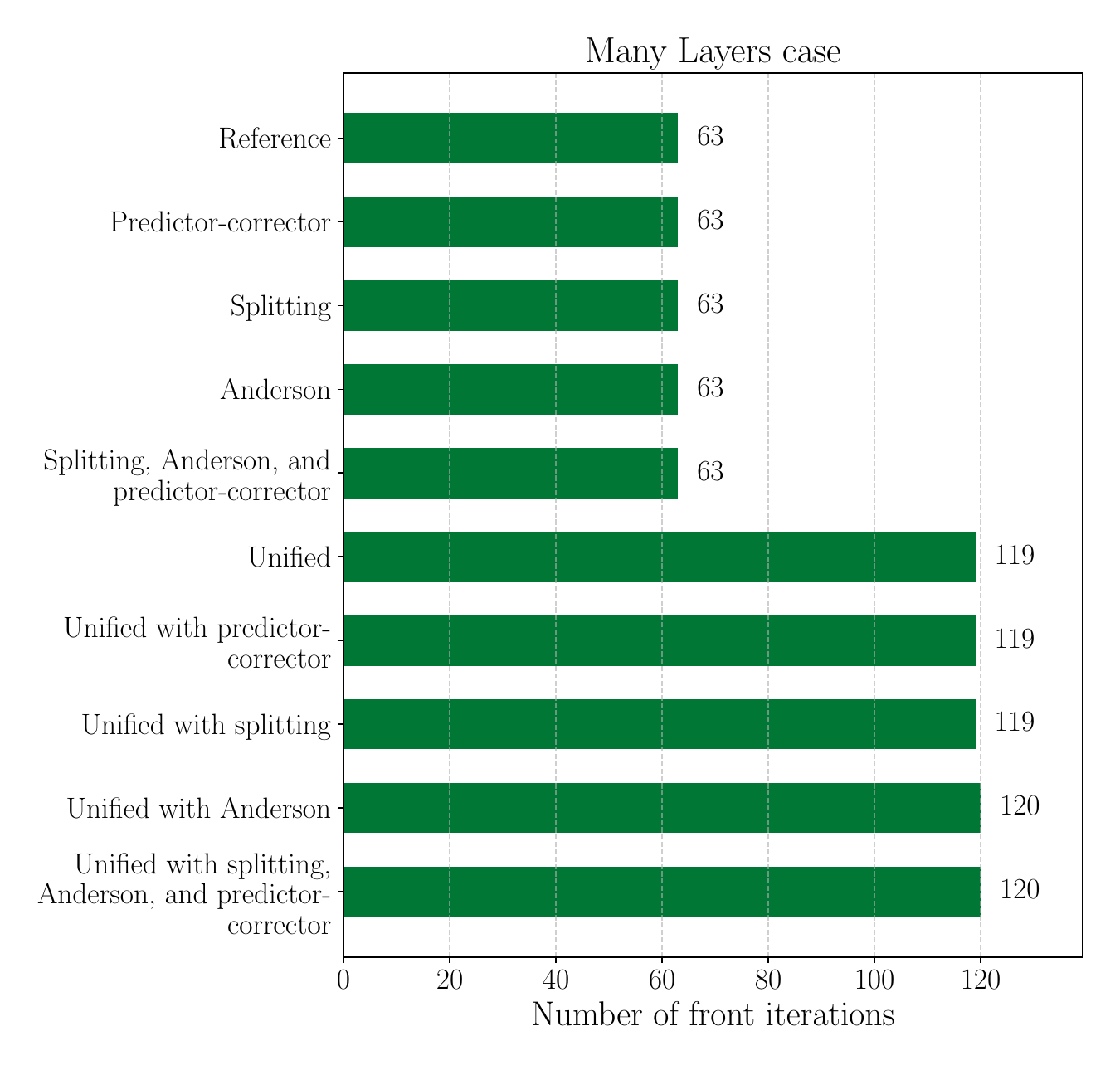}
                \caption{Total number of front iterations accumulated over the closure phase for the Many Layers case.}
                \label{fig:closure_front_iterations_ml}
            \end{figure}

             The number of front iterations accumulated over the closure phase is shown in~\Cref{fig:closure_front_iterations_p3d,fig:closure_front_iterations_ml}.

             For the variants based on the reference scheme, the count corresponds to approximately one front iteration per closure time step in both cases. Once the front is arrested, this iteration merely verifies that the front position remains unchanged: the criterion~\eqref{front_convergence_criteria} is satisfied immediately. The predictor--corrector variant reproduces exactly the same front-iteration count as the reference variant for both cases, because the predictor--corrector scheme plays no role once the front is arrested.

             The variants based on the unified scheme perform about twice as many front iterations, which follows from the update rule of the unified scheme. The front is updated as soon as the contact set has converged in the sense of~\eqref{eq:contact_set_convergence} and the relative aperture change has fallen to the order of unity, which usually happens before the elastohydrodynamic solver has converged. A second update is then performed after the convergence of the elastohydrodynamic solver in order to verify the front convergence criterion~\eqref{front_convergence_criteria}. Neither of the two updates moves the front. The unified variants that include Anderson acceleration perform one additional front iteration, consistently in both cases. This stems from a coincidental collapse of the two front-verification checks into one at the last closure time step for the variants without Anderson acceleration, and carries no practical significance. All other combinations reproduce exactly the front-iteration count of their base variant.

\begin{figure}[H]
                \centering
                \includegraphics[width=0.99\textwidth]{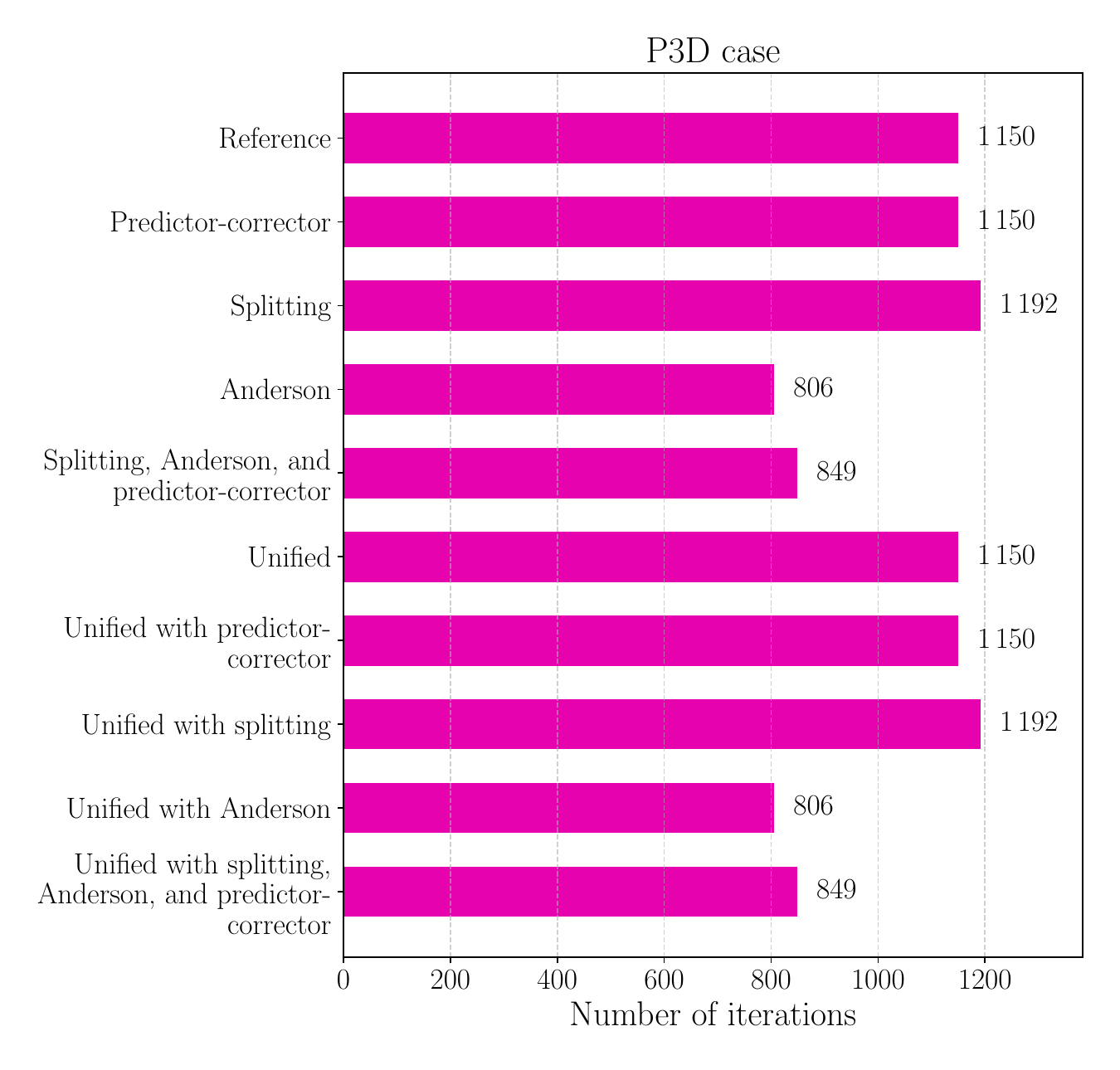}
                \caption{Total number of fixed-point iterations for the variants based on the reference planar 3D ILSA scheme and of unified iterations for the variants based on the unified scheme, accumulated over the closure phase for the P3D case.}
                \label{fig:closure_iterations_p3d}
            \end{figure}

\begin{figure}[H]
                \centering
                \includegraphics[width=0.99\textwidth]{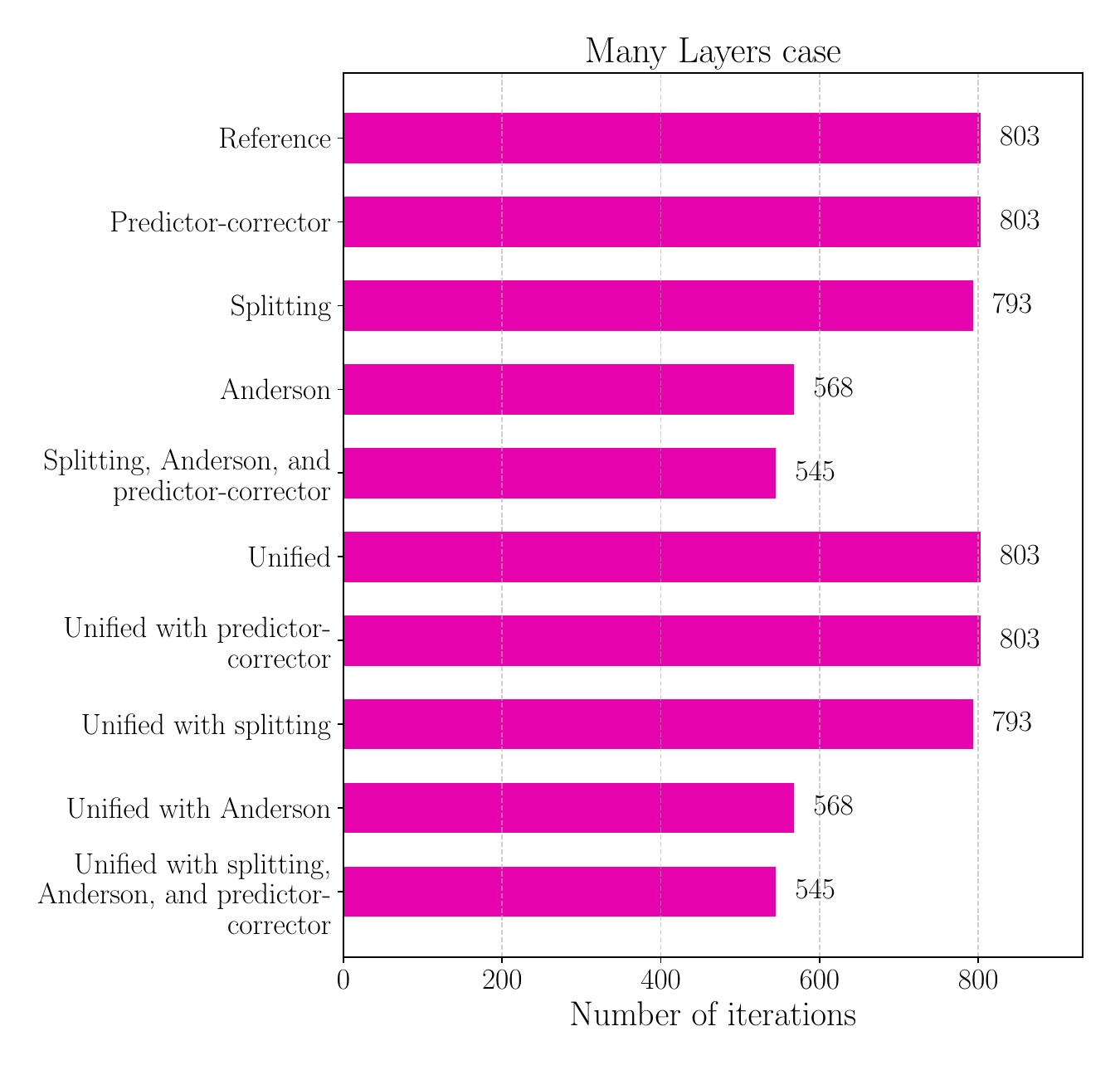}
                \caption{Total number of fixed-point iterations for the variants based on the reference planar 3D ILSA scheme and of unified iterations for the variants based on the unified scheme, accumulated over the closure phase for the Many Layers case.}
                \label{fig:closure_iterations_ml}
            \end{figure}

             The corresponding numbers of fixed-point and unified iterations are shown in~\Cref{fig:closure_iterations_p3d,fig:closure_iterations_ml}. The reference scheme performs markedly more fixed-point iterations per elastohydrodynamic solve during closure than during propagation, about 2.6 times more for the P3D case and 3.4 times more for the Many Layers case. The reason is that the fixed-point iterations must additionally resolve the contact set: elements enter the contact set as the aperture decreases, and every change of the set modifies the block structure of the linearized system~\eqref{elastohydrodynamic_system}. The unified scheme and its variants require exactly the same number of elastohydrodynamic iterations as the reference scheme in both cases: once the front is arrested, the two formulations solve the same fixed-point problem, and only the number of front-verification passes, described above, distinguishes them.

            Anderson acceleration reduces the iteration count by about 30\% for both cases. During propagation the same technique gives a reduction of at most 6\% for these two cases (\Cref{sec:anderson_acceleration_results}), because the elastohydrodynamic solver converges there in about four iterations, leaving little scope for improvement. During closure the number of iterations needed per solve rises to ten and above, which provides greater potential for acceleration.

            The matrix splitting leaves the iteration count during closure practically unchanged: it increases the count by about 4\% for the P3D case and decreases it by about 1\% for the Many Layers case. This is in contrast to the propagation stage, where it significantly raises the fixed-point iteration count (\Cref{fig:iterations_nonlinear_split_in_orig}). A possible reason is that the channel set shrinks as the fracture closes. The elements that enter the contact set keep a prescribed aperture and enter the right-hand side of~\eqref{eq:modified_elastohydrodynamic_system} exactly, so that the portion of the elastic operator treated by the lagged contribution $\bbE^{C C}_{\text{expl}} \mathbf{w}^C_{k-1}$ becomes smaller, and the stencil of the implicit part covers a larger share of the elastic interaction between the remaining open elements than during propagation.

        \subsubsection{Performance during closure}

\begin{figure}[H]
                \centering
                \includegraphics[width=0.99\textwidth]{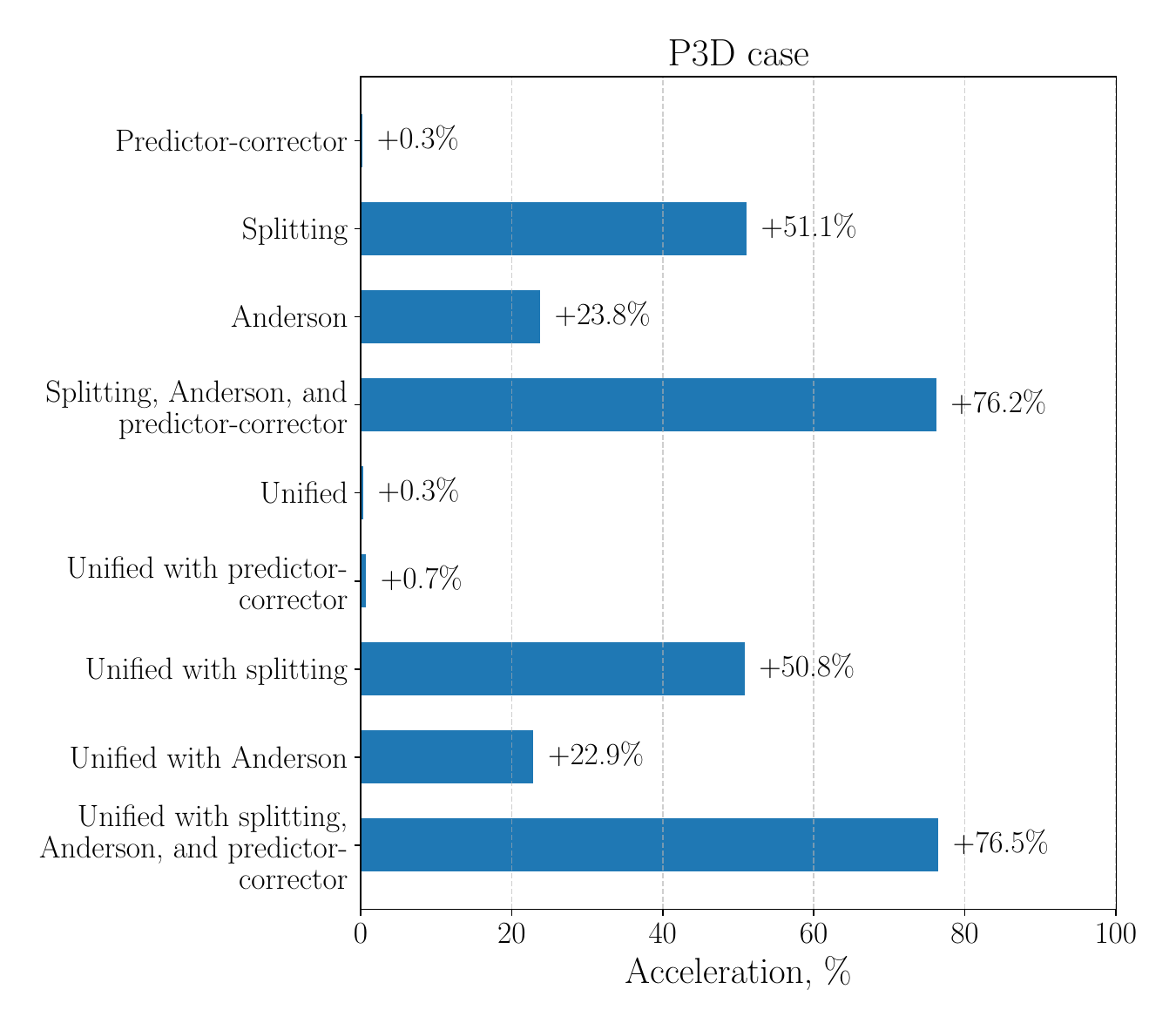}
                \caption{Acceleration of the closure phase with respect to the reference planar 3D ILSA scheme for the P3D case.}
                \label{fig:closure_acceleration_p3d}
            \end{figure}

\begin{figure}[H]
                \centering
                \includegraphics[width=0.99\textwidth]{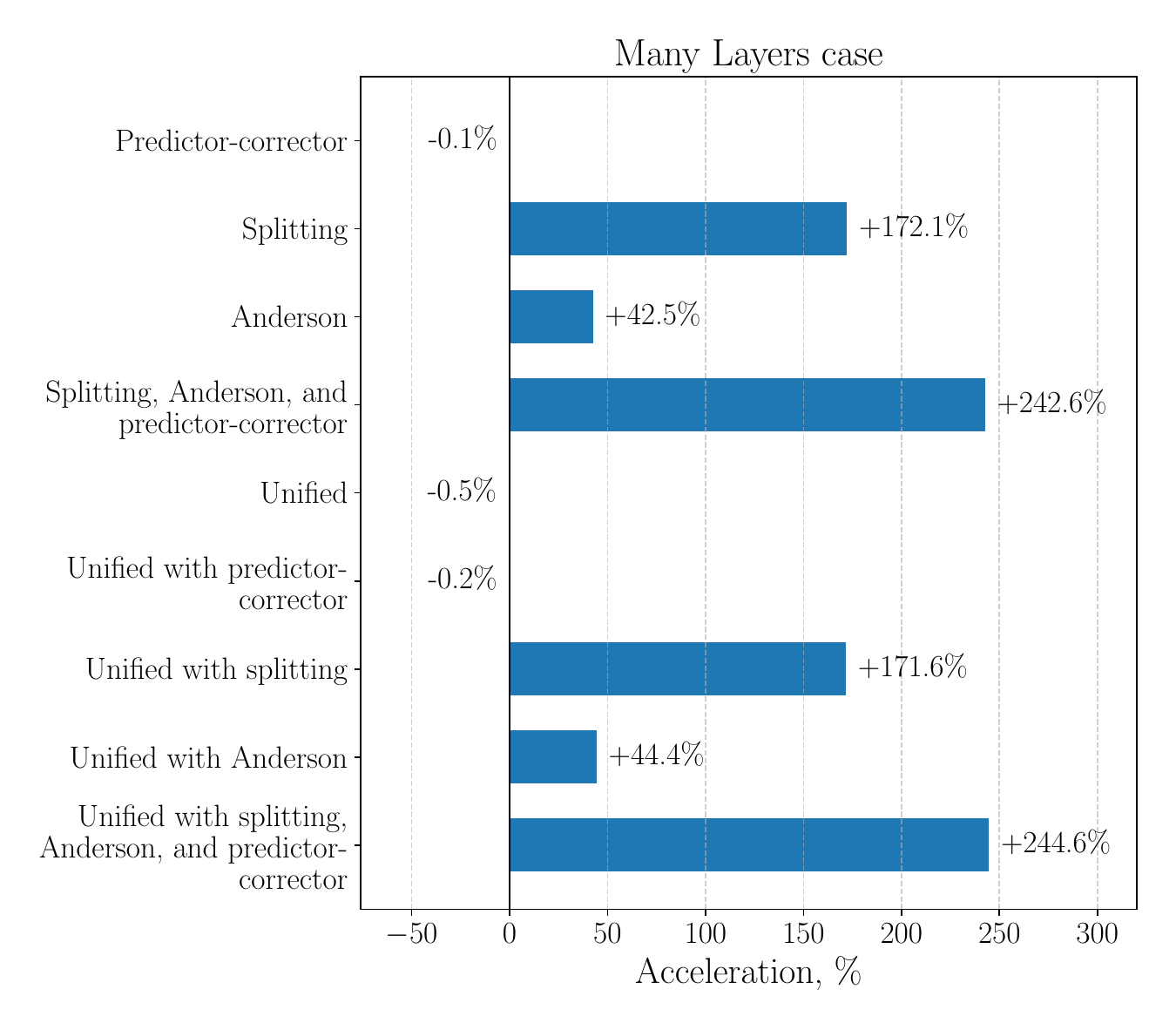}
                \caption{Acceleration of the closure phase with respect to the reference planar 3D ILSA scheme for the Many Layers case.}
                \label{fig:closure_acceleration_ml}
            \end{figure}

            The acceleration of the closure phase relative to the reference planar 3D ILSA scheme is presented in~\Cref{fig:closure_acceleration_p3d,fig:closure_acceleration_ml}. The unified scheme alone yields no appreciable gain during closure. Once the front is arrested, the front tracking loop reduces to a single verification pass per time step. The central idea of the unified scheme, namely the elimination of repeated elastohydrodynamic solves inside the front loop, therefore has nothing left to eliminate, while the unified scheme still performs twice as many front updates. Since the unified scheme provides no closure-specific benefit and tracks the reference scheme almost exactly in this regime, the remainder of this section reports results for the reference scheme only. The unified-scheme values are omitted for brevity, as they differ from the reference scheme by at most a few percent throughout.

            The predictor--corrector scheme is not expected to act during closure, and this is confirmed by the measured closure execution time, which changes by less than 1\% relative to the corresponding baseline. The matrix splitting is the most effective single technique in this regime, providing an acceleration of 51\% for the P3D case and 172\% for the Many Layers case. The gain for the Many Layers case exceeds that observed during propagation (see~\Cref{fig:split_acceleration}), which may be attributed to the absence of the increase in iteration count observed during propagation. Anderson acceleration gives an acceleration of 24\% and 43\%, which is consistent with the reduction of the iteration count reported above. The combined application of the three techniques gives 76\% for the P3D case and 243\% for the Many Layers case, and the entire gain is produced by the matrix splitting and Anderson acceleration, since the predictor--corrector scheme has no effect once the front is arrested.

             The accuracy of the modified algorithms during the closure phase is verified against the reference scheme using the relative $L_2$ and $L_\infty$ aperture errors. The errors are reported in Section~S4 of the Supplementary Material. The predictor--corrector variants and the unified scheme alone reproduce the reference aperture field exactly, with zero error at every time step. For the matrix splitting, Anderson acceleration, and their combination, the relative $L_2$ error does not exceed 0.22\%, and the relative $L_\infty$ error does not exceed 2.33\%.

             Overall, the closure stage changes the ranking of the acceleration techniques established for the propagation stage. The reorganization of the iteration loops, which is the main source of the gain while the fracture propagates, brings no benefit once the front is arrested. The matrix splitting and Anderson acceleration, in contrast, become more effective, while the predictor--corrector scheme has no measurable effect either way. Given the share of the closure phase in the total runtime quoted above, its contribution to the overall speed-up is limited in the present setup, but it would grow in applications dominated by the shut-in period, such as the interpretation of diagnostic mini-frac tests.

\section{Conclusions}

    This paper proposes and evaluates the acceleration strategies for the planar 3D Implicit Level Set Algorithm (ILSA) scheme. The unified planar 3D ILSA scheme, which restructures the iteration loops of the reference planar 3D ILSA scheme into a single iterative process, is introduced. In addition, a matrix splitting of the elastohydrodynamic system is presented, and Anderson acceleration is adapted to its fixed-point solution, with the predictor--corrector scheme of~\cite{zia2019explicit} incorporated as well. Each technique is applied individually and in combination to both the reference and the unified formulations, and evaluated across a range of benchmark cases and mesh resolutions.

    The results show that the unified formulation is a robust acceleration strategy across a wide range of fracture geometries and mesh resolutions. By eliminating nested iterative loops, it provides consistent speed-ups, averaging 130\% ($2.3\times$) and reaching up to 195\% ($3.0\times$) for the Radial, PKN, P3D, and Many Layers cases. The gain is largest for the Sandglass case, reaching 280\% ($3.8\times$). The comparison demonstrates that the unified planar 3D ILSA scheme is a practical and efficient alternative to the reference planar 3D ILSA scheme.

    The acceleration techniques applied to the reference planar 3D ILSA scheme act in a complementary fashion. The matrix splitting technique reduces the per-iteration cost by replacing the dense system matrix with a sparse one, but it may increase the number of fixed-point iterations. Anderson acceleration effectively mitigates this growth by utilizing information from previous iterations. Additionally, the predictor--corrector scheme reduces the number of elastohydrodynamic solves by improving the initial front estimate. Together, these techniques yield an acceleration of up to 363\% ($4.6\times$) for the Radial, PKN, P3D, and Many Layers benchmark cases and up to 523\% ($6.2\times$) for the Sandglass case.

    For the unified planar 3D ILSA scheme, the effect of the individual techniques differs from that observed for the reference scheme. The matrix splitting yields a smaller and more case-dependent acceleration. Anderson acceleration, in contrast, becomes more effective, as the higher baseline iteration count of the unified scheme provides it with greater scope to reduce the number of iterations. The predictor--corrector scheme, on the other hand, provides a smaller benefit, since the front is corrected within the coupled iteration, which makes the scheme less sensitive to the initial guess. The largest performance gain is obtained when the unified scheme is combined with all three acceleration techniques, which act complementarily to counterbalance the iteration increase caused by the matrix splitting. This configuration yields substantial acceleration, reaching up to 1340\% ($14.4\times$) for the Sandglass case and an average of 320\% ($4.2\times$) across the remaining four cases.

    The behavior of the schemes during fracture recession and closure is examined separately for the P3D and Many Layers configurations, with the pumping schedule extended by a shut-in stage. Closure is handled by a contact algorithm on a fixed grid, in a similar manner to the contact-based scheme analyzed in~\cite{talebkeikhah2025numerical}. Once the front is arrested, the front tracking loop reduces to a single verification pass per time step, so that the unified formulation and the predictor--corrector scheme lose their advantage and perform on par with the reference scheme. The matrix splitting and Anderson acceleration, in contrast, benefit from the closure regime, since the changing active-constraint set raises the number of fixed-point iterations per elastohydrodynamic solve, while the shrinking channel set reduces the portion of the elastic operator that the matrix splitting has to lag. Their combined application accelerates the closure phase by 76\% ($1.8\times$) for the P3D case and by 243\% ($3.4\times$) for the Many Layers case.

    The performance gains reported in this work are achieved while maintaining numerical accuracy and physical consistency with respect to the reference solutions: the relative $L_2$ aperture error does not exceed 1\%. Although these methods were developed within the ILSA framework, they do not rely on features exclusive to this specific approach. This suggests that the proposed acceleration strategies can be adapted to other planar 3D hydraulic fracturing simulators to enhance their computational efficiency.

\section*{CRediT authorship contribution statement}

    \textbf{V.I. Shukalo:} Writing -- original draft, Writing -- review \& editing, Visualization, Validation, Software, Methodology, Investigation, Conceptualization.
    \textbf{A.V. Valov:} Writing -- review \& editing, Methodology, Conceptualization.
    \textbf{A.N. Baykin:} Writing -- review \& editing, Methodology, Conceptualization, Project administration, Supervision, Funding acquisition.

\section*{Declaration of competing interest}
    The authors declare that they have no known competing financial interests or personal relationships that could have appeared to influence the work reported in this paper.

\section*{Data availability}
    Data will be made available on request.

\section*{Acknowledgments}
    The work was carried out within the framework of the state assignment of the Lavrentyev Institute of Hydrodynamics of the SB RAS (Project No.~FWGG-2026-0017).

\bibliographystyle{elsarticle-num}
\bibliography{References}
\end{document}


\begin{center}
    \Large \textbf{Supplementary Material} \\
    \vspace{0.3cm}
    \normalsize for the manuscript: \textit{``Acceleration methods for the planar 3D ILSA hydraulic fracturing model''}
\end{center}

\vspace{0.5cm}

This document collects the accuracy verification of the accelerated planar 3D ILSA algorithms against the reference planar 3D ILSA scheme, together with supporting iteration-count figures referenced in the main manuscript. \Cref{sec:supp_metrics} defines the error metrics used throughout the manuscript and in this document. \Cref{sec:supp_propagation} reports the errors for the propagation stage and examines the localized deviations observed for the Sandglass case, \Cref{sec:supp_front_iterations} reports the front iteration counts for the matrix splitting and Anderson acceleration applied to the reference planar 3D ILSA scheme, and \Cref{sec:supp_closure} reports the errors for the recession and closure stage.

\section{Error metrics}\label{sec:supp_metrics}

    Let $w_i$ and $w_i^{\text{ref}}$ denote the fracture aperture in cell $i$ obtained with a modified algorithm and with the reference planar 3D ILSA scheme, respectively, and let $N$ be the total number of fracture elements. Both solutions are computed on the same mesh and are compared at the same instant of time, so that the comparison is performed cell by cell. The relative $L_2$ error is defined as
    \begin{equation}
        \label{eq:supp_l2}
        \delta_{L_2}(t) =
        \frac{
            \sqrt{ \displaystyle \sum_{i=1}^{N} \left( w_i(t) - w_i^{\text{ref}}(t) \right)^2 }
        }{
            \sqrt{ \displaystyle \sum_{i=1}^{N} \left( w_i^{\text{norm}} \right)^2 }
        } \times 100\%,
    \end{equation}
    and the relative $L_\infty$ error as
    \begin{equation}
        \label{eq:supp_linf}
        \delta_{L_\infty}(t) =
        \frac{
            \displaystyle \max_{1 \leq i \leq N} \left| w_i(t) - w_i^{\text{ref}}(t) \right|
        }{
            \displaystyle \max_{1 \leq i \leq N} \left| w_i^{\text{norm}} \right|
        } \times 100\%.
    \end{equation}

    The $L_2$ error characterizes the discrepancy averaged over the fracture footprint, whereas the $L_\infty$ error is governed by the single cell with the largest deviation and is therefore the more conservative of the two. Both quantities are evaluated at every time step, which yields the error histories reported below.

    The normalizing aperture field $w^{\text{norm}}$ depends on the stage of the simulation. During propagation it is the reference solution at the same instant, $w_i^{\text{norm}} = w_i^{\text{ref}}(t)$. During recession and closure the reference aperture decreases toward the minimum admissible value $w_{\mathrm{min}}$, so that the normalizing quantity would itself tend to zero, which makes the relative error sensitive to the choice of $w_{\mathrm{min}}$ and harder to analyze. The errors reported for the closure phase are therefore normalized by the reference aperture field taken at the instant of front arrest $t_a$, that is, $w_i^{\text{norm}} = w_i^{\text{ref}}(t_a)$, which is fixed throughout the phase. A comparable convention is adopted in~\cite{talebkeikhah2025numerical}, where the aperture during recession is scaled by its value at the recession initiation time.

\section{Accuracy during the propagation stage}\label{sec:supp_propagation}

    This section reports the relative $L_2$ and $L_\infty$ aperture errors of the accelerated algorithms with respect to the reference planar 3D ILSA scheme for the propagation-stage benchmark cases of the main manuscript, namely the Radial, PKN, P3D, Sandglass, and Many Layers configurations. Every variant is compared with the reference scheme computed on the same mesh, so that the comparison is performed cell by cell as described in \Cref{sec:supp_metrics}. The errors are evaluated according to~\eqref{eq:supp_l2} and~\eqref{eq:supp_linf} at every time step, with the normalizing field taken at the same instant, $w_i^{\text{norm}} = w_i^{\text{ref}}(t)$.

    A single mesh is used for each case. The corresponding resolutions are listed in \Cref{tab:supp_meshes}; all remaining simulation parameters are those reported in the main manuscript, where the benchmark setup is described. The error histories are reported from $t = 100$~s onward, which is more representative of the accuracy of the acceleration techniques: over the initial growth the fracture occupies few elements with small apertures, so that the relative measures are sensitive, as they involve division by small values.

    \begin{table}[H]
        \centering
        \resizebox{\textwidth}{!}{%
        \begin{tabular}{lccccc}
        \toprule
        \textbf{Parameter} & \textbf{Radial} & \textbf{PKN} & \textbf{P3D} & \textbf{Sandglass} & \textbf{Many Layers} \\
        \midrule
        \textbf{Cell size} $\Delta x \times \Delta y$, m & $3.90 \times 3.89$ & $0.83 \times 0.83$ & $4.17 \times 4.17$ & $2.00 \times 2.00$ & $2.22 \times 2.22$ \\
        \bottomrule
        \end{tabular}%
        }
        \caption{Meshes used for the error histories of this section.}
        \label{tab:supp_meshes}
    \end{table}

\begin{figure}[H]
        \centering
        \includegraphics[width=0.9\textwidth]{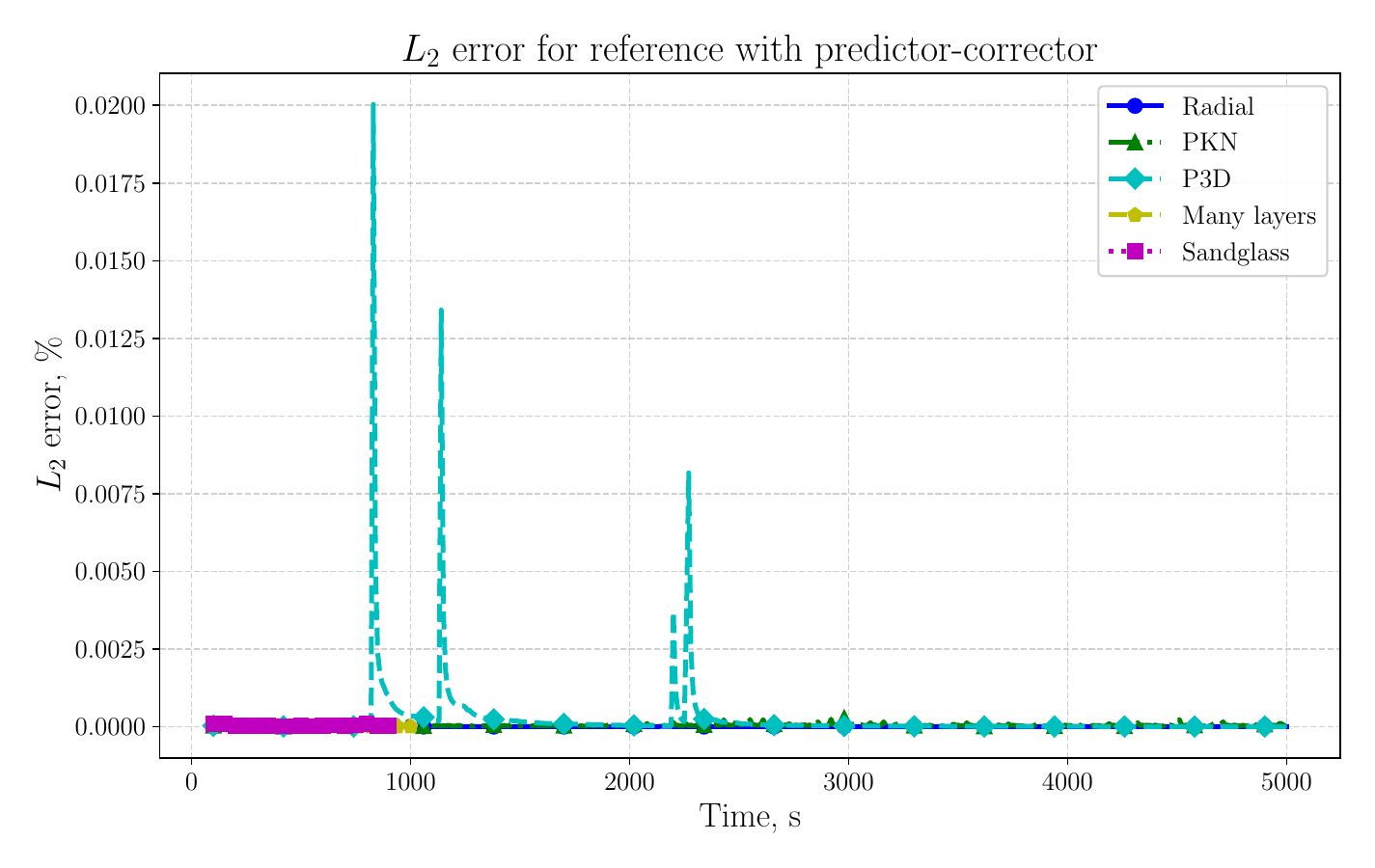}
        \caption{Relative $L_2$ aperture error of the reference planar 3D ILSA scheme with the predictor--corrector scheme during the propagation stage.}
        \label{fig:supp_open_l2_ref_pc}
    \end{figure}

\begin{figure}[H]
        \centering
        \includegraphics[width=0.9\textwidth]{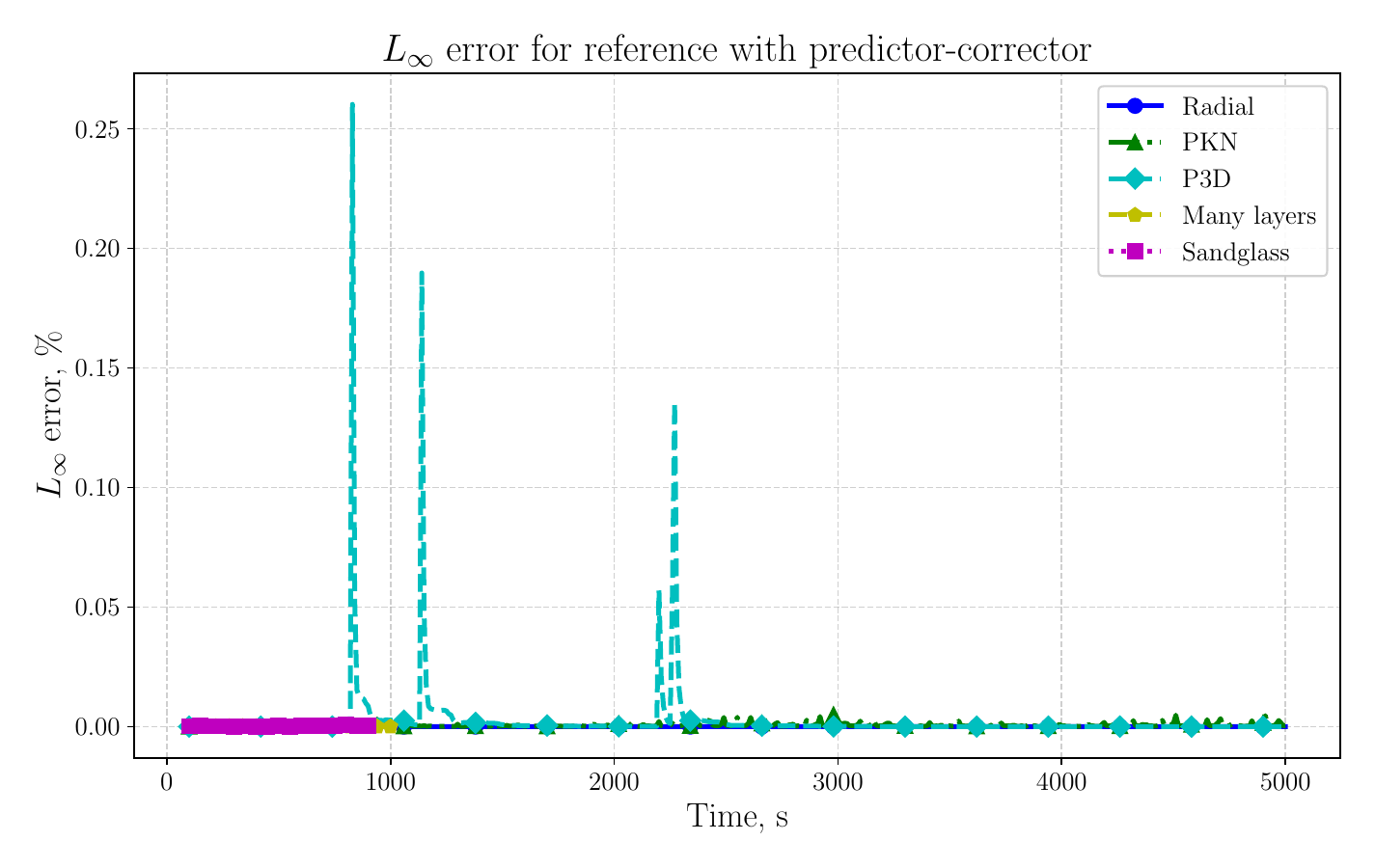}
        \caption{Relative $L_\infty$ aperture error of the reference planar 3D ILSA scheme with the predictor--corrector scheme during the propagation stage.}
        \label{fig:supp_open_linf_ref_pc}
    \end{figure}

\begin{figure}[H]
        \centering
        \includegraphics[width=0.9\textwidth]{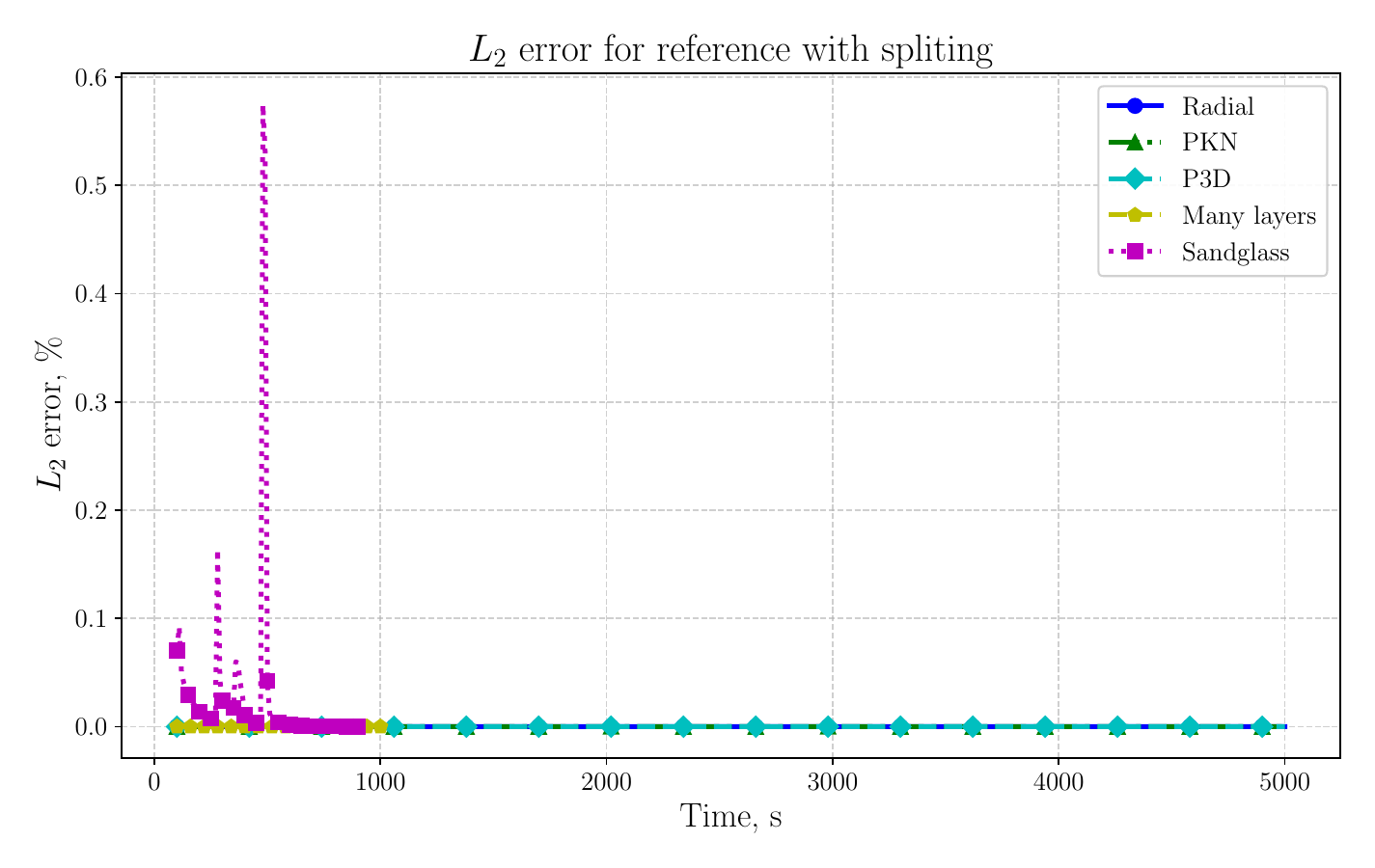}
        \caption{Relative $L_2$ aperture error of the reference planar 3D ILSA scheme with the matrix splitting during the propagation stage.}
        \label{fig:supp_open_l2_ref_split}
    \end{figure}

\begin{figure}[H]
        \centering
        \includegraphics[width=0.9\textwidth]{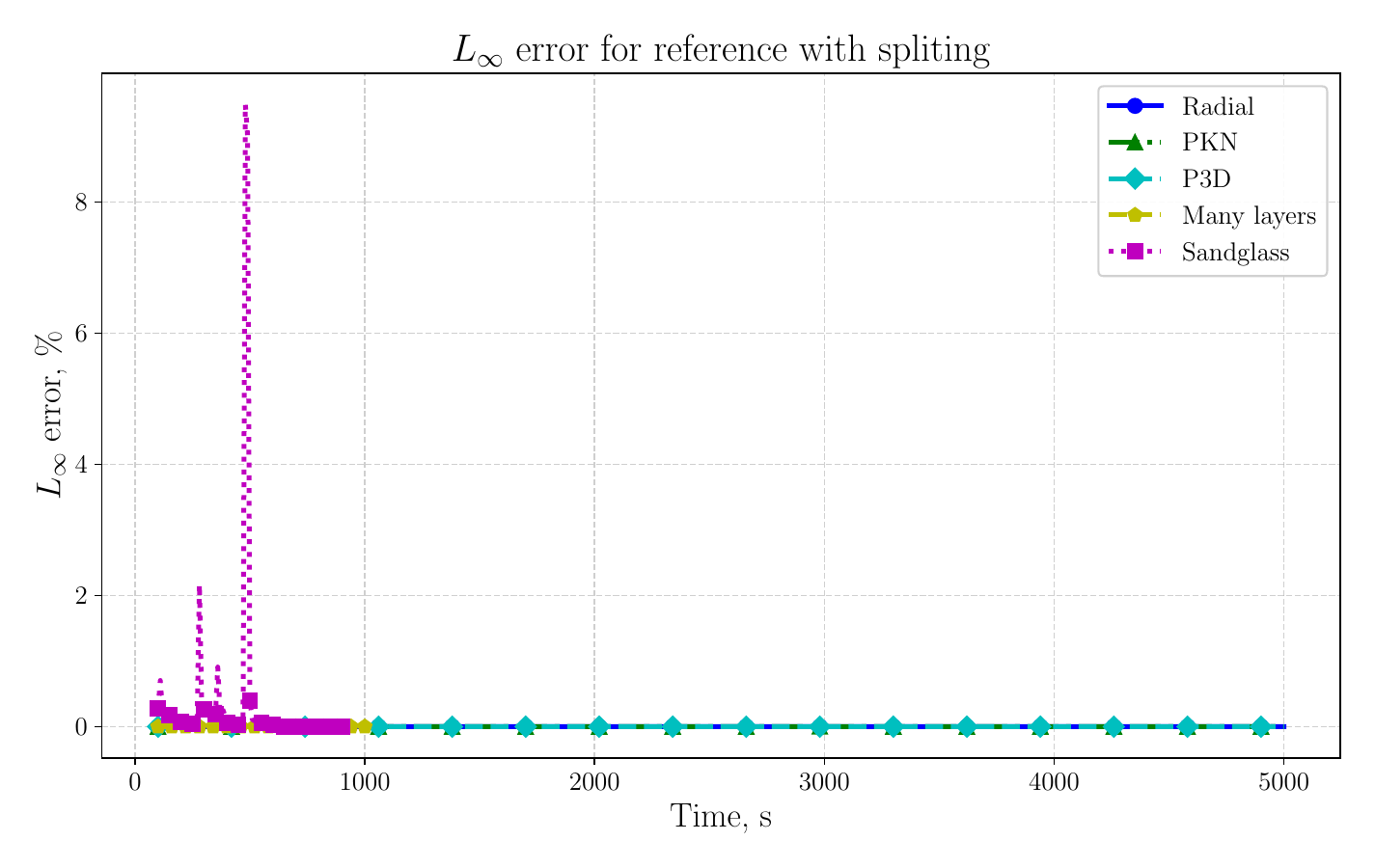}
        \caption{Relative $L_\infty$ aperture error of the reference planar 3D ILSA scheme with the matrix splitting during the propagation stage.}
        \label{fig:supp_open_linf_ref_split}
    \end{figure}

\begin{figure}[H]
        \centering
        \includegraphics[width=0.9\textwidth]{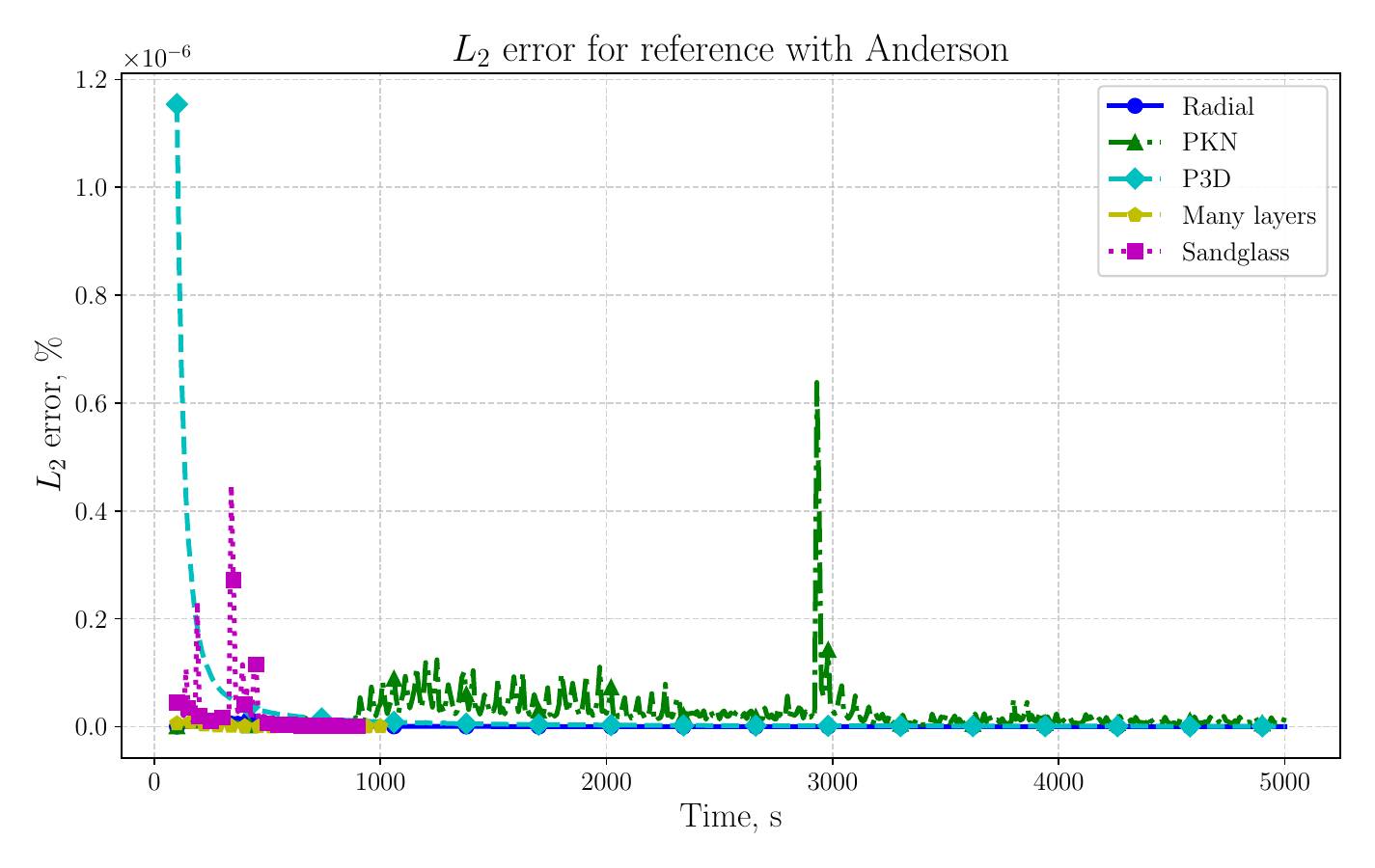}
        \caption{Relative $L_2$ aperture error of the reference planar 3D ILSA scheme with Anderson acceleration during the propagation stage.}
        \label{fig:supp_open_l2_ref_anderson}
    \end{figure}

\begin{figure}[H]
        \centering
        \includegraphics[width=0.9\textwidth]{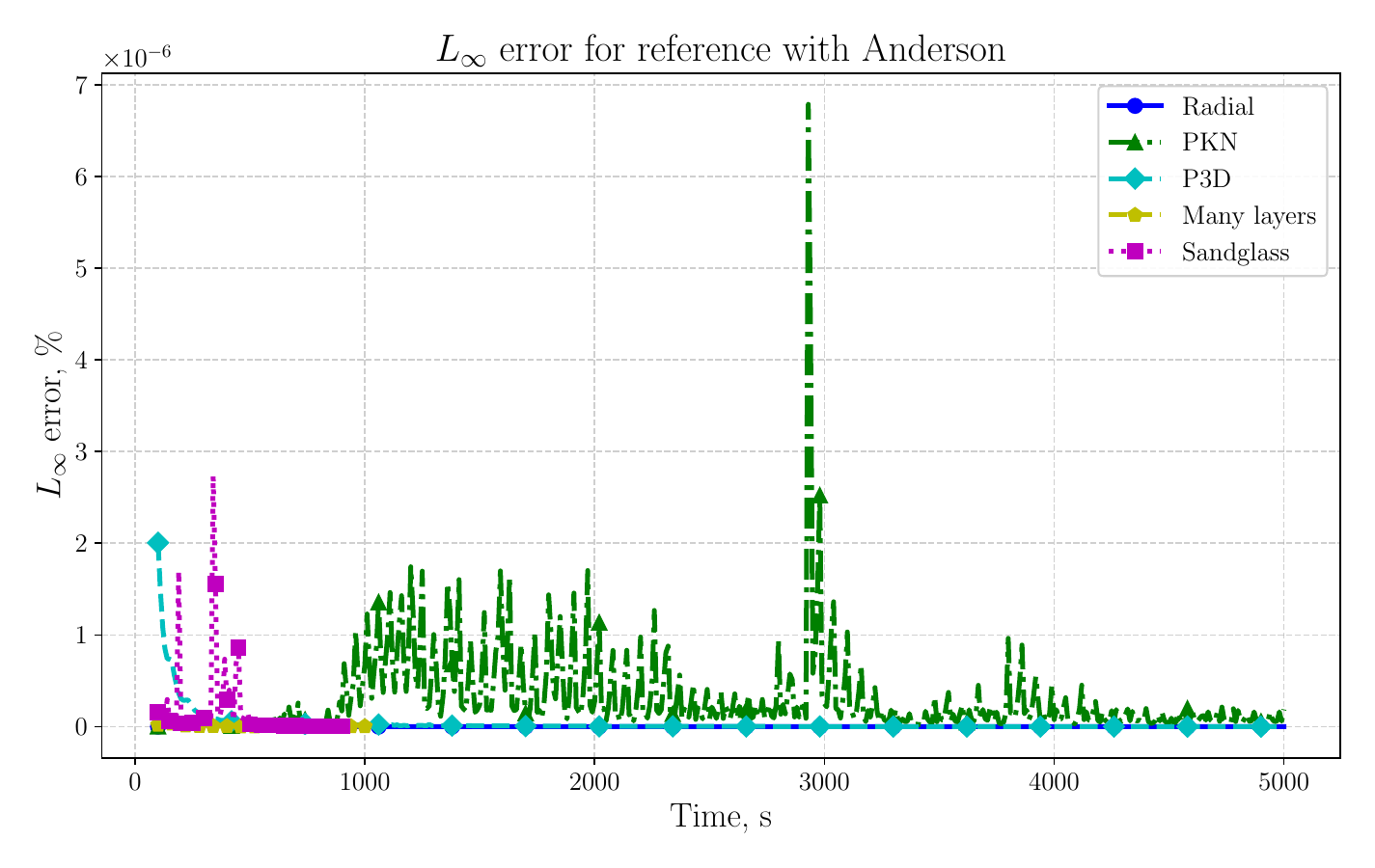}
        \caption{Relative $L_\infty$ aperture error of the reference planar 3D ILSA scheme with Anderson acceleration during the propagation stage.}
        \label{fig:supp_open_linf_ref_anderson}
    \end{figure}

\begin{figure}[H]
        \centering
        \includegraphics[width=0.9\textwidth]{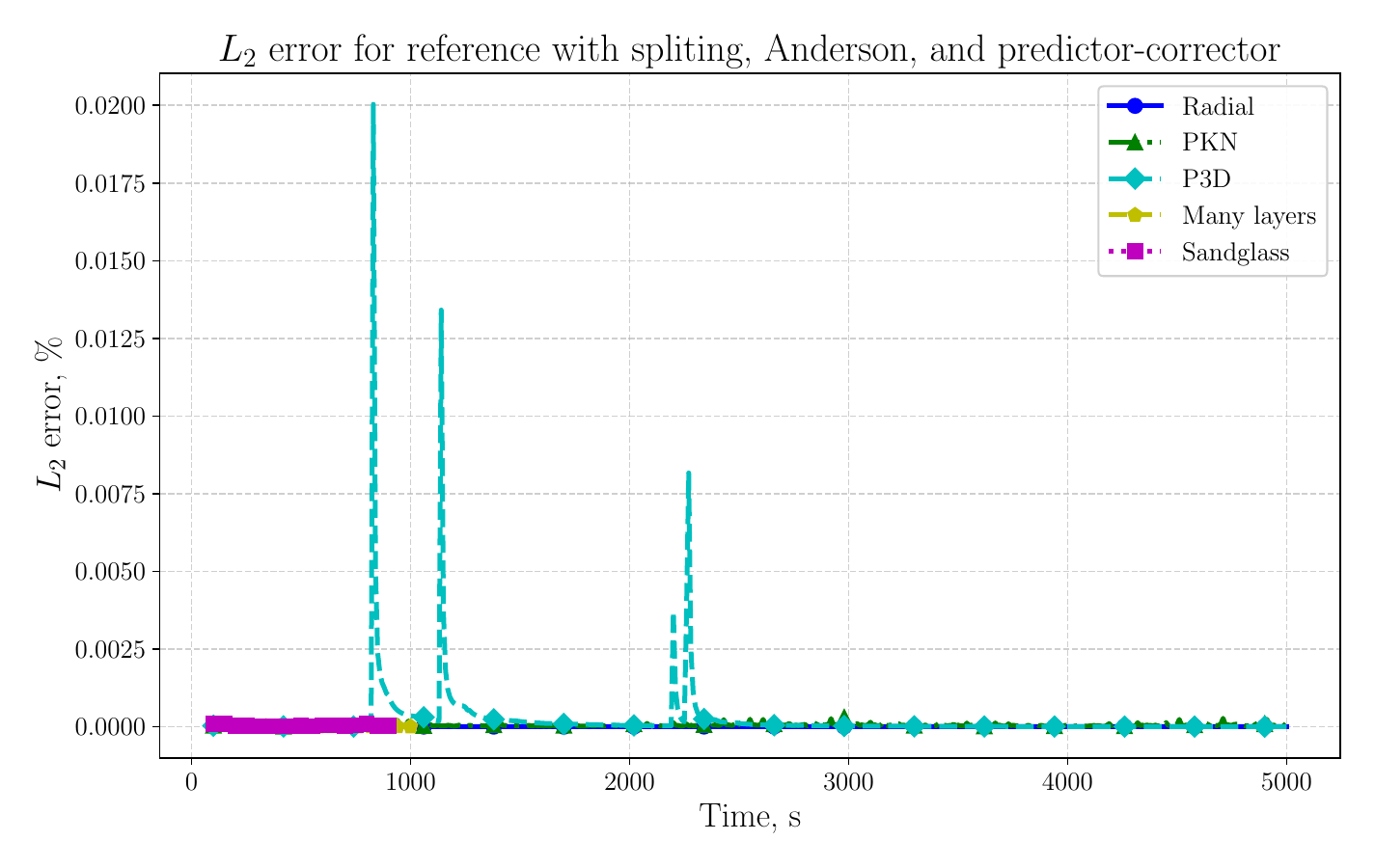}
        \caption{Relative $L_2$ aperture error of the reference planar 3D ILSA scheme with the matrix splitting, Anderson acceleration, and the predictor--corrector scheme during the propagation stage.}
        \label{fig:supp_open_l2_ref_all}
    \end{figure}

\begin{figure}[H]
        \centering
        \includegraphics[width=0.9\textwidth]{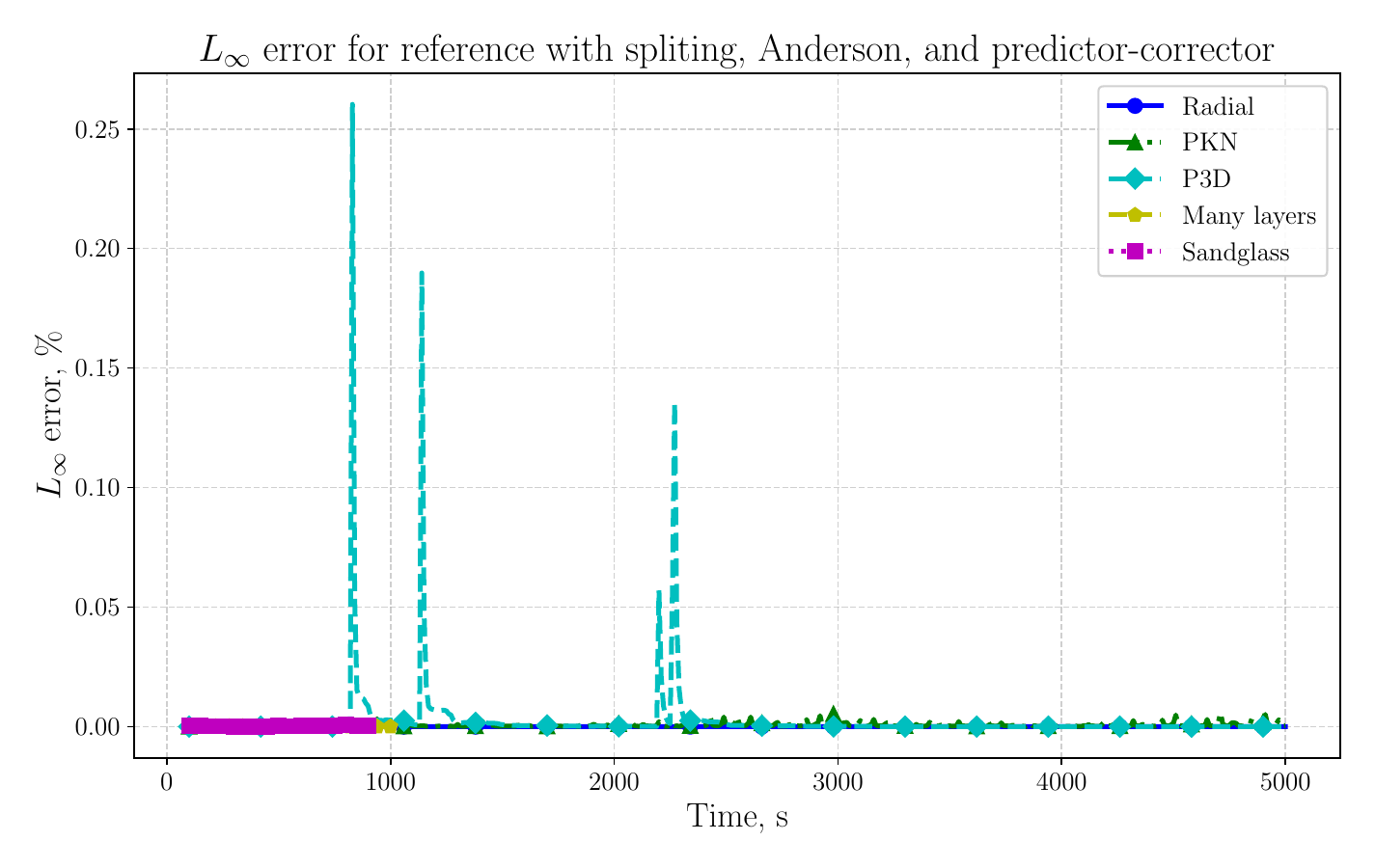}
        \caption{Relative $L_\infty$ aperture error of the reference planar 3D ILSA scheme with the matrix splitting, Anderson acceleration, and the predictor--corrector scheme during the propagation stage.}
        \label{fig:supp_open_linf_ref_all}
    \end{figure}

    \Cref{fig:supp_open_l2_ref_pc,fig:supp_open_linf_ref_pc,fig:supp_open_l2_ref_split,fig:supp_open_linf_ref_split,fig:supp_open_l2_ref_anderson,fig:supp_open_linf_ref_anderson,fig:supp_open_l2_ref_all,fig:supp_open_linf_ref_all} report the $L_2$ and $L_\infty$ errors of the acceleration techniques applied to the reference planar 3D ILSA scheme, and \Cref{fig:supp_open_l2_unified,fig:supp_open_linf_unified,fig:supp_open_l2_unified_pc,fig:supp_open_linf_unified_pc,fig:supp_open_l2_unified_split,fig:supp_open_linf_unified_split,fig:supp_open_l2_unified_anderson,fig:supp_open_linf_unified_anderson,fig:supp_open_l2_unified_all,fig:supp_open_linf_unified_all} report the same quantities for the unified planar 3D ILSA scheme and for the techniques applied to it. The benchmark cases are distinguished within each figure as indicated in its legend. Over the reported interval the relative $L_2$ error does not exceed $1\%$ for all variants and cases. For the Radial, PKN, P3D, and Many Layers cases the relative $L_\infty$ error does not exceed $3\%$, and its largest values are attained in the first part of the reported interval and decrease as the fracture grows. The Sandglass case is the only configuration for which the $L_\infty$ error exceeds this level. It does so at isolated instants, reaching up to $9.5\%$, which are examined in \Cref{sec:supp_sandglass}.

    \begin{figure}[H]
        \centering
        \includegraphics[width=0.9\textwidth]{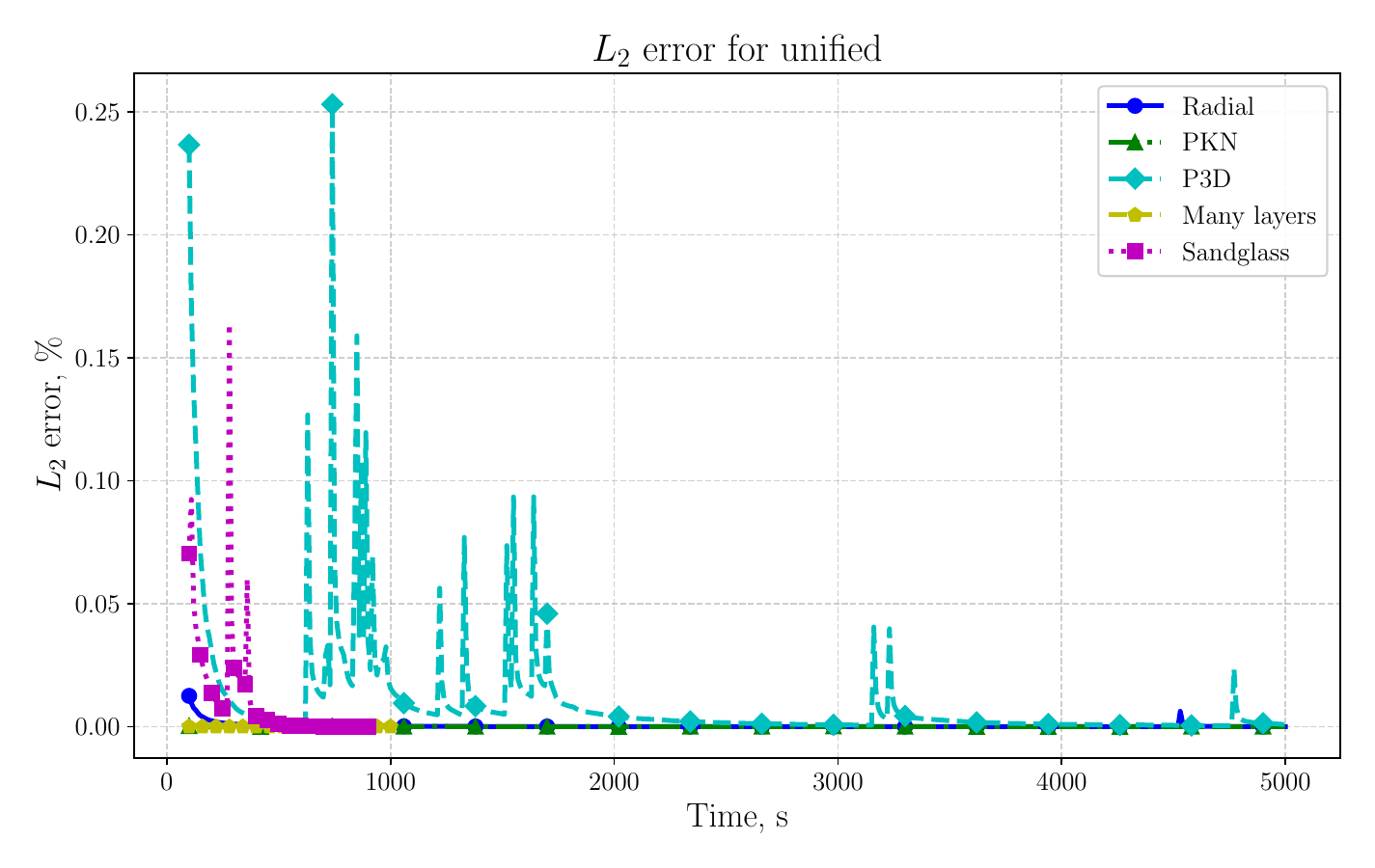}
        \caption{Relative $L_2$ aperture error of the unified planar 3D ILSA scheme during the propagation stage.}
        \label{fig:supp_open_l2_unified}
    \end{figure}

    \begin{figure}[H]
        \centering
        \includegraphics[width=0.9\textwidth]{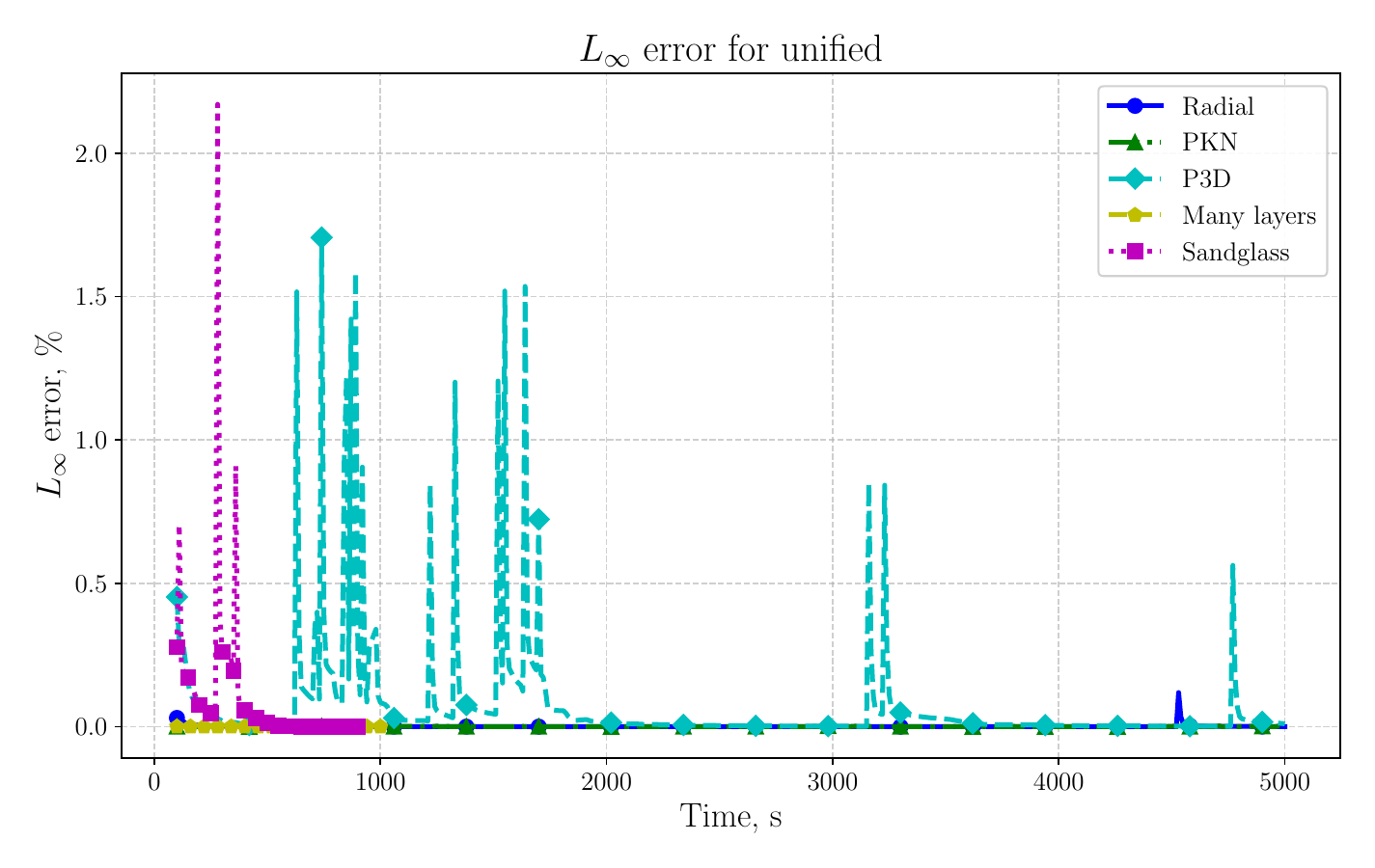}
        \caption{Relative $L_\infty$ aperture error of the unified planar 3D ILSA scheme during the propagation stage.}
        \label{fig:supp_open_linf_unified}
    \end{figure}

    \begin{figure}[H]
        \centering
        \includegraphics[width=0.9\textwidth]{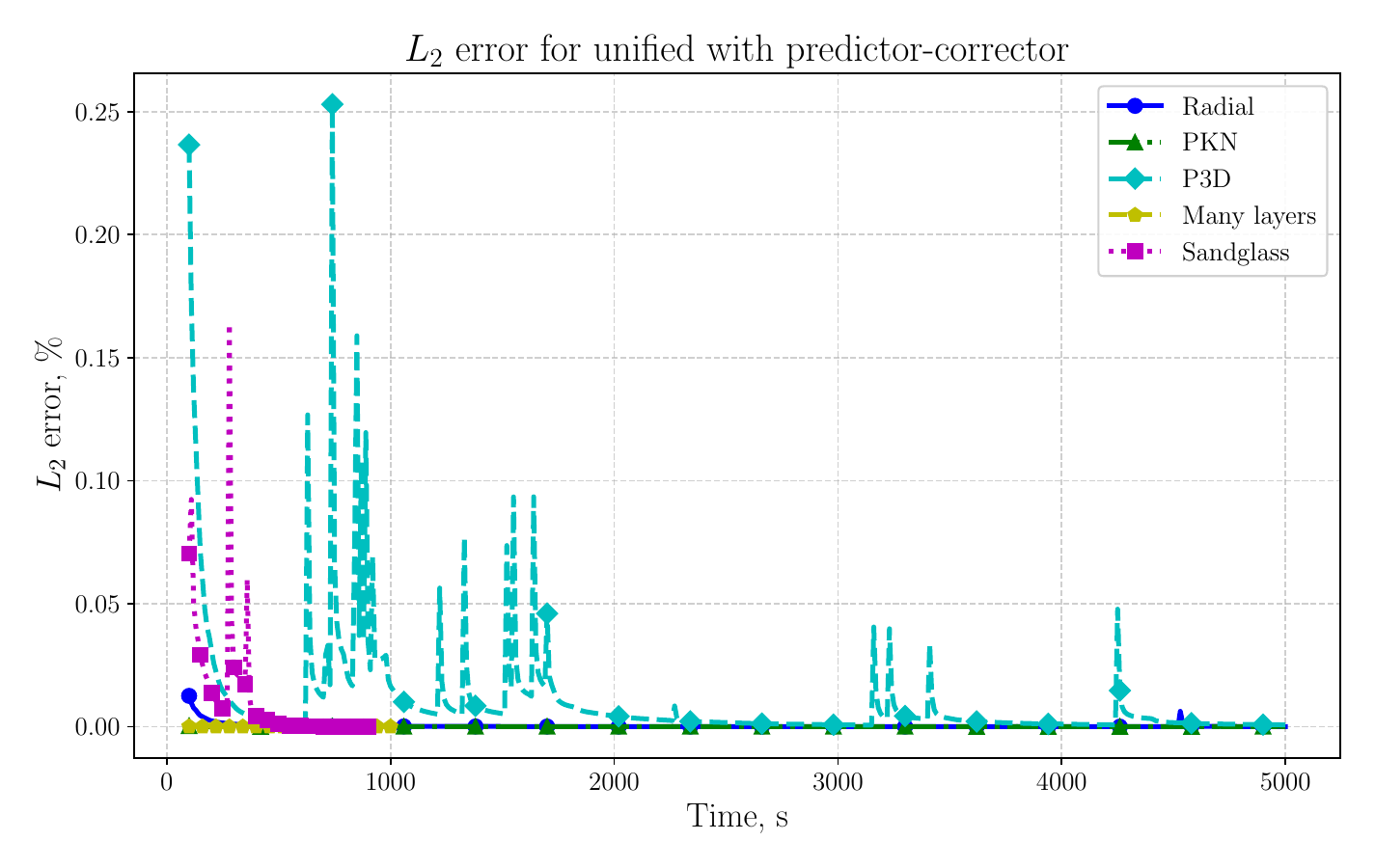}
        \caption{Relative $L_2$ aperture error of the unified planar 3D ILSA scheme with the predictor--corrector scheme during the propagation stage.}
        \label{fig:supp_open_l2_unified_pc}
    \end{figure}

    \begin{figure}[H]
        \centering
        \includegraphics[width=0.9\textwidth]{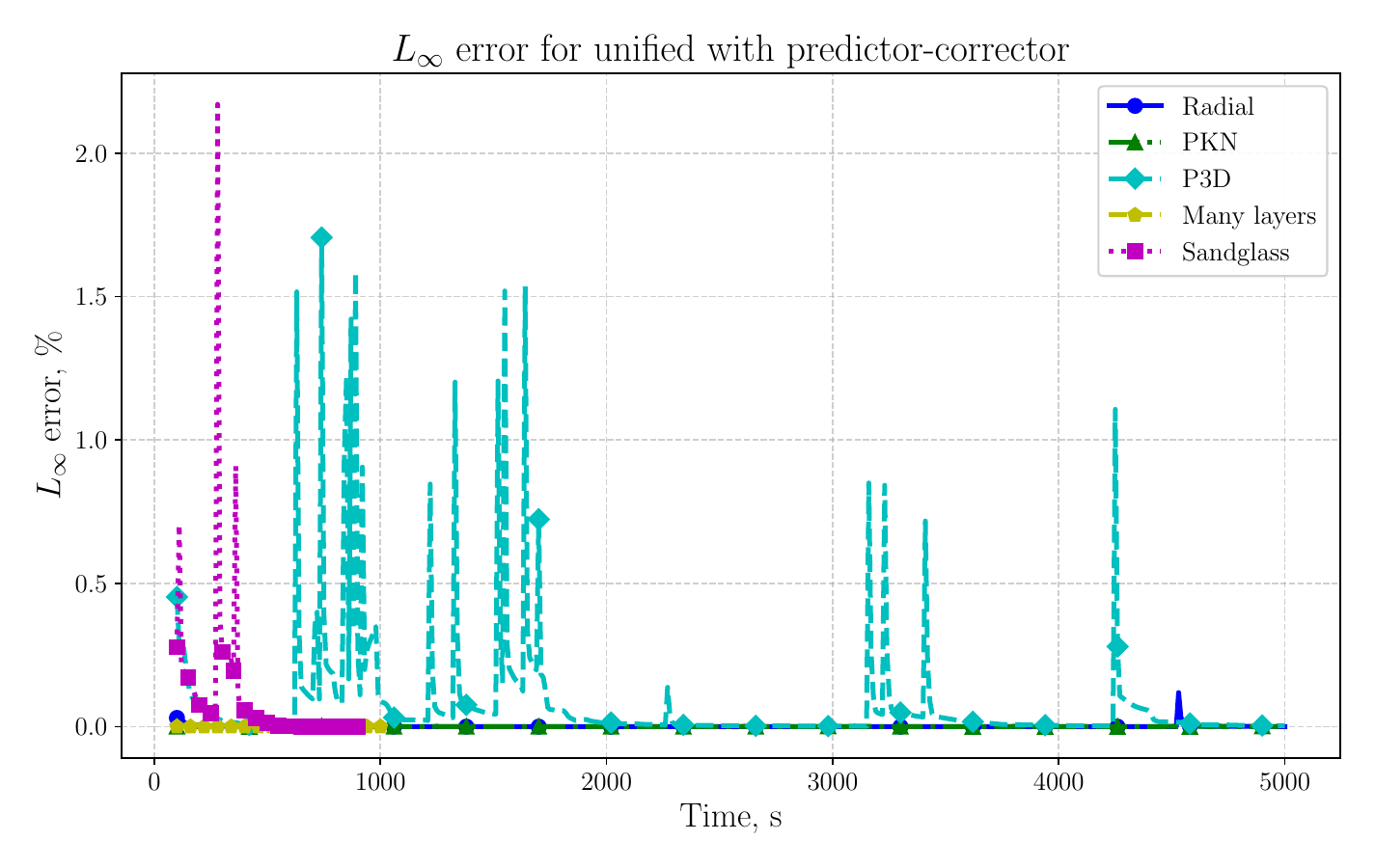}
        \caption{Relative $L_\infty$ aperture error of the unified planar 3D ILSA scheme with the predictor--corrector scheme during the propagation stage.}
        \label{fig:supp_open_linf_unified_pc}
    \end{figure}

    \begin{figure}[H]
        \centering
        \includegraphics[width=0.9\textwidth]{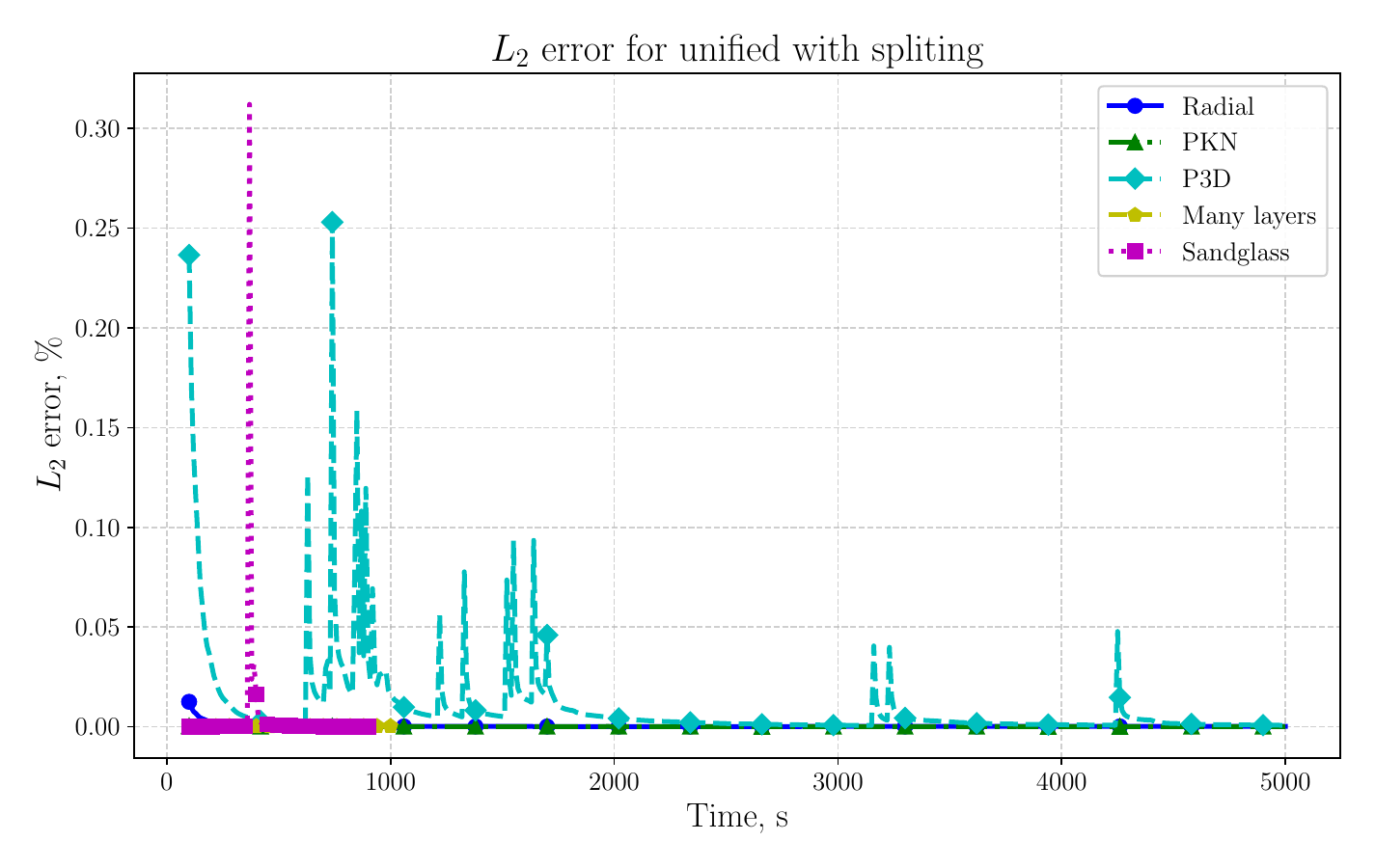}
        \caption{Relative $L_2$ aperture error of the unified planar 3D ILSA scheme with the matrix splitting during the propagation stage.}
        \label{fig:supp_open_l2_unified_split}
    \end{figure}

    \begin{figure}[H]
        \centering
        \includegraphics[width=0.9\textwidth]{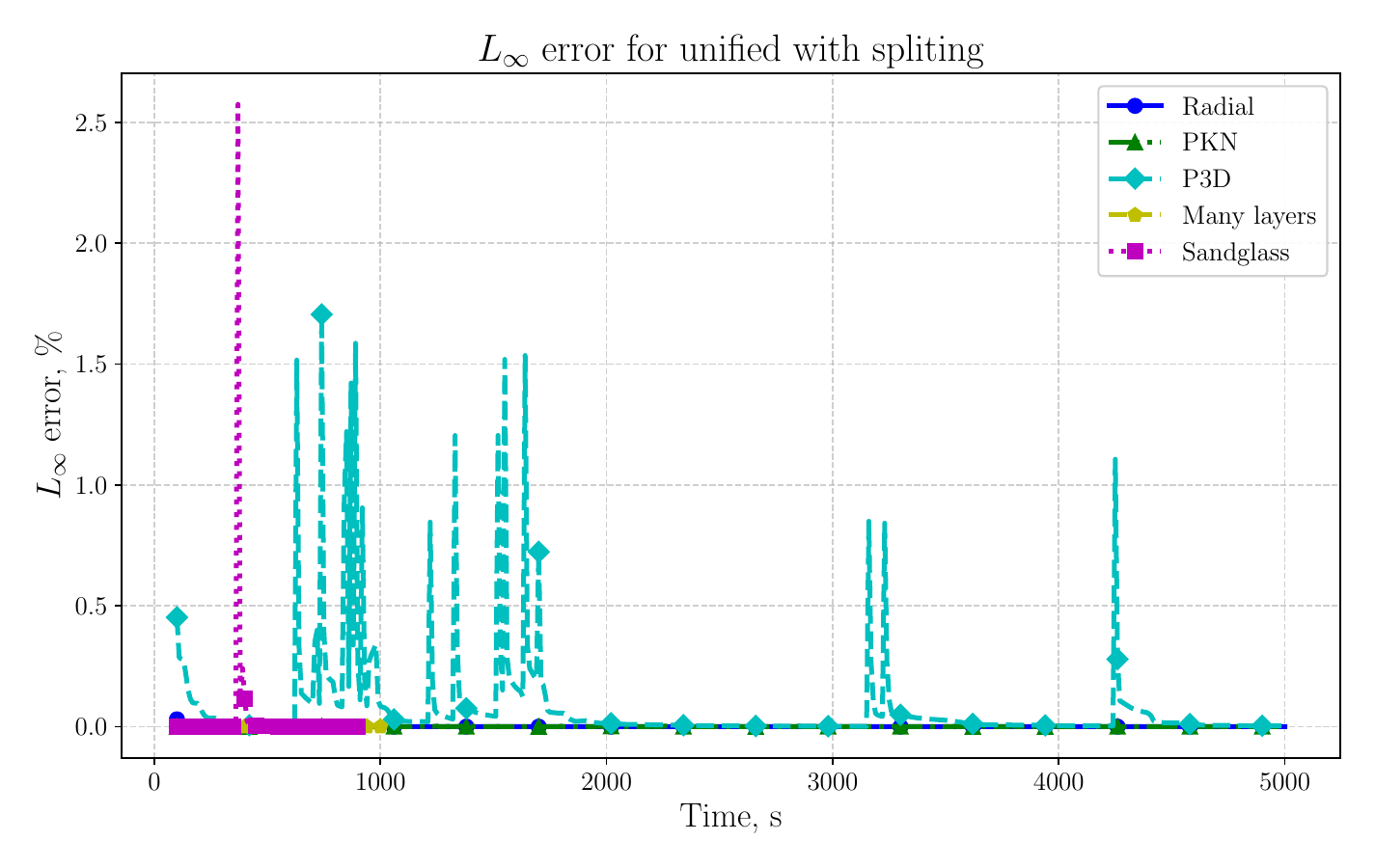}
        \caption{Relative $L_\infty$ aperture error of the unified planar 3D ILSA scheme with the matrix splitting during the propagation stage.}
        \label{fig:supp_open_linf_unified_split}
    \end{figure}

    \begin{figure}[H]
        \centering
        \includegraphics[width=0.9\textwidth]{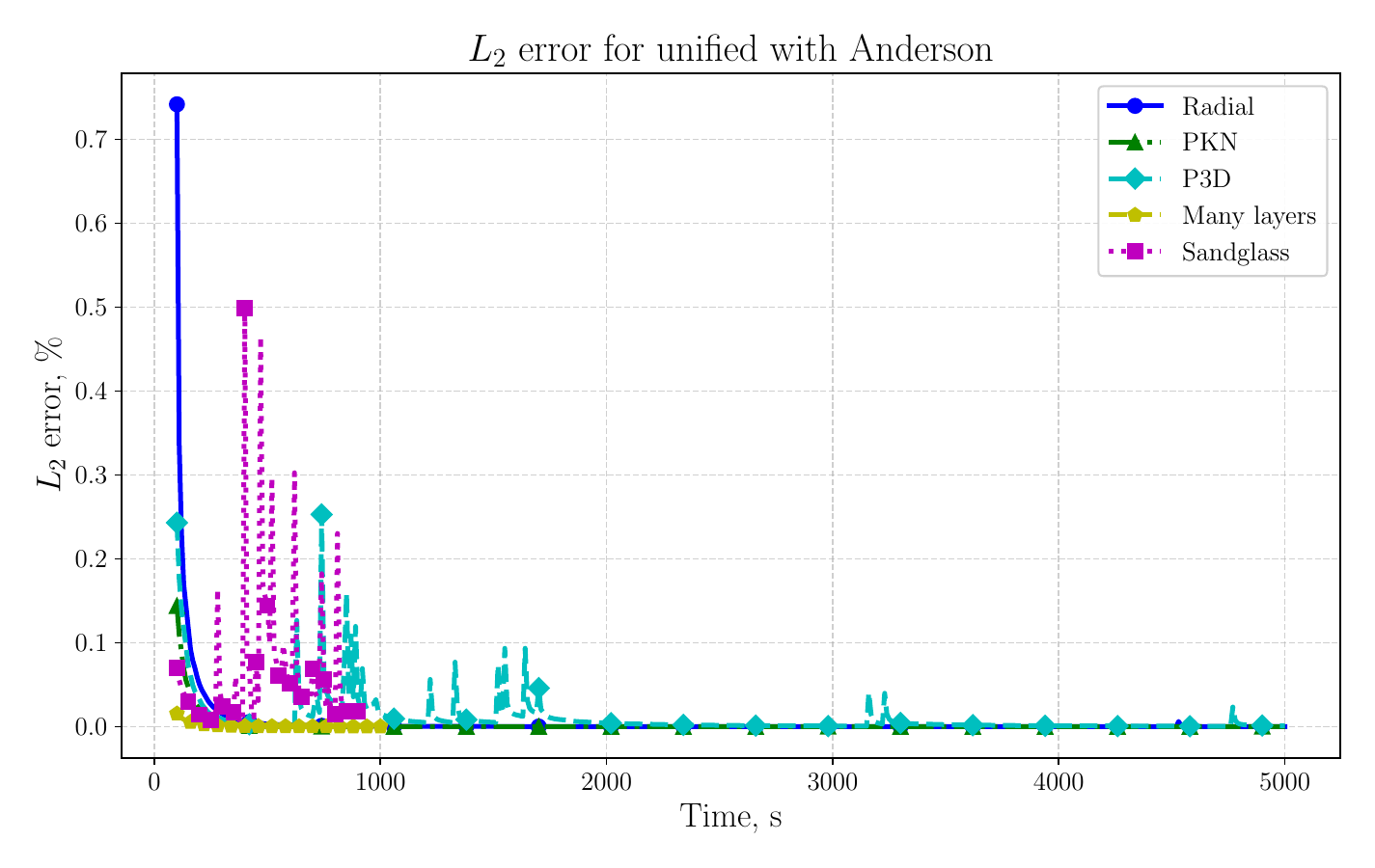}
        \caption{Relative $L_2$ aperture error of the unified planar 3D ILSA scheme with Anderson acceleration during the propagation stage.}
        \label{fig:supp_open_l2_unified_anderson}
    \end{figure}

    \begin{figure}[H]
        \centering
        \includegraphics[width=0.9\textwidth]{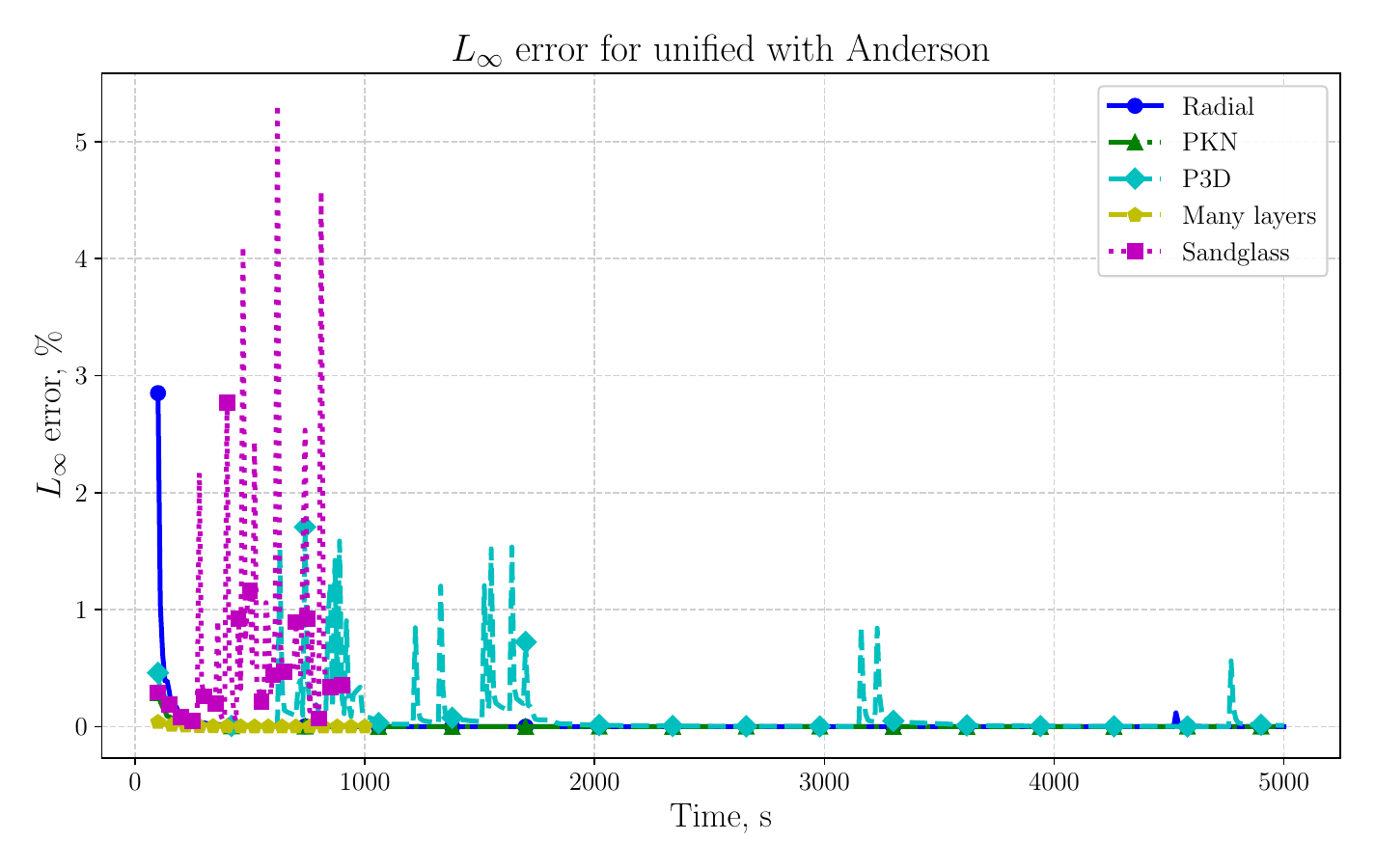}
        \caption{Relative $L_\infty$ aperture error of the unified planar 3D ILSA scheme with Anderson acceleration during the propagation stage.}
        \label{fig:supp_open_linf_unified_anderson}
    \end{figure}

    \begin{figure}[H]
        \centering
        \includegraphics[width=0.9\textwidth]{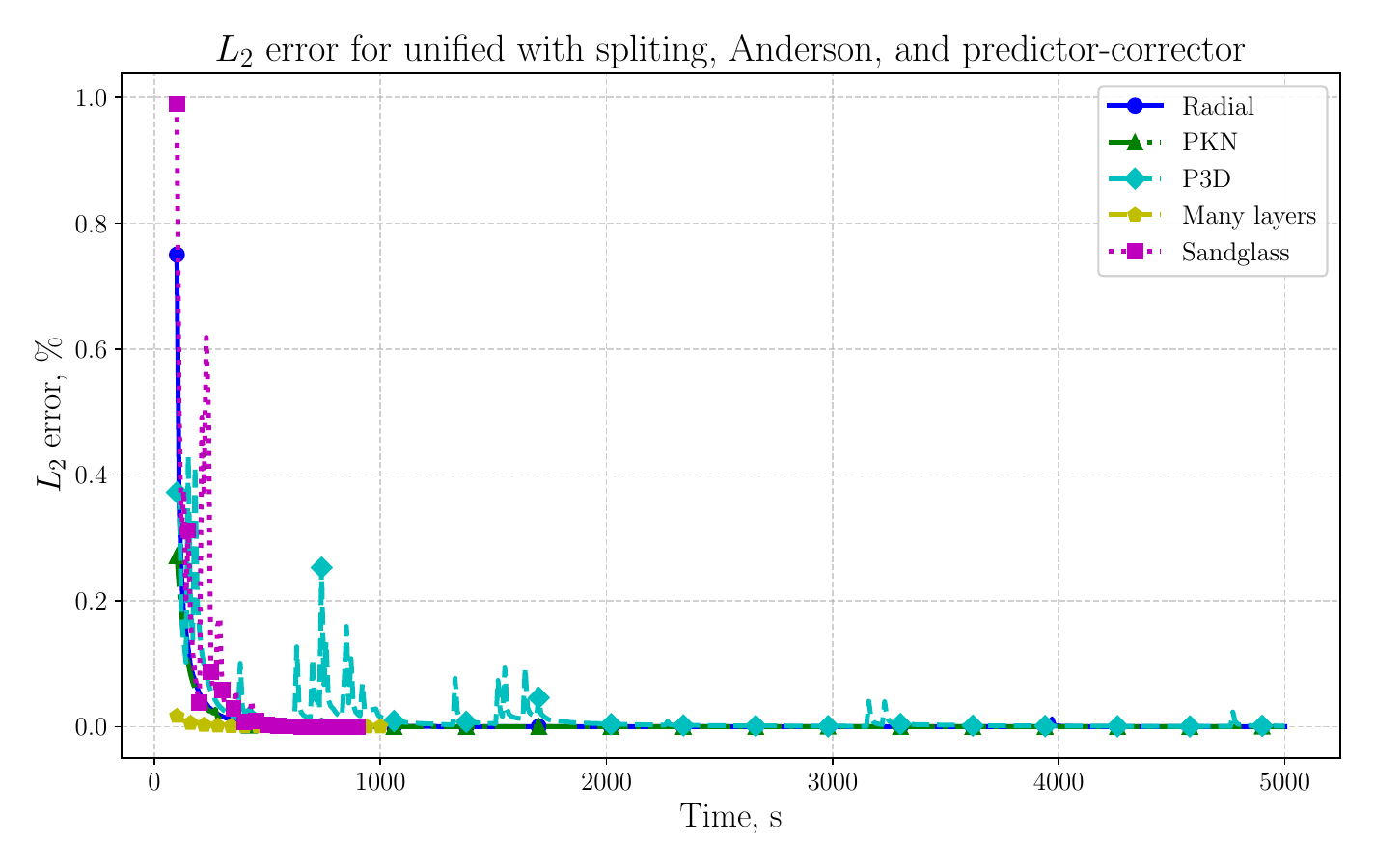}
        \caption{Relative $L_2$ aperture error of the unified planar 3D ILSA scheme with the matrix splitting, Anderson acceleration, and the predictor--corrector scheme during the propagation stage.}
        \label{fig:supp_open_l2_unified_all}
    \end{figure}

    \begin{figure}[H]
        \centering
        \includegraphics[width=0.9\textwidth]{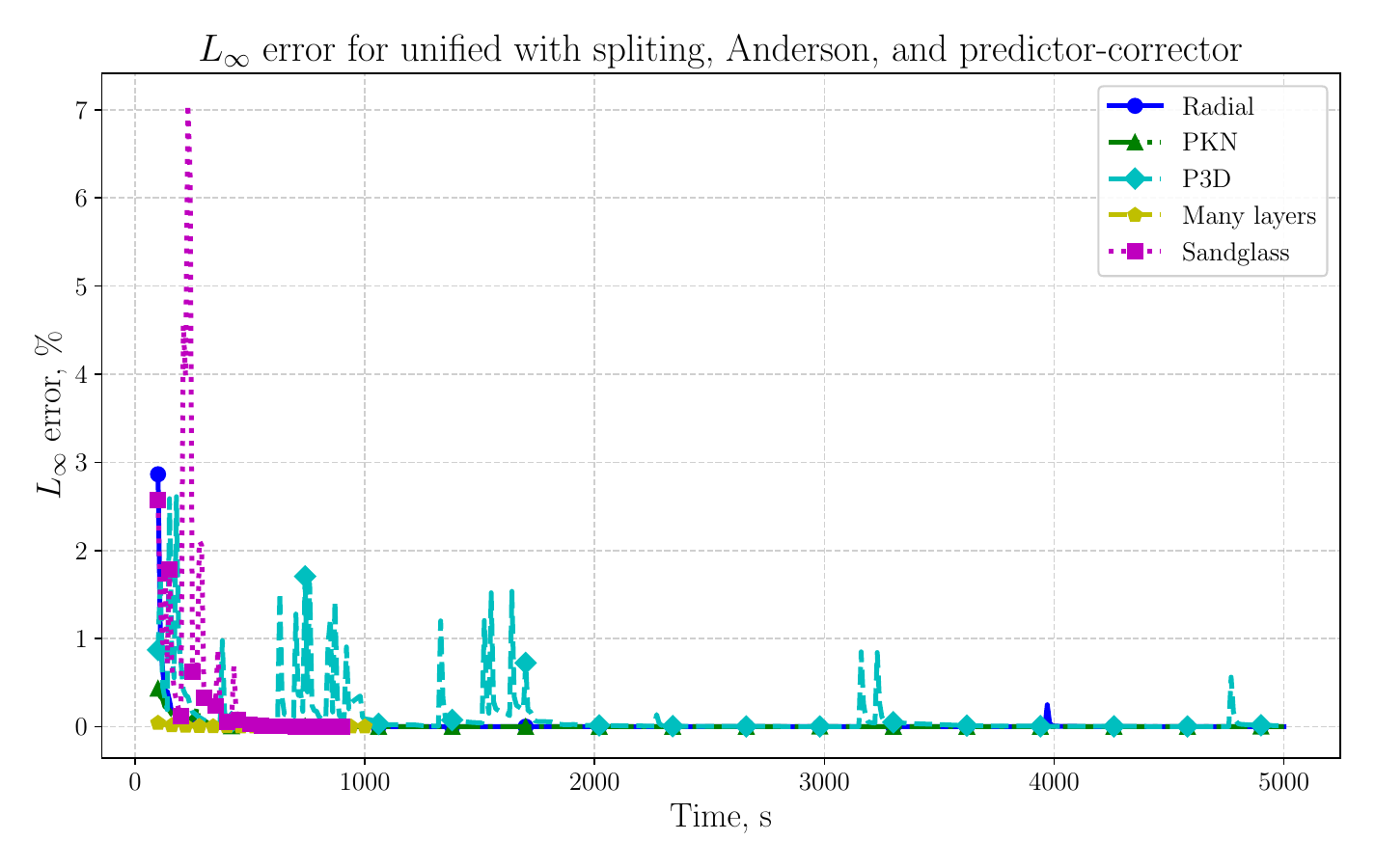}
        \caption{Relative $L_\infty$ aperture error of the unified planar 3D ILSA scheme with the matrix splitting, Anderson acceleration, and the predictor--corrector scheme during the propagation stage.}
        \label{fig:supp_open_linf_unified_all}
    \end{figure}

    \subsection{Localized deviations for the Sandglass case}\label{sec:supp_sandglass}

        For the Sandglass case the relative $L_\infty$ error stays below $3\%$ at most time steps but attains isolated peaks. Five of the reported instants exceed $5\%$: $9.5\%$ and $9.0\%$ at $t = 480$ and $490$~s for the reference scheme with the matrix splitting, $7.1\%$ and $6.2\%$ at $t = 230$ and $240$~s for the unified scheme with all three techniques, and $5.3\%$ at $t = 620$~s for the unified scheme with Anderson acceleration. Each peak spans one or two time steps, after which the error returns to its previous level. The relative $L_2$ error stays below $1\%$ at the same instants, which indicates that the deviation is not distributed over the fracture footprint.

\begin{figure}[H]
            \centering
            \includegraphics[width=0.9\textwidth]{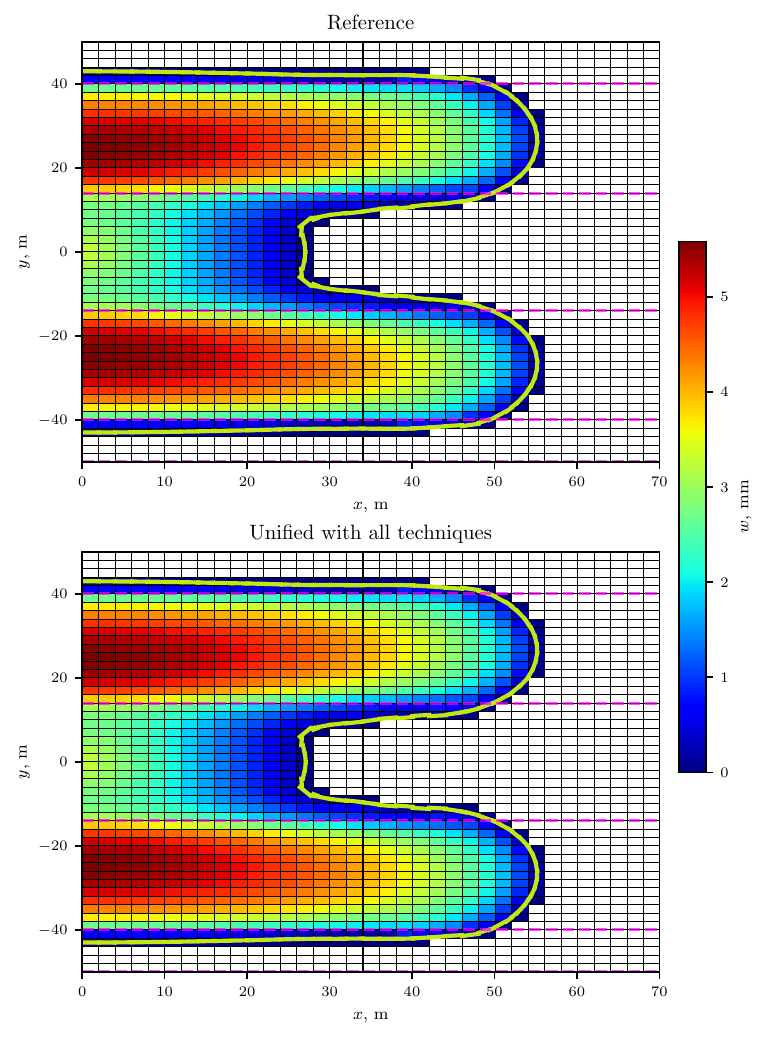}
            \caption{Fracture aperture at $t = 230$~s for the Sandglass case obtained with the reference planar 3D ILSA scheme and with the unified scheme with the matrix splitting, Anderson acceleration, and the predictor--corrector scheme. Both panels share the same color scale. The solid yellow line indicates the fracture front, the dashed magenta lines indicate the boundaries of the stress layers, and the thin black lines indicate the mesh.}
            \label{fig:supp_sandglass_apertures}
        \end{figure}

\begin{figure}[H]
            \centering
            \includegraphics[width=0.9\textwidth]{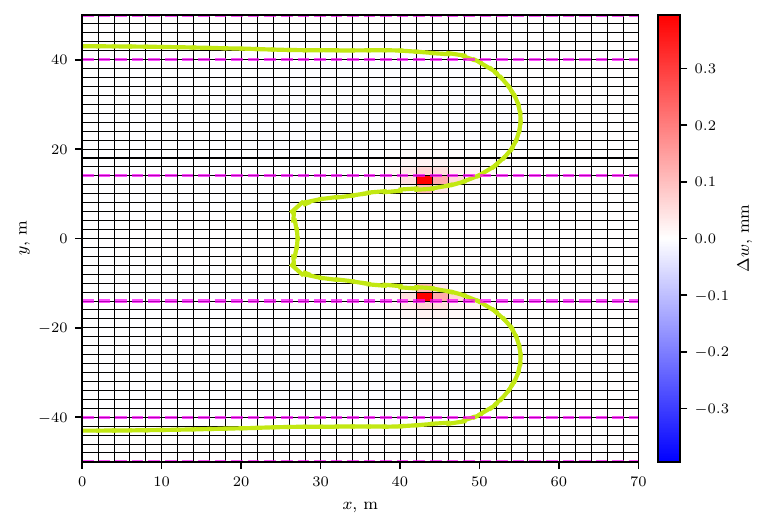}
            \caption{Difference between the two aperture fields of \Cref{fig:supp_sandglass_apertures}, computed as the unified scheme with all three techniques minus the reference planar 3D ILSA scheme. The discrepancy is confined to a single element adjacent to the fracture front. The solid yellow line indicates the fracture front, the dashed magenta lines indicate the boundaries of the stress layers, and the thin black lines indicate the mesh.}
            \label{fig:supp_sandglass_difference}
        \end{figure}

        One of these peaks, that of the unified scheme with all three techniques at $t = 230$~s, is examined in \Cref{fig:supp_sandglass_apertures,fig:supp_sandglass_difference}. \Cref{fig:supp_sandglass_apertures} shows the aperture fields of the reference scheme and of the unified scheme with all techniques applied at this instant: the two footprints coincide and the distributions are visually indistinguishable. \Cref{fig:supp_sandglass_difference} shows the difference between the same two fields, which is confined to a single element adjacent to the front. This element alone accounts for the value $\delta_{L_\infty} = 7.1\%$, whereas $\delta_{L_2} = 0.62\%$ at the same instant. The remaining four peaks listed above were inspected in the same manner and display the same pattern: a single element adjacent to the front, with the rest of the aperture field unaffected.

        Deviations of this kind may be produced by the treatment of the near-tip region. The aperture in a tip element is prescribed by a near-tip asymptotic solution, and the relation employed depends on whether the front is locally propagating or locally arrested. The two regimes are separated by a criterion on the local front velocity, so in an element whose front advances marginally a difference between two converged solutions may be sufficient to place them in different regimes. The aperture in that element is then obtained from different relations, and a discrepancy localized in a single cell appears. Since $\delta_{L_\infty}$ is governed by the cell with the largest deviation, such an event is reflected in the $L_\infty$ history in full, whereas its imprint on $\delta_{L_2}$ remains limited.

        The Sandglass configuration is the only benchmark in which the front advances at strongly different rates along its length, since the fracture is initiated in a high-stress layer and remains confined there while it propagates in the neighboring layers. Consequently, segments of the front advancing at extremely low rates are present at all stages of the simulation, which may explain why deviations of this kind are observed for this case alone. On this evidence the peaks reported above are attributed to the sensitivity of the near-tip treatment to the converged solution rather than to a loss of accuracy of the acceleration techniques themselves.

\section{Front iteration counts for the matrix splitting and Anderson acceleration}\label{sec:supp_front_iterations}

\begin{figure}[H]
        \centering
        \includegraphics[width=0.9\textwidth]{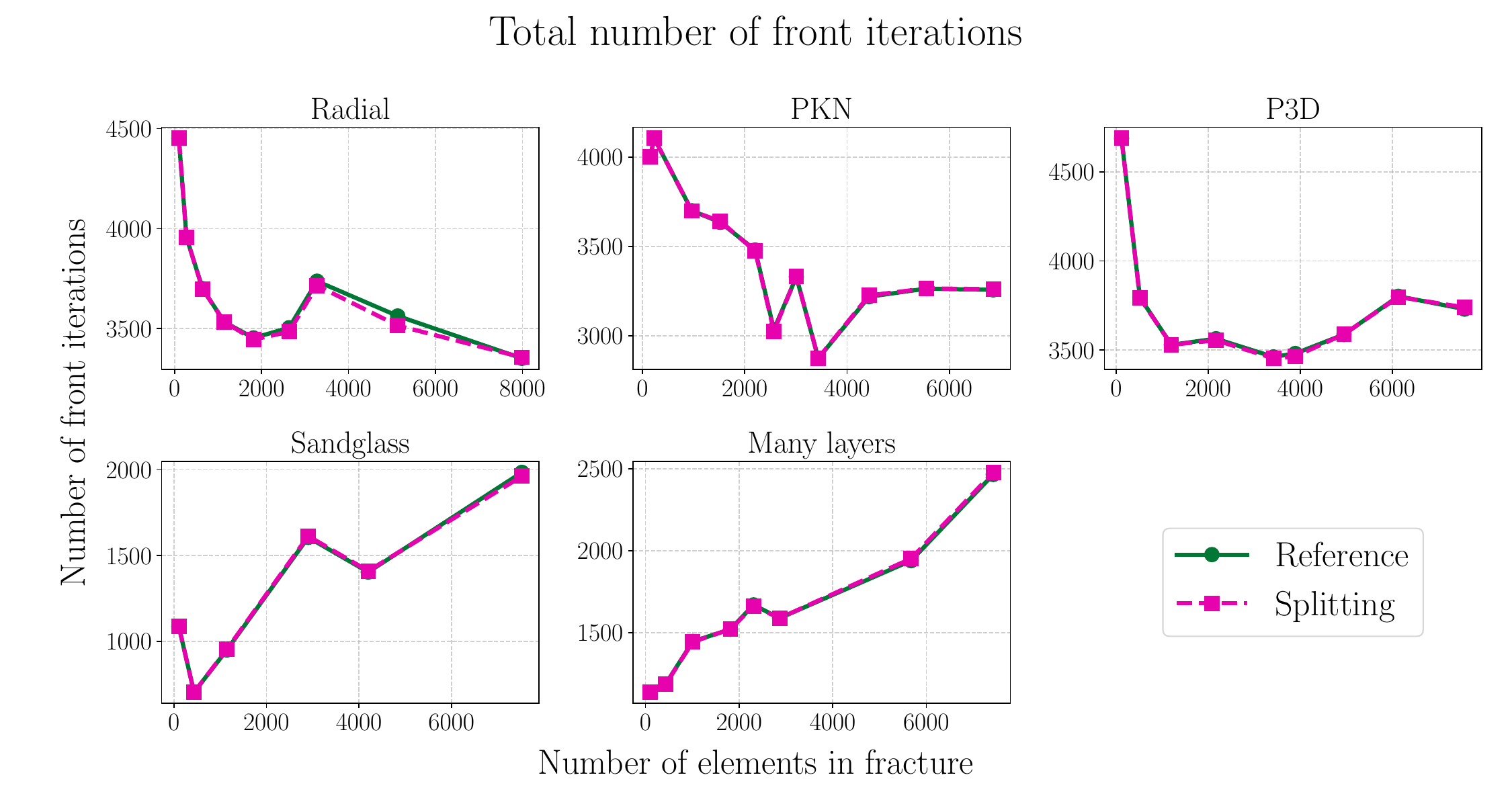}
        \caption{Total front iteration count of the reference planar 3D ILSA scheme and the reference scheme with the matrix splitting against the number of fracture elements for the benchmark cases.}
        \label{fig:supp_iterations_front_split}
    \end{figure}

\begin{figure}[H]
        \centering
        \includegraphics[width=0.9\textwidth]{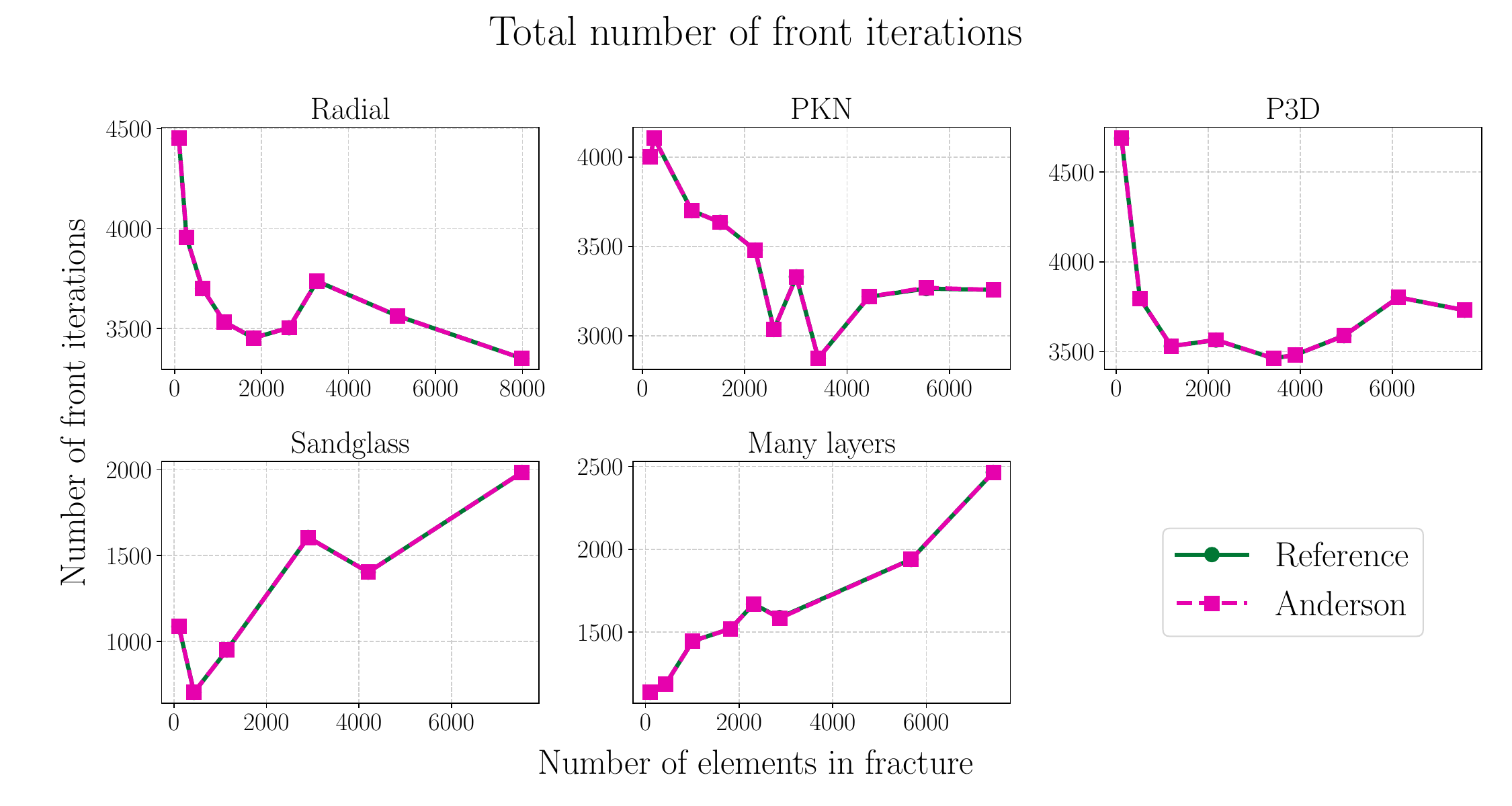}
        \caption{Total front iteration count of the reference planar 3D ILSA scheme and the reference scheme with Anderson acceleration against the number of fracture elements for the benchmark cases.}
        \label{fig:supp_iterations_front_anderson}
    \end{figure}

    This section reports the total front iteration count of the reference planar 3D ILSA scheme with the matrix splitting (\Cref{fig:supp_iterations_front_split}) and with Anderson acceleration (\Cref{fig:supp_iterations_front_anderson}), referenced in the main manuscript. In both cases the front iteration count stays close to that of the reference scheme, confirming that neither technique has a noticeable effect on the front tracking process.

\section{Accuracy during the recession and closure stage}\label{sec:supp_closure}

    This section reports the accuracy of the accelerated algorithms during the recession and closure stage described in the main manuscript. Two configurations are considered, P3D and Many Layers, both with the pumping schedule extended by a shut-in stage. As in the main manuscript, every variant employs the reference planar 3D ILSA scheme throughout propagation, and the corresponding technique is switched on only from the arrest of the front onward, so that the comparison isolates the effect of each technique on the closure phase itself. The errors are evaluated at every time step of the closure phase, from the arrest of the fracture front at $t_a = 5330$~s for the P3D case and $t_a = 2120$~s for the Many Layers case to the end of the simulation. In every figure of this section the P3D case is shown by a solid blue line and the Many Layers case by a dashed orange line.

\begin{figure}[H]
        \centering
        \includegraphics[width=0.9\textwidth]{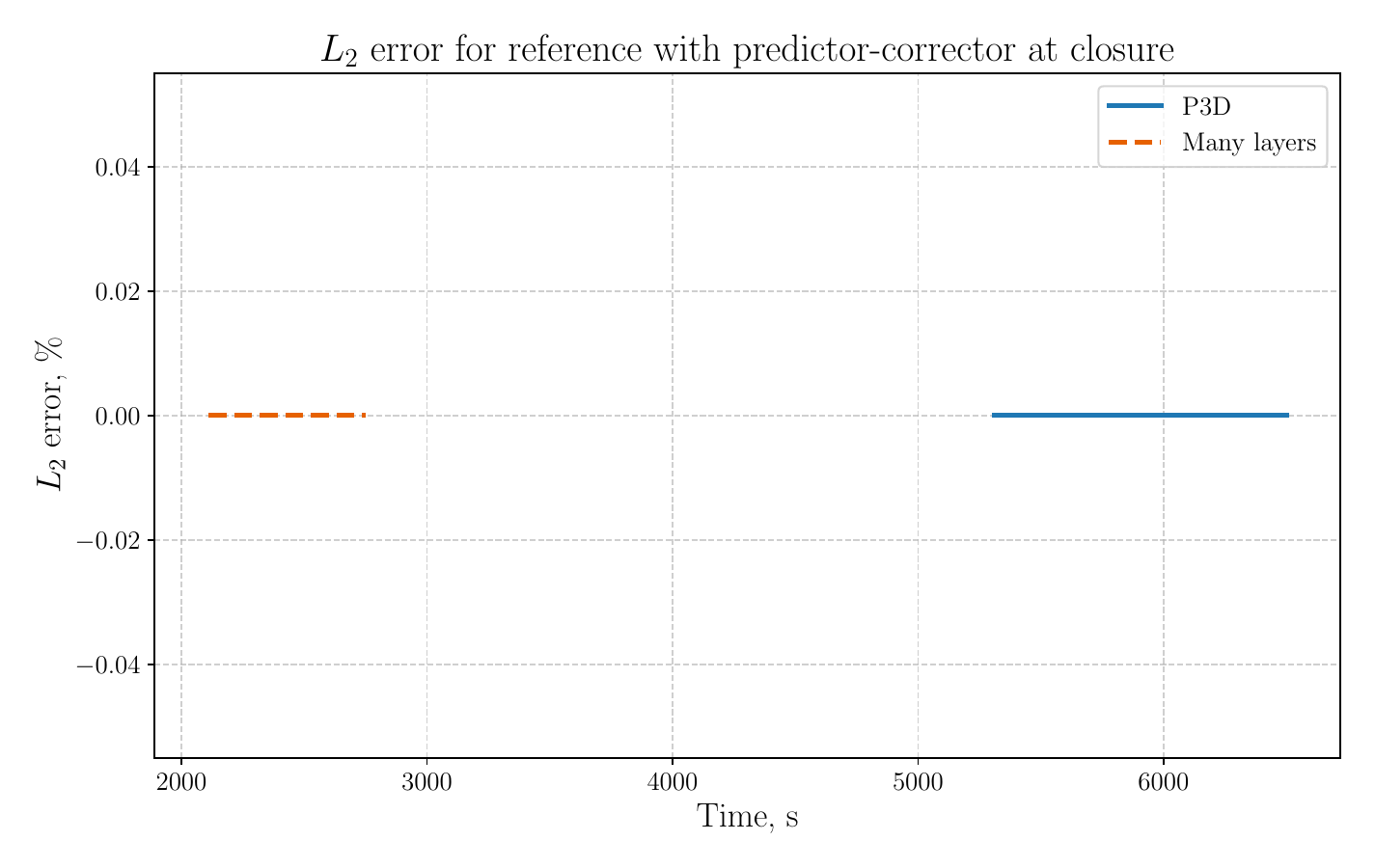}
        \caption{Relative $L_2$ aperture error of the reference planar 3D ILSA scheme with the predictor--corrector scheme during the closure phase.}
        \label{fig:supp_l2_ref_pc}
    \end{figure}

\begin{figure}[H]
        \centering
        \includegraphics[width=0.9\textwidth]{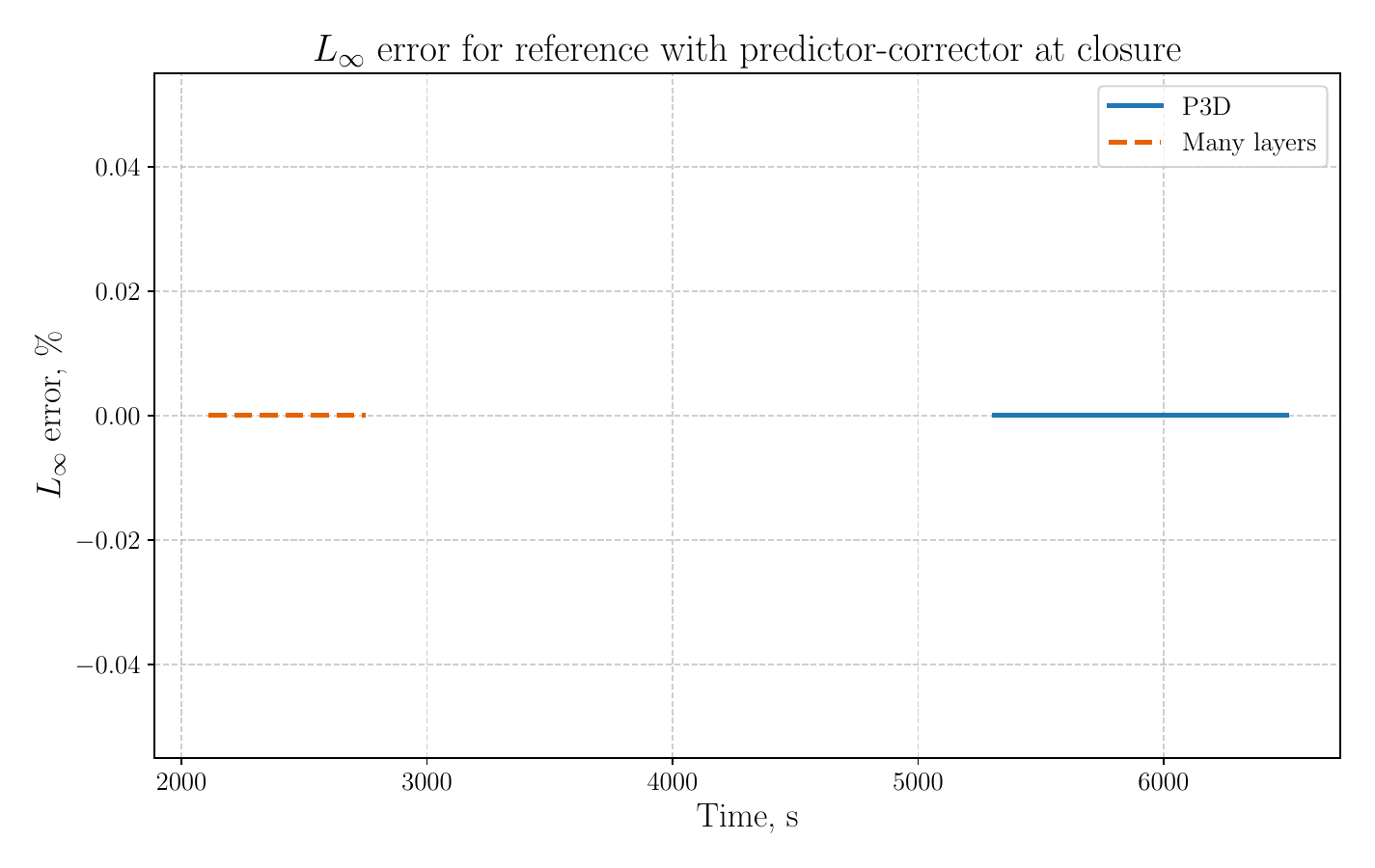}
        \caption{Relative $L_\infty$ aperture error of the reference planar 3D ILSA scheme with the predictor--corrector scheme during the closure phase.}
        \label{fig:supp_linf_ref_pc}
    \end{figure}

\begin{figure}[H]
        \centering
        \includegraphics[width=0.9\textwidth]{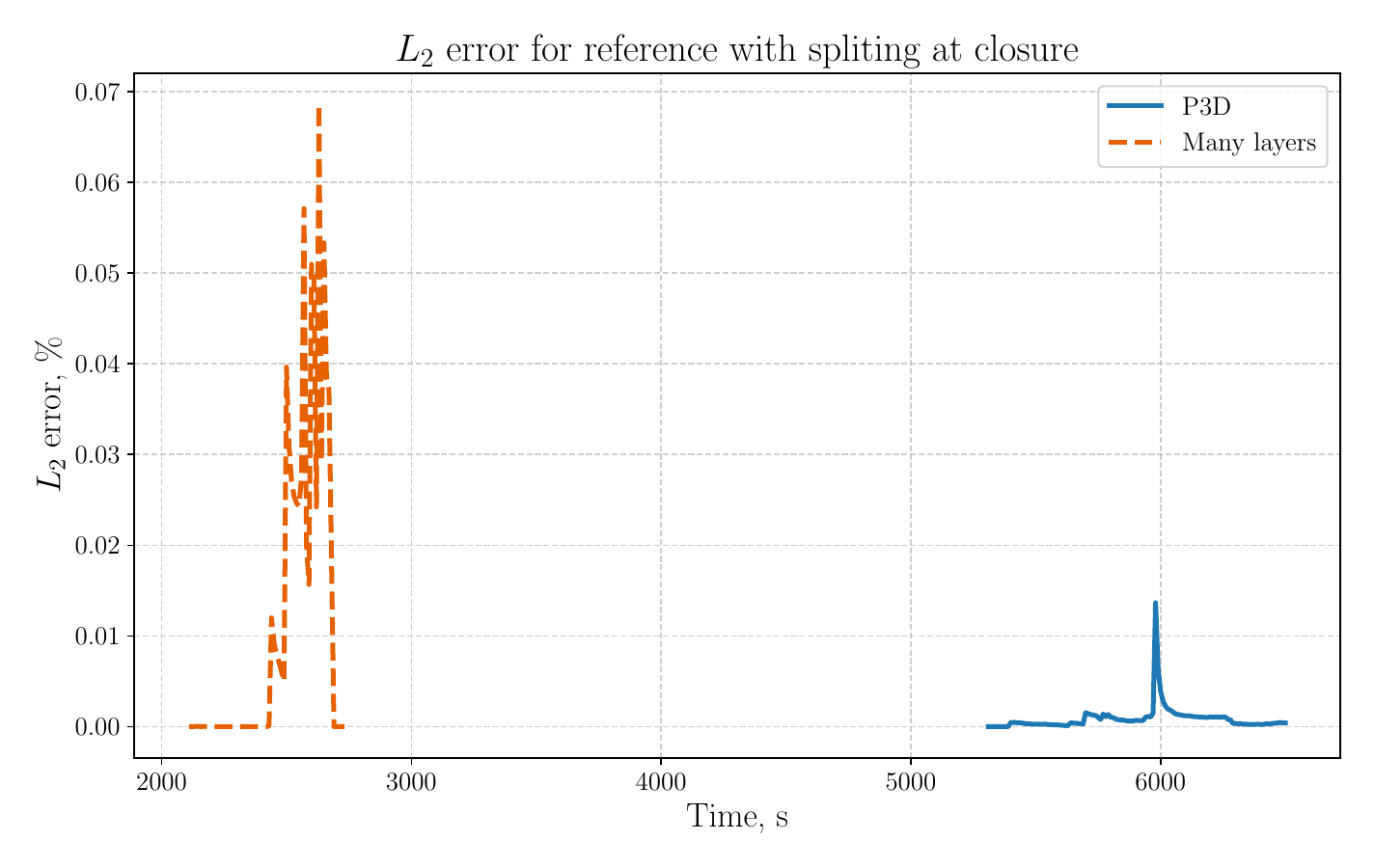}
        \caption{Relative $L_2$ aperture error of the reference planar 3D ILSA scheme with the matrix splitting during the closure phase.}
        \label{fig:supp_l2_ref_split}
    \end{figure}

\begin{figure}[H]
        \centering
        \includegraphics[width=0.9\textwidth]{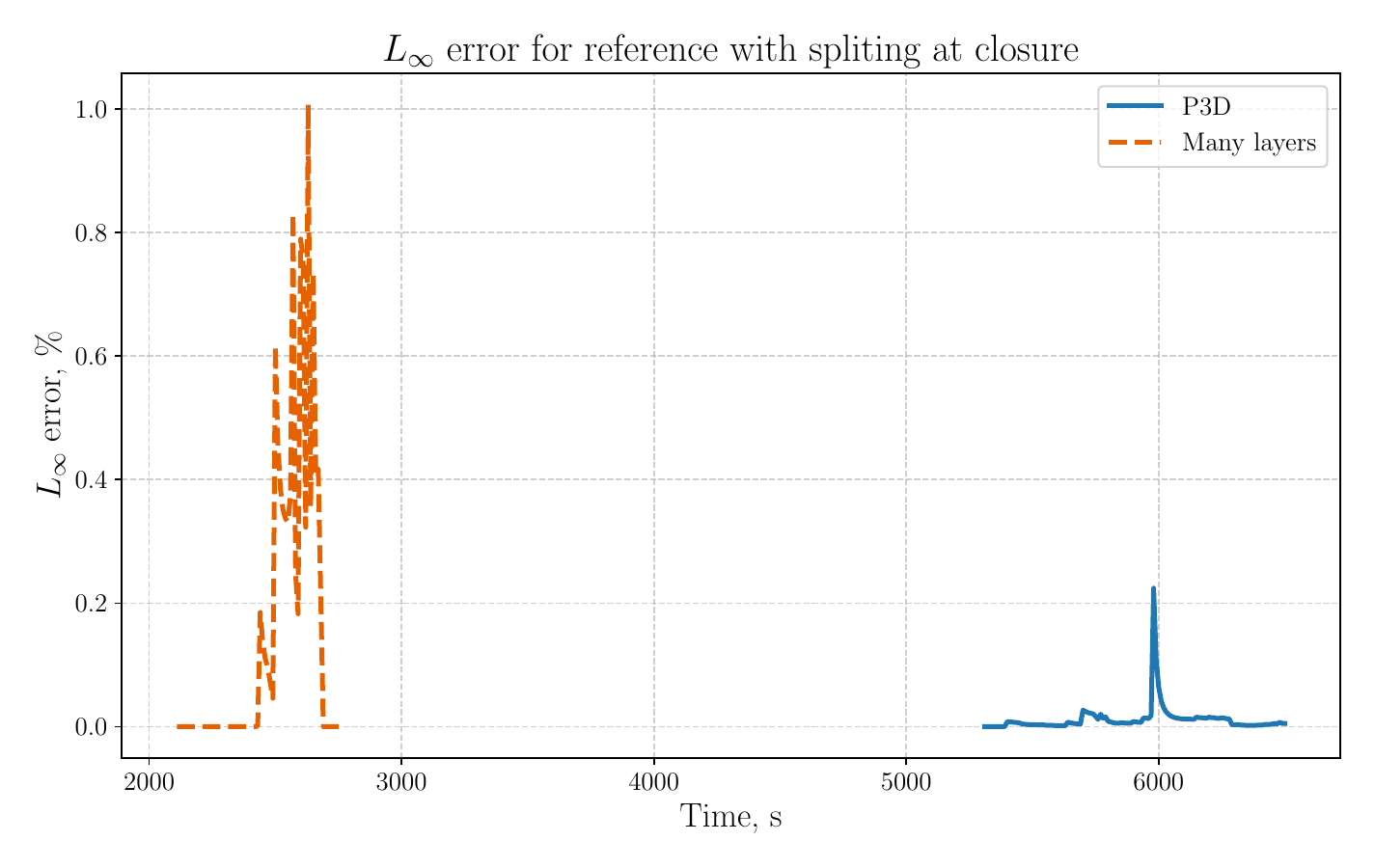}
        \caption{Relative $L_\infty$ aperture error of the reference planar 3D ILSA scheme with the matrix splitting during the closure phase.}
        \label{fig:supp_linf_ref_split}
    \end{figure}

\begin{figure}[H]
        \centering
        \includegraphics[width=0.9\textwidth]{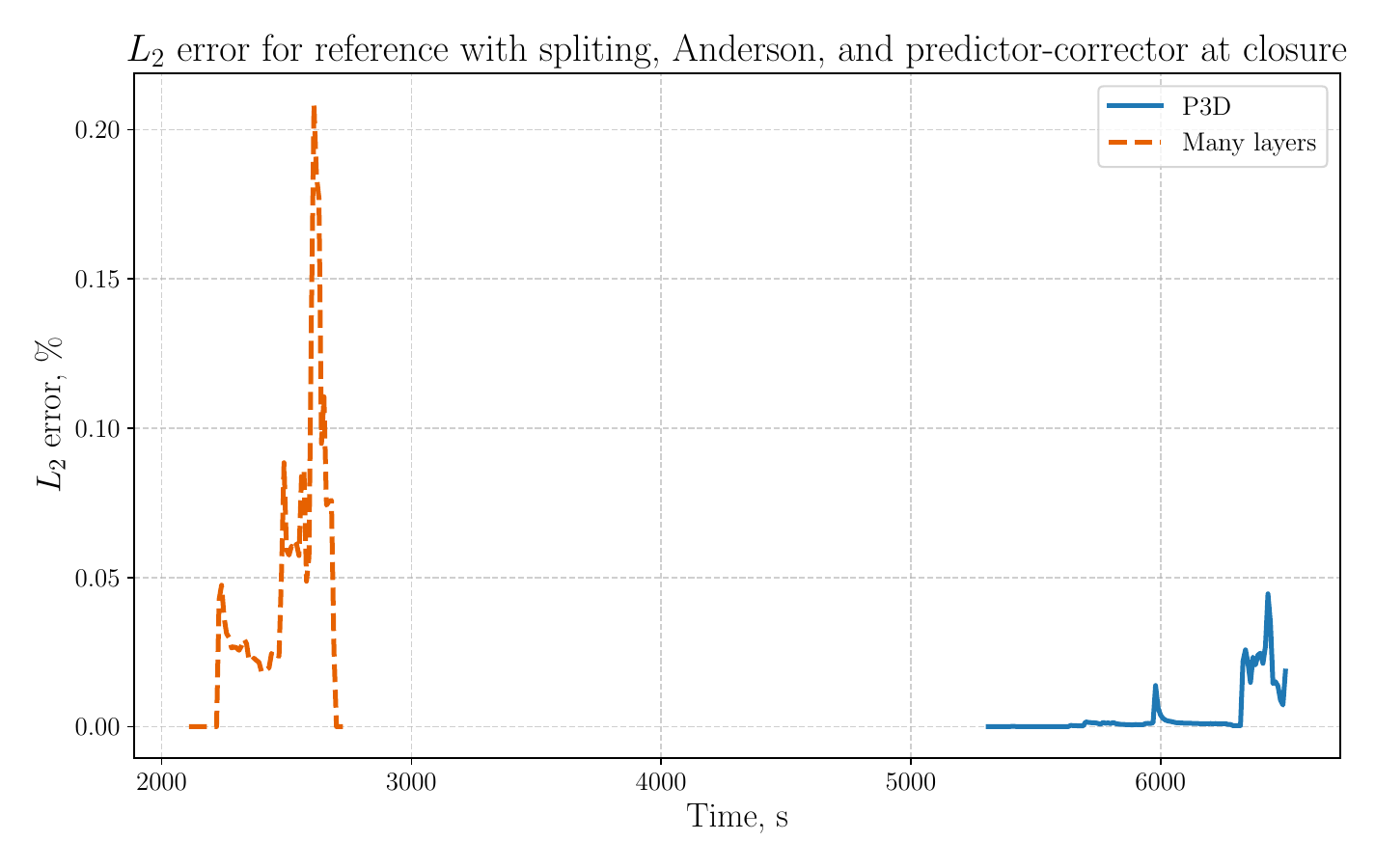}
        \caption{Relative $L_2$ aperture error of the reference planar 3D ILSA scheme with the matrix splitting, Anderson acceleration, and the predictor--corrector scheme during the closure phase.}
        \label{fig:supp_l2_ref_all}
    \end{figure}

\begin{figure}[H]
        \centering
        \includegraphics[width=0.9\textwidth]{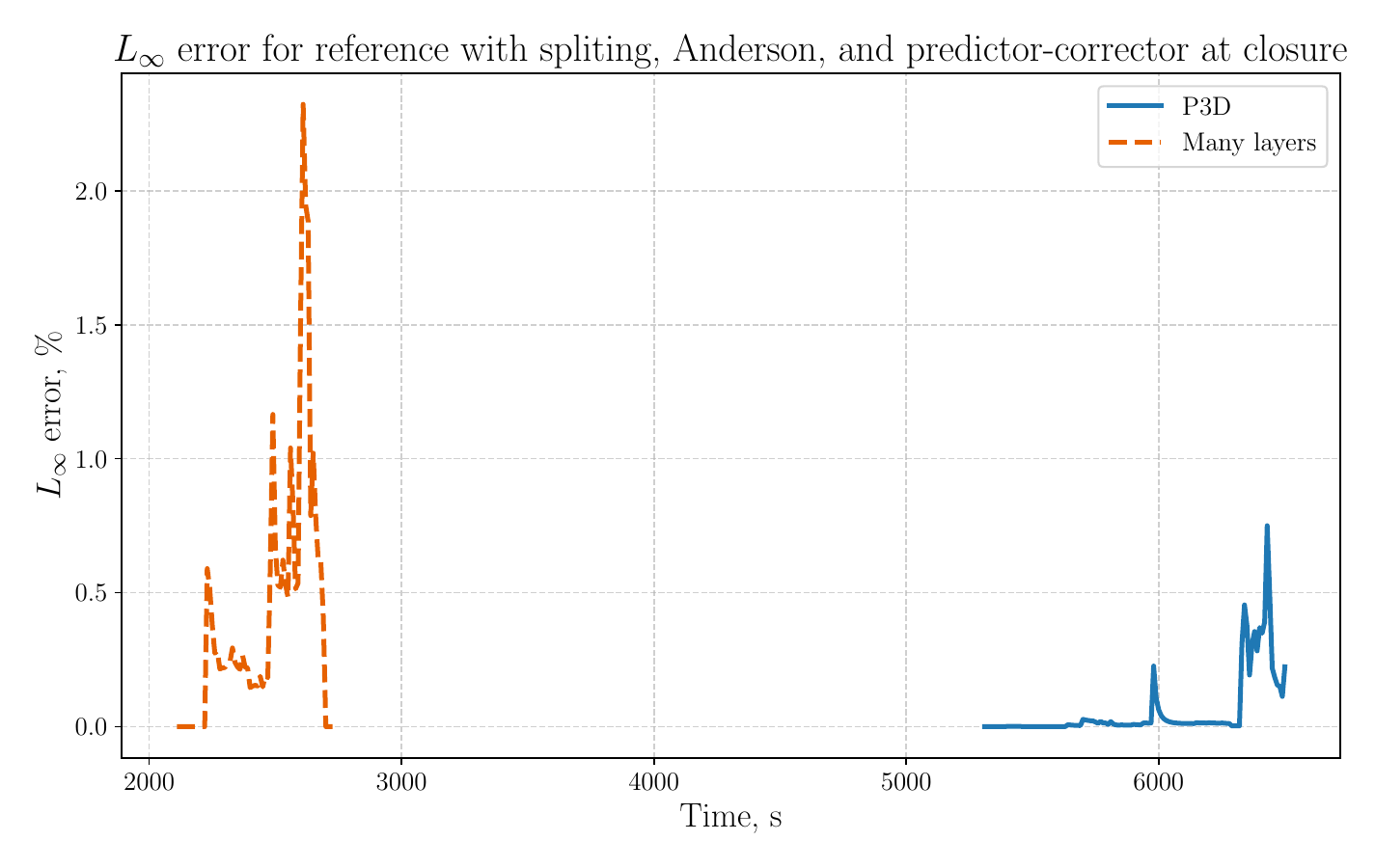}
        \caption{Relative $L_\infty$ aperture error of the reference planar 3D ILSA scheme with the matrix splitting, Anderson acceleration, and the predictor--corrector scheme during the closure phase.}
        \label{fig:supp_linf_ref_all}
    \end{figure}

\begin{figure}[H]
        \centering
        \includegraphics[width=0.9\textwidth]{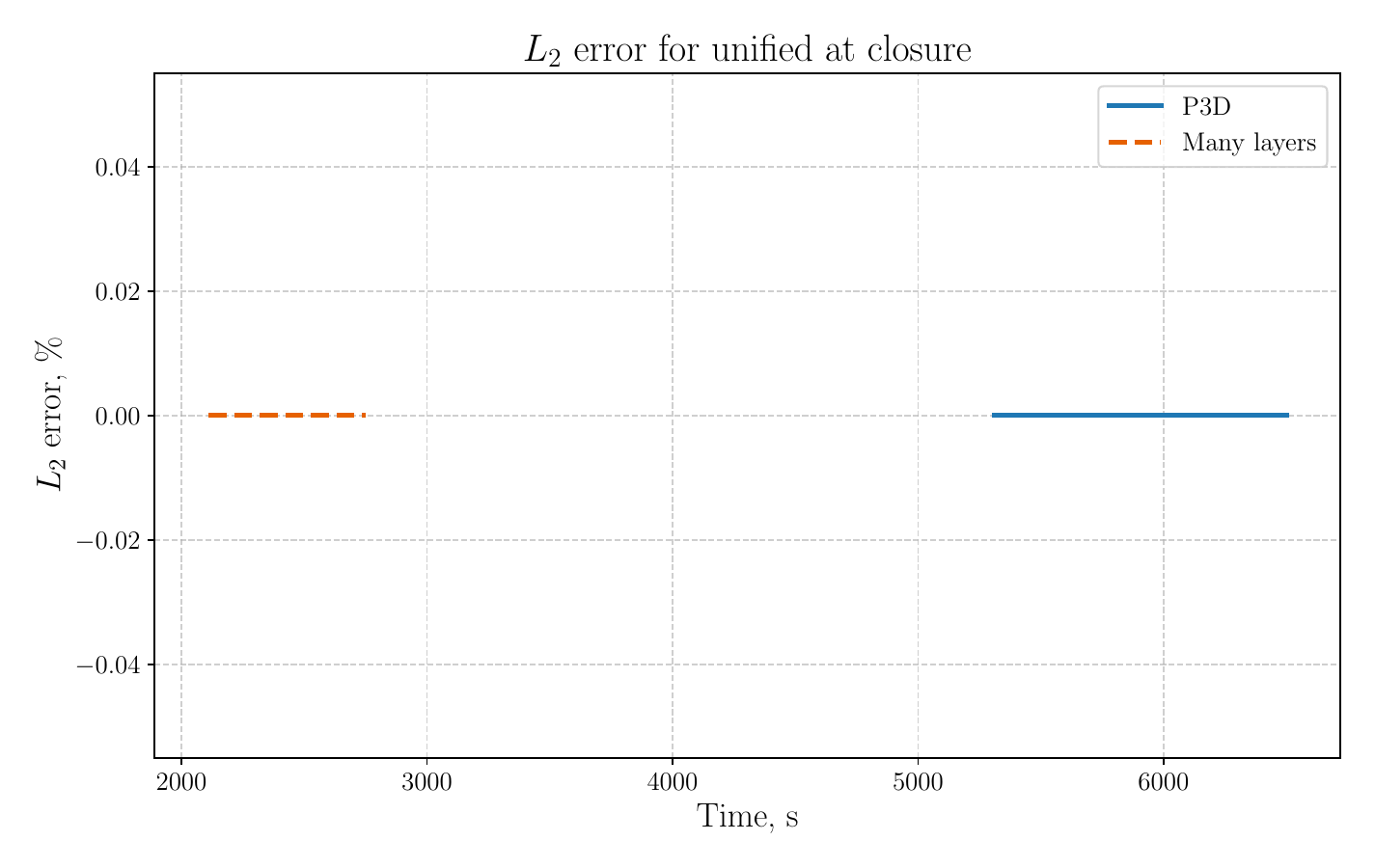}
        \caption{Relative $L_2$ aperture error of the unified planar 3D ILSA scheme during the closure phase.}
        \label{fig:supp_l2_unified}
    \end{figure}

\begin{figure}[H]
        \centering
        \includegraphics[width=0.9\textwidth]{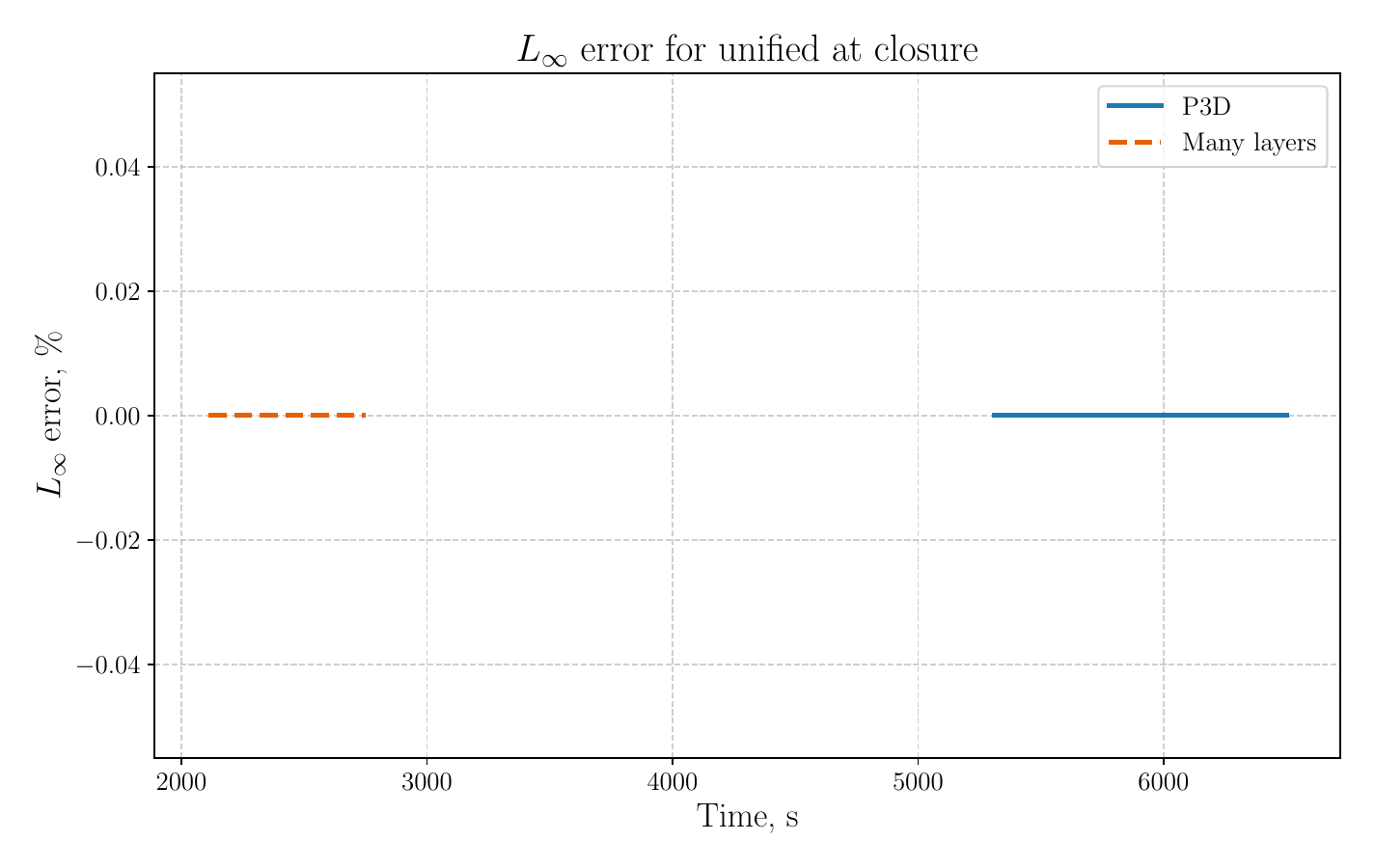}
        \caption{Relative $L_\infty$ aperture error of the unified planar 3D ILSA scheme during the closure phase.}
        \label{fig:supp_linf_unified}
    \end{figure}

\begin{figure}[H]
        \centering
        \includegraphics[width=0.9\textwidth]{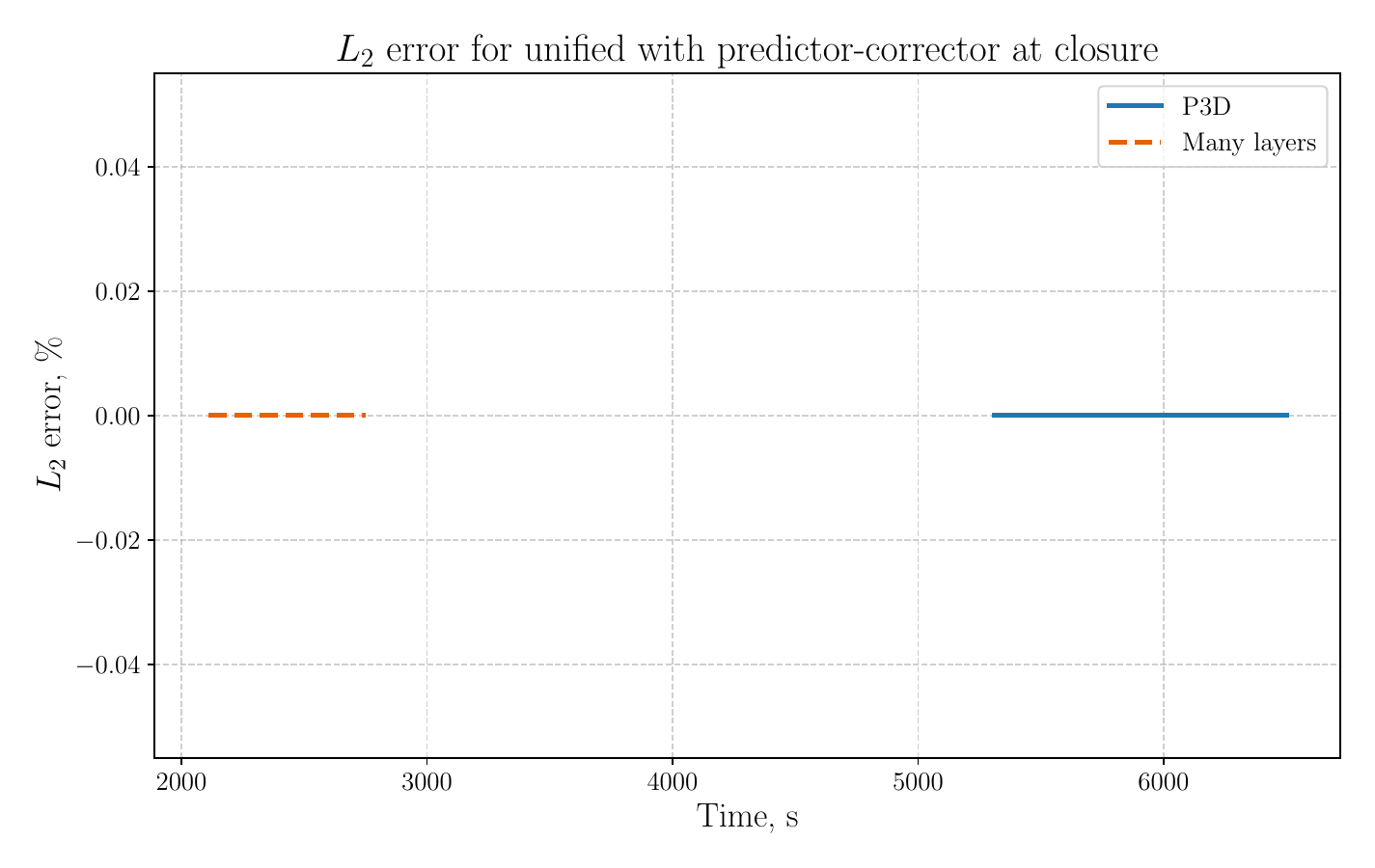}
        \caption{Relative $L_2$ aperture error of the unified planar 3D ILSA scheme with the predictor--corrector scheme during the closure phase.}
        \label{fig:supp_l2_unified_pc}
    \end{figure}

\begin{figure}[H]
        \centering
        \includegraphics[width=0.9\textwidth]{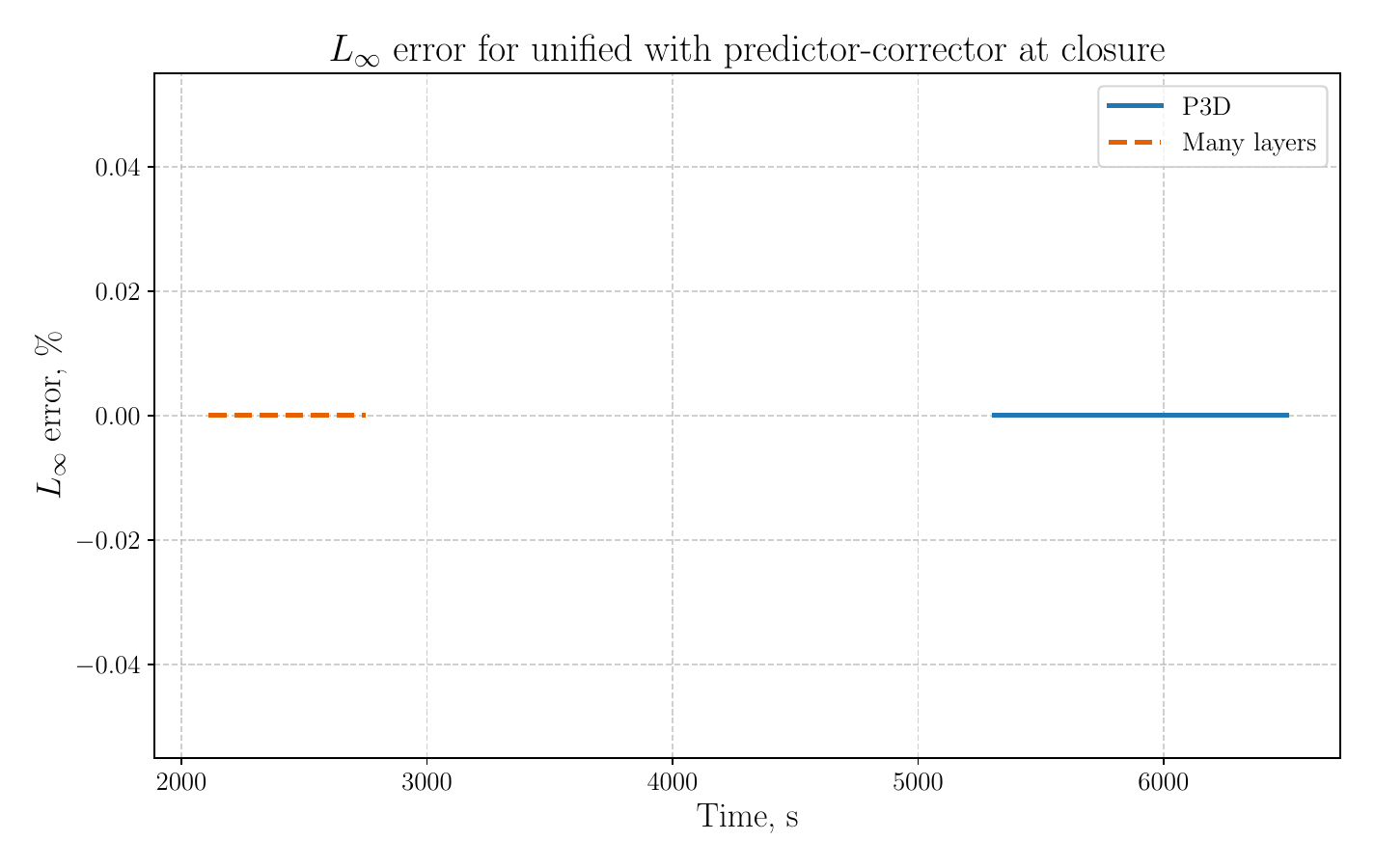}
        \caption{Relative $L_\infty$ aperture error of the unified planar 3D ILSA scheme with the predictor--corrector scheme during the closure phase.}
        \label{fig:supp_linf_unified_pc}
    \end{figure}

\begin{figure}[H]
        \centering
        \includegraphics[width=0.9\textwidth]{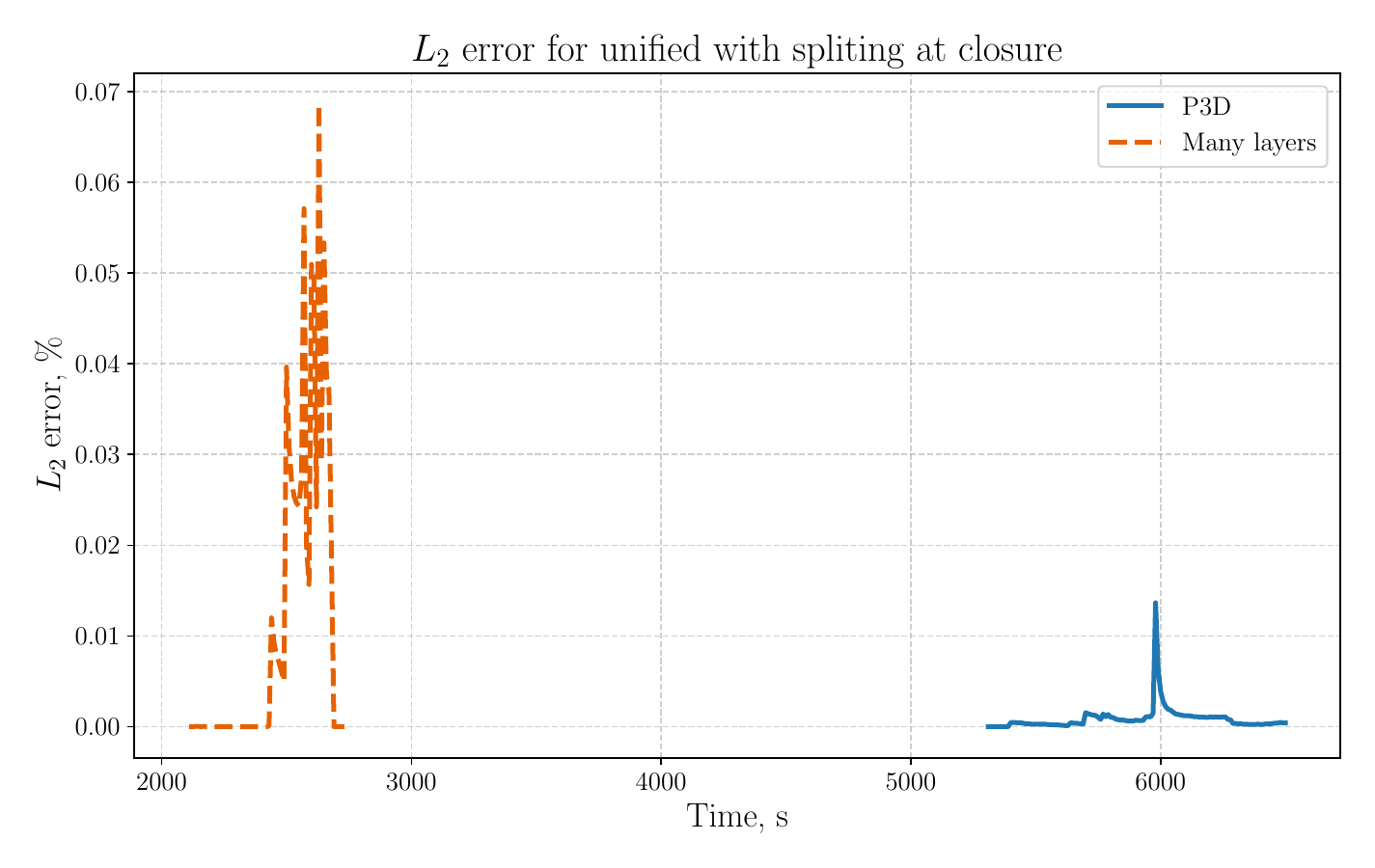}
        \caption{Relative $L_2$ aperture error of the unified planar 3D ILSA scheme with the matrix splitting during the closure phase.}
        \label{fig:supp_l2_unified_split}
    \end{figure}

\begin{figure}[H]
        \centering
        \includegraphics[width=0.9\textwidth]{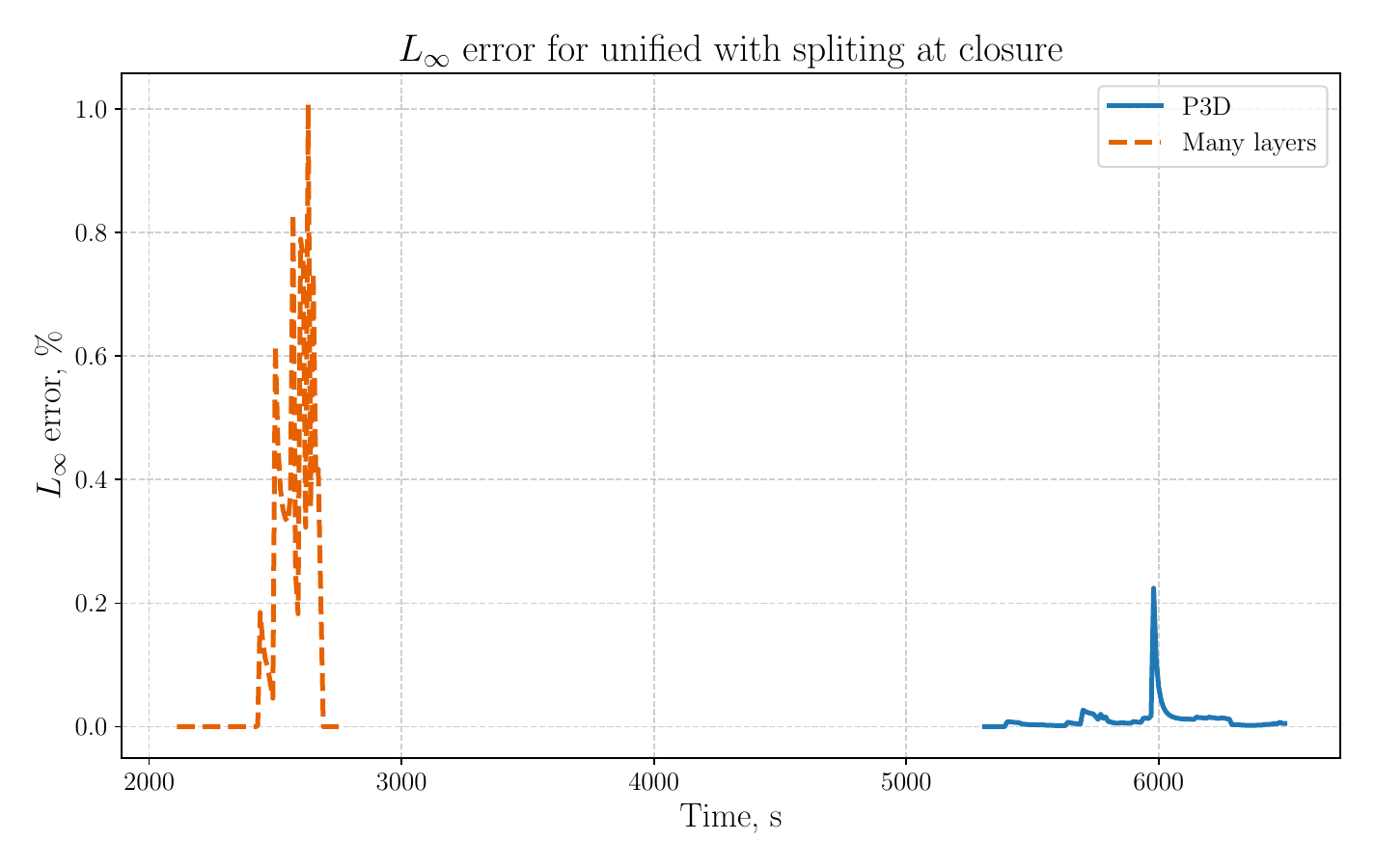}
        \caption{Relative $L_\infty$ aperture error of the unified planar 3D ILSA scheme with the matrix splitting during the closure phase.}
        \label{fig:supp_linf_unified_split}
    \end{figure}

\begin{figure}[H]
        \centering
        \includegraphics[width=0.9\textwidth]{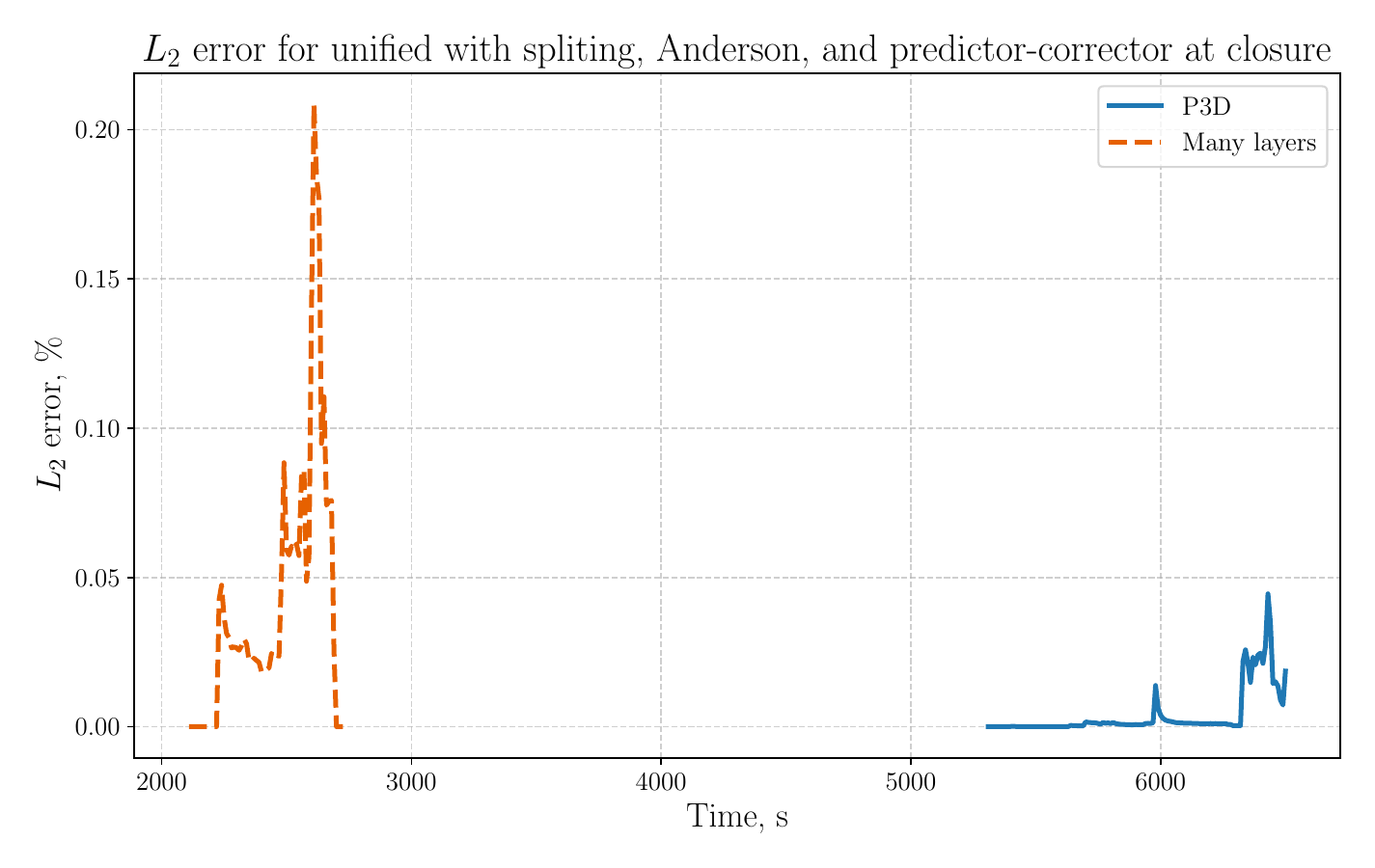}
        \caption{Relative $L_2$ aperture error of the unified planar 3D ILSA scheme with the matrix splitting, Anderson acceleration, and the predictor--corrector scheme during the closure phase.}
        \label{fig:supp_l2_unified_all}
    \end{figure}

\begin{figure}[H]
        \centering
        \includegraphics[width=0.9\textwidth]{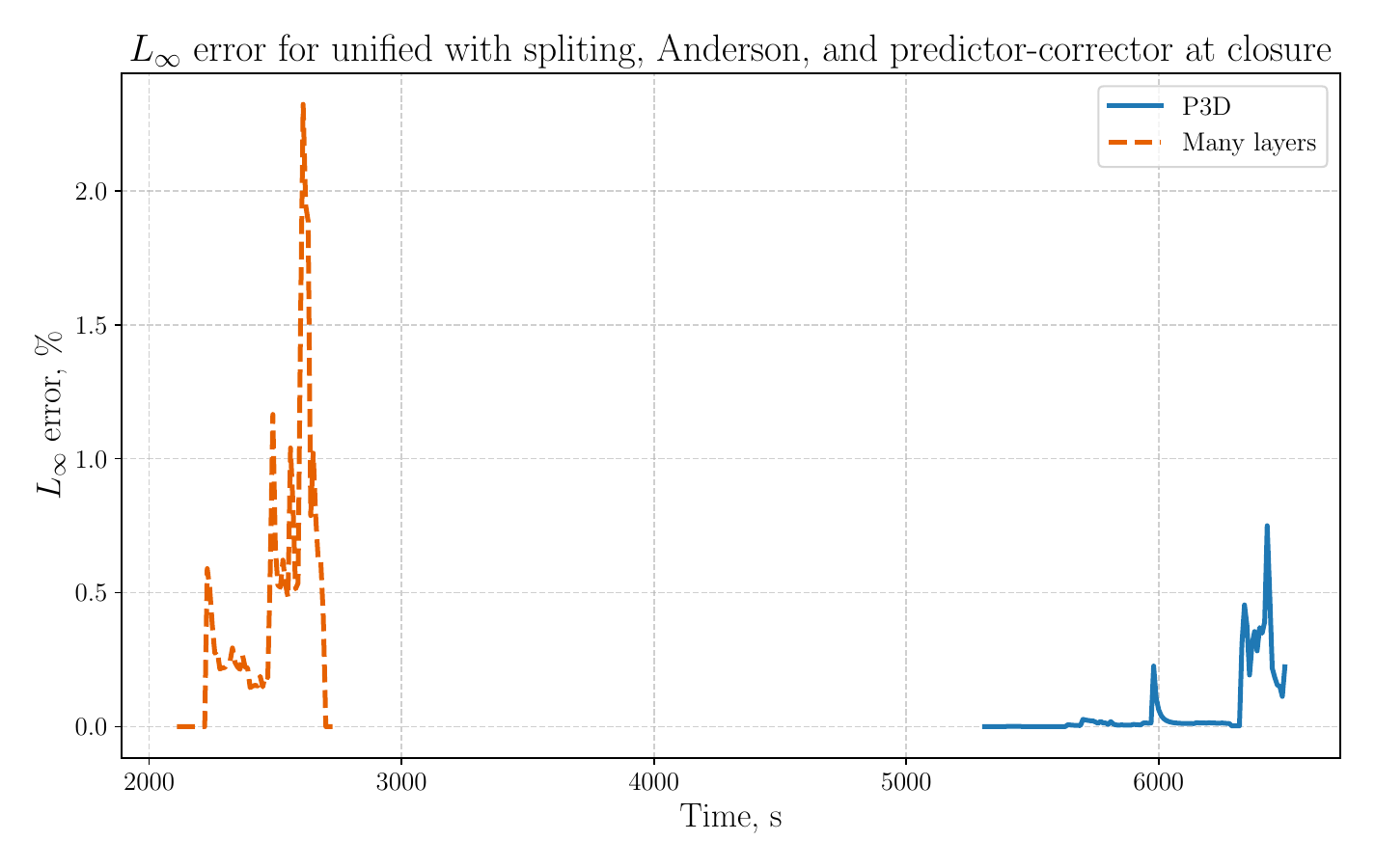}
        \caption{Relative $L_\infty$ aperture error of the unified planar 3D ILSA scheme with the matrix splitting, Anderson acceleration, and the predictor--corrector scheme during the closure phase.}
        \label{fig:supp_linf_unified_all}
    \end{figure}

    \Cref{fig:supp_l2_ref_pc,fig:supp_linf_ref_pc,fig:supp_l2_ref_split,fig:supp_linf_ref_split,fig:supp_l2_ref_all,fig:supp_linf_ref_all} report the $L_2$ and $L_\infty$ errors of the acceleration techniques applied to the reference planar 3D ILSA scheme, and \Cref{fig:supp_l2_unified,fig:supp_linf_unified,fig:supp_l2_unified_pc,fig:supp_linf_unified_pc,fig:supp_l2_unified_split,fig:supp_linf_unified_split,fig:supp_l2_unified_all,fig:supp_linf_unified_all} report the same quantities for the unified planar 3D ILSA scheme and the techniques applied to it. The predictor--corrector variants of both schemes, as well as the unified scheme on its own, reproduce the reference aperture field exactly, with zero $L_2$ and $L_\infty$ error at every time step and for both cases. The predictor--corrector scheme has no effect once the front is arrested, and the unified scheme reduces during closure to the same fixed-point problem as the reference scheme, so no discrepancy can arise. For the matrix splitting, Anderson acceleration, and their combination, over the whole closure phase the relative $L_2$ error does not exceed $0.05\%$ for the P3D case and $0.22\%$ for the Many Layers case, and the relative $L_\infty$ error does not exceed $0.76\%$ for the P3D case and $2.33\%$ for the Many Layers case, identically for the reference and unified formulations of each technique.

\bibliographystyle{elsarticle-num}
\bibliography{References}